\documentclass[aps,prl,twocolumn,showpacs,psfig,superscriptaddress,longbibliography]{revtex4-1}
\usepackage{textcomp}
\usepackage{times}
\usepackage{graphicx}
\usepackage{float}
\usepackage{latexsym,amsmath,amssymb,bm,euscript}
\usepackage{color}
\usepackage{subfigure}
\usepackage{epstopdf}
\usepackage{comment}
\usepackage[colorlinks=true,linkcolor=blue,citecolor=blue,urlcolor=blue]{hyperref}

\usepackage[normalem]{ulem}
\usepackage{mathrsfs}
\usepackage{amsmath,amssymb}
\usepackage{lettrine}
\usepackage{xfrac}
\usepackage{xspace}
\usepackage{textcomp}
\usepackage{wasysym}
\usepackage{xr}
\usepackage{booktabs}
\usepackage{tabularx,array}
\newcolumntype{C}{>{\centering\arraybackslash}X}
\usepackage{diagbox}
\usepackage{lineno}

\usepackage{multirow}
\graphicspath{{./Figs/}}

\begin{document}

\title{Collective excitations in chiral spin liquid: chiral roton and long-wavelength nematic mode
}
\author{Hongyu Lu}
\affiliation{New Cornerstone Science Lab, Department of Physics, The University of Hong Kong, Hong Kong, China}
\affiliation{HK Institute of Quantum Science \& Technology, The University of Hong Kong, Hong Kong, China}
\affiliation{State Key Laboratory of Optical Quantum Materials, The University of Hong Kong, Hong Kong, China}

\author{Wei Zhu}
\affiliation{Department of Physics, School of Science, Westlake University, Hangzhou 310030, China}

\author{Wang Yao}
\affiliation{New Cornerstone Science Lab, Department of Physics, The University of Hong Kong, Hong Kong, China}
\affiliation{HK Institute of Quantum Science \& Technology, The University of Hong Kong, Hong Kong, China}
\affiliation{State Key Laboratory of Optical Quantum Materials, The University of Hong Kong, Hong Kong, China}

\begin{abstract}
Chiral spin liquid (CSL) is a magnetic analogue of the fractional quantum Hall (FQH) liquid. 
Collective excitations play a vital role in shaping our understanding of these exotic quantum phases of matter and their quantum phase transitions. While the magneto-roton and long-wavelength chiral graviton modes in the FQH and fractional Chern insulator (FCI) liquids  have been extensively explored, 
whether CSLs host analogous or qualitatively different modes remains elusive.
Here we explore the collective excitations in the SU(2) symmetric CSL phase.
Combining exact diagonalization and time-dependent variational principle calculations, we identify
two spin-singlet collective modes: a chiral p-wave roton mode at finite momentum, and a elliptically polarized d-wave nematic mode at zero momentum, both of which are prominent across the CSL phase.
The chiral p-wave singlet roton has no counterpart in  FQH of FCI systems, and the $q=0$ d-wave mode also exhibits fingerprint distinct from those of FQH/FCI liquids.
We also elucidate that both singlet modes are general for CSLs on various lattice models.
By tuning $J_2$, we find the nematic mode to be pronouncedly soft, together with the spin-triplet two-spinon bound states, potentially promoting strong nematic and spin stripe instabilities.
Our work paves the way for further understanding CSL from the dynamical perspective and provides new spectroscopic signatures for future experiments  of CSL candidates.

\end{abstract}

\date{\today }
\maketitle

\noindent{\textcolor{blue}{\it Introduction.}---}
The study of quantum spin liquid (QSL) is a central theme in strongly correlated systems, with long-range entanglement, emergent gauge fields, and fractionalized excitations~\cite{Savary2017_QSL,Zhou2017_QSL,Broholm2020_QSL}.
One particularly striking class is the chiral spin liquids (CSLs) that break time-reversal symmetry (TRS) and realize topological order~\cite{Kalmeyer1987_RVB_FQH,Wen1989_chiral_spin_states}, providing a close magnetic analogue of fractional quantum Hall (FQH) liquids~\cite{Stormer1999_FQH_review,Yang1993_spin_liquid,Haldane1995_spin_currents}. 
The topological nature and ground-state (GS) properties of such CSLs are well established~\cite{nielsen2013,He2014_CSL,Gong2014_CSL,bauer2014,Wietek2017_CSL_triangular,Gong2017_QSL_triangle,Hu_2015}. 
However, the dynamical properties, especially the bulk collective excitations, still remain elusive, even for the simplest Kalmeyer–Laughlin (KL) CSL (magnetic realization of the bosonic 1/2 Laughlin state~\cite{Laughlin1983_FQH}), despite the understanding of fractionalized spinons and related spin-triplet bound states~\cite{Laughlin1989_spin_hamiltonian, Balatsky1991_csl_spin_rigidity,Zhu2019_spinon_DSF,Pichler2024_CSL_DSSF,drescher2025_dssf_triangle}.
This is an essential and fundamental question and would also provide insights to future experimental probes of such long-sought CSLs.

In FQH systems, the prototypical example of a
neutral collective excitation is the magneto-roton mode, normally with energy minima 
at finite wave vectors~\cite{Girvin1985,GMP1986,BoYang2013}. In the long-wavelength limit, a distinct collective mode emerges at higher energy: a chiral and quadrupolar orbital spin-2 excitation, also known as the chiral graviton mode~\cite{Haldane_2011,Yang2012_collective_modes,Golkar_2016}. These modes have been numerically identified \cite{ZhaoLiu2018,Liou2019_graviton,Kumar2022_neutral_excitations_FQH,Liu2024_geometric_excitations_fqh} and experimentally detected in two-dimensional electron gas~\cite{Igor2009_dispersion_FQH, Liang2024_chiral_graviton}. And their nature in fractional Chern insulators (FCIs) -- the lattice-based cousins of FQH states -- has also spurred intensive theoretical investigations~\cite{Repellin2014_SMA_FCI, Lu2024_fqah_neutral_excitations, Bishoy2025_magneotoroton, Long2025_spectra_FCI,Wang2025_geometric_excitations_moire,shen2025magnetorotonsmoirefractionalchern,Xavier2025_graviton, Long2026_graviton_FCI}.  
CSLs share profound similarities with FQH liquids in their ground-state topology. 
Yet, the ground-state topological equivalence by itself does not determine the massive bulk collective excitations, and whether CSLs (particularly in SU(2)-invariant quantum magnets) host analogous or qualitatively different modes remains elusive.
Unlike FCIs, where the FQH connection is rooted in the single-particle Chern-band–Landau-level correspondence, such a microscopic connection is generally absent in CSLs.

\begin{figure}[htp!]
	\centering		
	\includegraphics[width=0.5\textwidth]{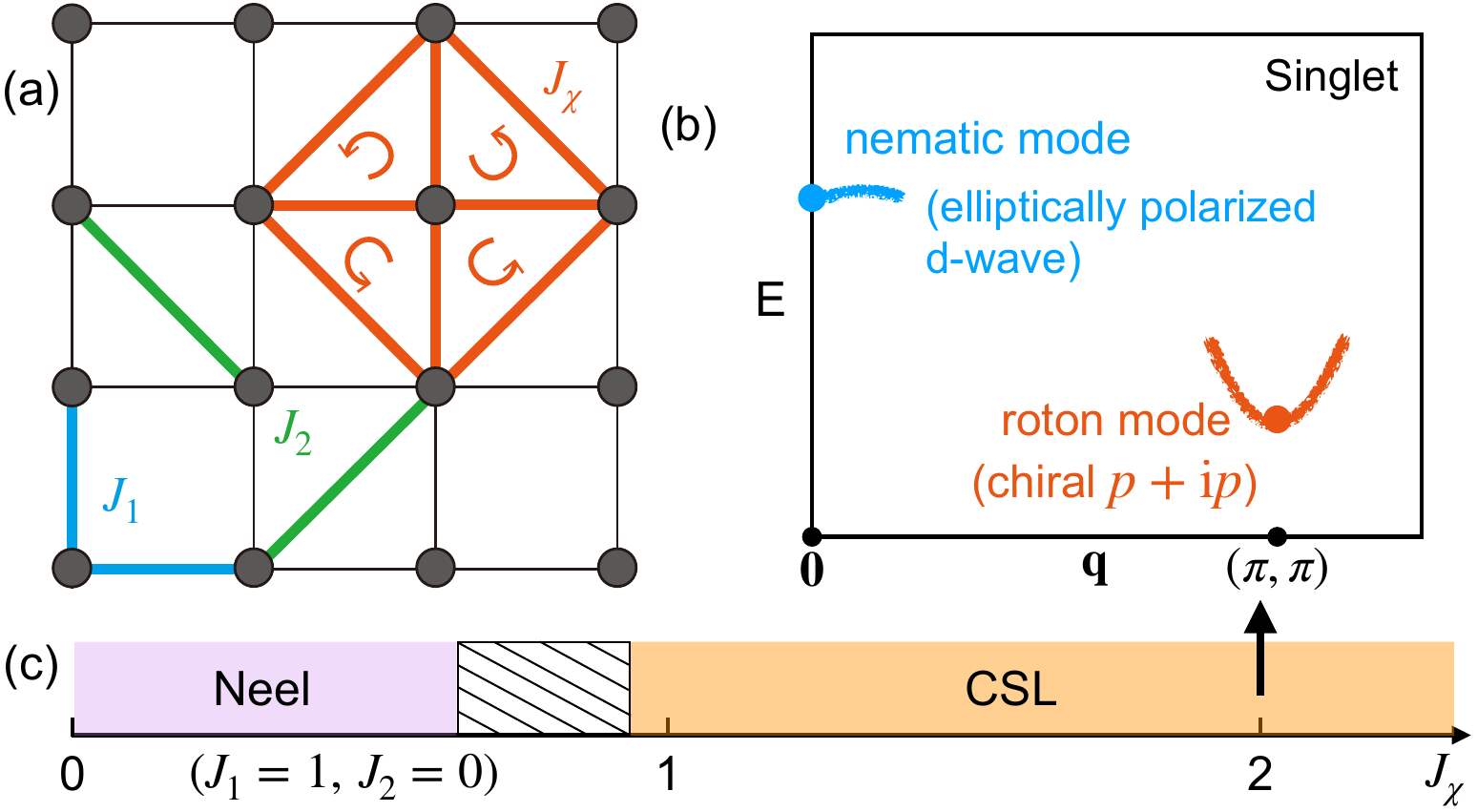}
	\caption{\textbf{Lattice model and phase diagram.} (a) The square-lattice model. 
    (b) Schematic plot of the two exotic spin-singlet collective modes deep inside the CSL phase. For positive $J_\chi$ the chiral roton mode lies in the $p+\mathrm{i}p$ channel, and the elliptically polarized nematic mode exhibits stronger $d+\mathrm{i}d$ component.
    (c) The quantum phase diagram with fixed $J_1=1,\ J_2=0$ and tuning $J_\chi$. The GS in the shaded region is still under debate and is not the focus of this work.
    }
	\label{fig_lattice_phase_diagram}
\end{figure}

\begin{figure*}[htp!]
	\centering		
	\includegraphics[width=\textwidth]{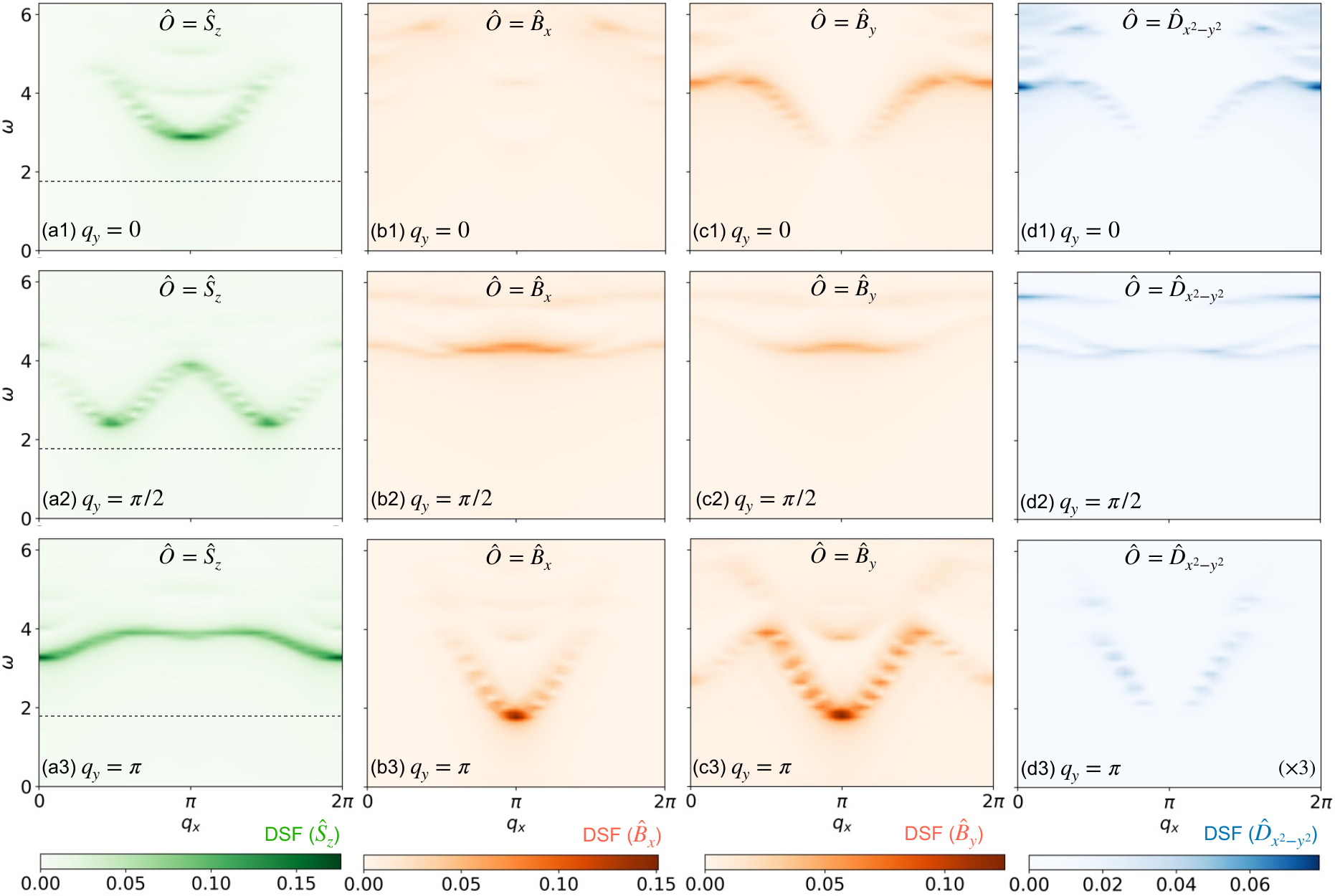}
	\caption{\textbf{Dynamical structure factors of CSL.} Different panels show dynamical response of (a1-a3) spin $S_z$, (b1-b3) x-direction bond operator $B_x$, (c1-c3) y-direction bond operator $B_y$ and (d) d-wave bond operator $D$ (details see main text).   
    The calculations are performed by the TDVP for $J_2=0$ and $J_\chi=2$ on $L_y\times L_x=4\times32$ cylinders. The horizontal/vertical axis is the energy/$q_x$ in each panel with different fixed $q_y$.
		The dashed lines in (a1-a3) correspond to the energy scale of the singlet roton mode ($\omega/J_1\sim1.76$).
		In (d3), the DSF along $q_y=\pi$ are multiplied by a factor of three for visibility. 
	}
	\label{fig_CSL_spectrum}
\end{figure*}

Here, we systematically investigate the collective excitations of the SU(2) symmetric CSL, combining exact diagonalization (ED),
density matrix renormalization group (DMRG),
and time-dependent variational principle (TDVP) calculations.
For the first time, we discover a chiral p-wave roton mode at finite momentum, and an elliptically polarized d-wave nematic mode at $q=0$, in the spin-singlet sector of CSL [Fig.~\ref{fig_lattice_phase_diagram}(b)].
The chiral p-wave roton, with chirality locked to the GS chirality, has no counterpart in FCI or FQH systems, and remains prominent even near the phase boundaries. 
The $q=0$ d-wave mode also tracks the GS chirality, but exhibits only partial helicity selectivity, in contrast to the chiral graviton, and it softens substantially at finite $J_2$, suggesting the nematic instability.
In addition, the magnetoroton-like two-spinon bound states are also observed in the triplet sector.

\noindent{\textcolor{blue}{\it Dynamical structure factors of the CSL.}---}
To investigate the collective excitations of the CSL, we probe the dynamical responses  of the CSL living in a spin-1/2 Heisenberg model on the square lattice~\cite{Samajdar2019_thermal_hall_Square,Zhang2024_CSL_square,Yang2024_CSL_square,Jin2025_dirac_csl_square}. The model reads 
\begin{equation}\label{eq:ham}
    \hat{H}=J_1\sum_{\langle i,j\rangle}\hat{\mathbf{S}}_i\cdot\hat{\mathbf{S}}_j+J_2\sum_{\langle\langle i,j\rangle\rangle}\hat{\mathbf{S}}_i\cdot\hat{\mathbf{S}}_j
    +J_\chi\sum_{\triangle}\hat{\mathbf{S}}_i\cdot(\hat{\mathbf{S}}_j\times\hat{\mathbf{S}}_k),
\end{equation}
where $J_1$ (fixed $J_1=1$ as the energy scale throughout this work) and $J_2$ are the nearest-neighbor (NN) and next-NN antiferromagnetic (AFM) Heisenberg interactions. $J_\chi$ refers to the three-spin scalar chiral coupling, with $\triangle$ including four kinds of minimal triangles in each plaquette, and $i,j,k$ follow the counterclockwise order,
as shown in Fig.~\ref{fig_lattice_phase_diagram}(a).
The GS phase diagram of tuning $J_\chi$ with fixed $J_2=0$ has been studied~\cite{Zhang2024_CSL_square,Yang2024_CSL_square}, as illustrated in Fig.\ref{fig_lattice_phase_diagram} (c), where a robust CSL phase exists driven by a finite $J_{\chi}$.

\begin{figure}[htp!]
	\centering		
	\includegraphics[width=0.5\textwidth]{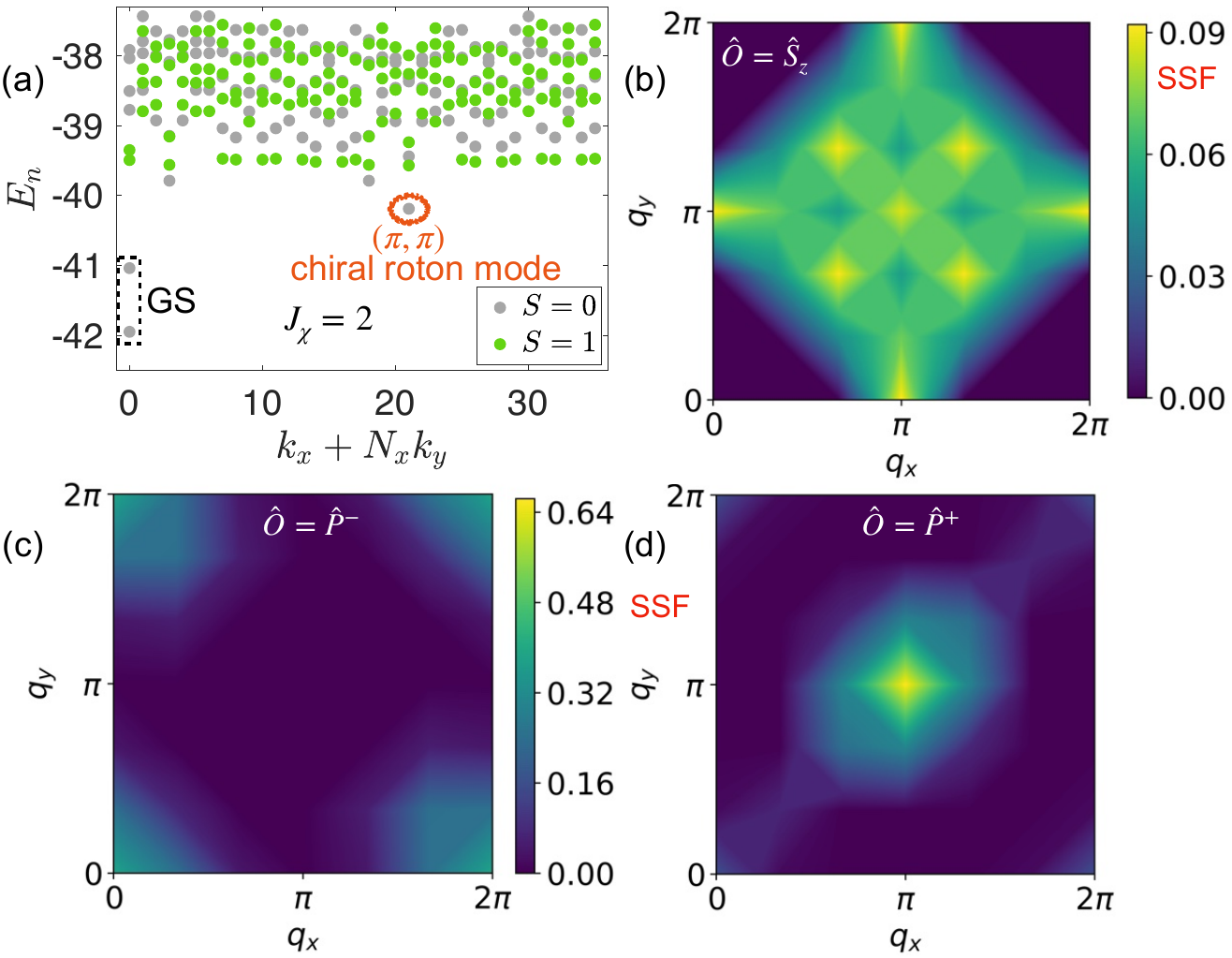}
	\caption{\textbf{Spectral property of CSL.} (a) Low-energy spectrum obtained on a $6\times6$ torus by setting $J_2=0$, $J_\chi=2$. A singlet roton mode appears at $(\pi,\pi)$. The states of larger $S$ are at higher energies.
		The static structure factors (SSFs) of $B_x,B_y$ operator are shown in (b-d) with panels (c) and (d) sharing the same color bar.
	}
	\label{fig_ED_spectrum_SSF}
\end{figure}

We probe the dynamical structure factors (DSFs) for several local operators ($\hat{O}$), 
and the results of DSFs (at $J_2=0,\ J_\chi=2$ for example)  are shown in Fig.~\ref{fig_CSL_spectrum}. 
The seemingly discrete behavior in some panels is due to the finite cylinder length that allows a discrete set of
accessible $q_x$ points~\cite{suppl}.

First, we begin with the triplet sector, as is customary in the study of spin liquids~\cite{ Halimeh2016_dsf_spin_csl,Zhu2019_spinon_DSF,Pichler2024_CSL_DSSF,Josef2025_spin_liquid_DSSF,drescher2025_dssf_triangle}, and present the spin DSFs by targeting $\hat O=S^z$ operator in Fig.~\ref{fig_CSL_spectrum}(a1-a3).
Rather than a purely featureless continuum, we find some finite-momentum bound states with relatively concentrated spectral weight. 
One notable feature is that these modes appear in relatively high-frequency regime.

Second, for the singlet sector, we naturally study the bond operators: $\hat{B}_{\mu}(\mathbf{r})=\hat{\mathbf{S}}_{\mathbf{r}}\cdot \hat{\mathbf{S}}_{\mathbf{r}+\mathbf{\mu}}$ 
($\mathbf{\mu}\in\{\mathbf{x},\mathbf{y}\}$ are the primitive vectors along two directions), as they are the simplest ones that conserve the total spin.
The spectra of $\hat O= \hat{B}_x$ and $\hat{B}_y$ are shown in Fig.~\ref{fig_CSL_spectrum}(b1-b3,c1-c3), respectively.
Interestingly, we observe a pronounced roton mode with the energy minimum at $(\pi,\pi)$ where the spectral weight peaks.
Importantly, in a large region of the CSL parameter space~\cite{suppl}, this $S=0$ roton mode at $(\pi,\pi)$ remains the lowest excitation, much lower than the spin triplet excitations.

Third, in the DSFs of $\hat{B}_y$ operator here, we also observe some non-vanishing signals at $\mathbf{q}\rightarrow0$, which are much weaker than the roton minimum and at higher energies (the signal of $\hat{B}_x$ is weaker possibly due to the cylinder geometry~\cite{suppl}).
Inspired by the orbital spin-2 nature of the graviton mode in FQH systems, we explore if the $\mathbf{q}=0$ signal here is related to any orbital spin-2 excitation~\cite{Yang2012_collective_modes,Liou2019_graviton,Liang2024_chiral_graviton,Yang2025_geometric_fluctuation_FQH}.
Therefore, we define the d-wave operator $\hat{D}_{x^2-y^2} \equiv \hat{B}_x-\hat{B}_y$, and the DSFs are shown in Fig.~\ref{fig_CSL_spectrum}(d1-d3).
In this rank-2 quadrupolar channel, the $\mathbf{q}=0$ signal indeed manifests as a pronounced spectral peak.
The d-wave nature is intrinsic, and consistently identified in symmetric tori and in cylinders with similar response in $\hat{B}_x$ and $\hat{B}_y$ channels at $q=0$~\cite{suppl}.
We refer to this higher energy mode at long-wavelength limit as the d-wave nematic mode. Notably, it can soften upon tuning parameters, making it more accessible, and we will discuss more later.
At finite momentum, the DSF spectral response in the d-wave channel is rather weak and fades away when approaching the roton mode.

\begin{table}[t]
	\centering
	\begin{tabular}{l|c|c|c|c}
		\toprule
        \hline 
        \hline
		Overlaps & \multicolumn{2}{c|}{$\langle\psi_0(\pi,\pi)|\hat{O}|\tilde{\psi}_0(0,0)\rangle$} & \multicolumn{2}{c}{$|\langle\psi_0(\pi,\pi)|\hat{O}|\tilde{\psi}_0(0,0)\rangle|$} \\
		\hline
		\cmidrule(lr){2-3}\cmidrule(lr){4-5}
		\diagbox[width=6em,height=3.6em]{Size:\\ $L_y\times L_x$}{\raisebox{-2.5ex}{\hspace{-1.6em}$\hat{O}$}} & $\hat{B}_x$ & $\hat{B}_y$ & $\hat{B}_x+i\hat{B}_y$  & $\hat{B}_x-i\hat{B}_y$   \\
		\hline
		\midrule
		\hspace{0.7em}$4\times4$ & $0.71+0.48\mathrm{i}$ & $0.48-0.71\mathrm{i}$ & 0.96  & 0.00 \\
		\hline
		\hspace{0.7em}$4\times6$ & $-0.13+0.84\mathrm{i}$ & $0.80+0.12\mathrm{i}$ & 0.93  & 0.06 \\
        \hline 
        \hline        
		\bottomrule
	\end{tabular}
	\caption{\textbf{Chiral nature of the roton mode}. The wave function overlap using the SMA reveals the lowest excited state $|\psi_0(\mathbf{k}=(\pi,\pi))\rangle$ is chiral.  The wave function overlap $\langle\psi_0(\pi,\pi)|\hat{O}|\tilde{\psi}_0(0,0)\rangle$ is properly normalized by imposing$\langle\tilde{\psi}_0(0,0)|\hat{O}^\dagger\hat{O}|\tilde{\psi}_0(0,0)\rangle=1$, i.e., the excited state $\hat{O}|\tilde{\psi}_0(0,0)\rangle$ is renormalized to unit norm. 
	}
	\label{tab_roton_overlap}
\end{table}

\noindent{\textcolor{blue}{\it Nature of the chiral roton.}---}
To provide further insight on the observed singlet roton mode, we show the energy spectrum of the CSL on torus in Fig.\ref{fig_ED_spectrum_SSF}. 
We find that the lowest excitation belongs to the spin singlet and is located at $(\pi,\pi)$. Its quantum number matches that of the roton mode we found in Fig. \ref{fig_CSL_spectrum}(b-c).
The precise nature of this low-energy excitation can be obtained by inspecting its content using the single-mode approximation (SMA).
As shown in Tab.\ref{tab_roton_overlap}, on the symmetric torus, we find very high overlaps between the target excited state and the variational states made of $\hat{B}_x$ and $\hat{B}_y$~\cite{overlap_size}. 
More interestingly, we find the relation $\hat{B_x}|\tilde{\psi}_0(0,0)\rangle=\mathrm{i}\hat{B_y}|\tilde{\psi}_0(0,0)\rangle$ exactly holds.
This motivates us to design the chiral p-wave operators $\hat{P}^{\pm}=\hat{B}_x\pm\mathrm{i}\hat{B}_y$.
The overlap of the new chiral states is either close to $1$ (larger than the $\hat{B}_{\mu}$ operators) or exactly $0$, and we have checked that the chirality is locked to the GS chirality, which could be simply flipped by the sign of $J_\chi$.
We have also checked that the chirality of this excited state is well retained even for the non-symmetric torus such as the $4\times6$ one [Tab.~\ref{tab_roton_overlap}]. 
Beside, we have checked the next-NN bond operators do not reveal this chiral roton, and
the aforementioned properties are consistently found across the CSL phase~\cite{suppl}.

\begin{figure}[htp!]
	\centering		
	\includegraphics[width=0.5\textwidth]{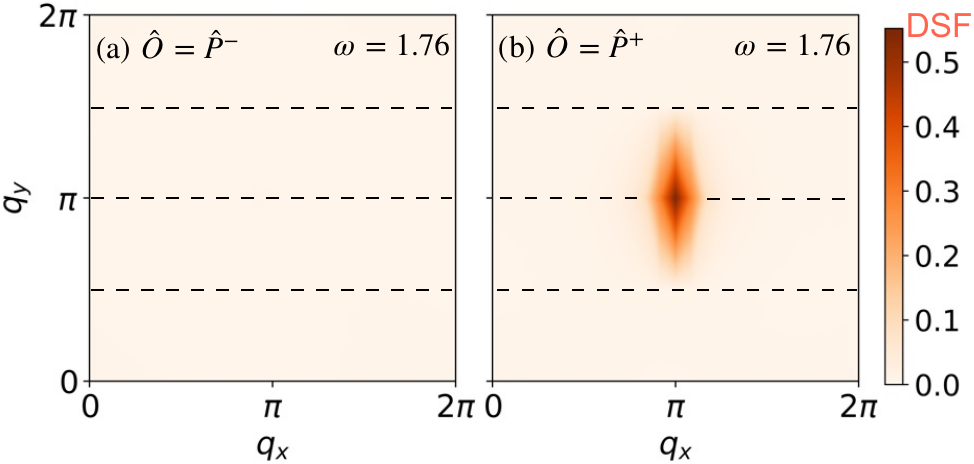}
	\caption{\textbf{The chiral roton.} DSFs of the chiral p-wave operators $\hat{P}^{\mp}$ at fixed $\omega=1.76$ (energy scale of the roton mode) for $J_2=0$ and $J_\chi=2$ are shown, respectively.
    The calculations are performed by the TDVP on $L_y\times L_x=4\times32$ cylinders. 
    The black dashed lines denote $q_y=\frac{\pi}{2},\pi,\frac{3\pi}{2}$, respectively.
	}
	\label{fig_chiral_roton}
\end{figure}

Furthermore, we test the SMA picture by computing the static structure factors (SSFs). 
For the $\hat{P}^+$ channel, there is a peak at ($\pi,\pi$),  which is stronger than those of the $\hat{B}_\mu$ operators even after renormalization~\cite{suppl}, implying the larger frequency-integrated dynamical spectral weight in this chiral channel.
Interestingly, for the $\hat{P}^-$ channel, the SSF at ($\pi,\pi$) is $\sim0$.
These results further support that the low-energy state at ($\pi,\pi$) we observe is consistent with a chiral p-wave mode, which we call the chiral roton mode.
To further confirm this chiral p-wave roton mode, we show the DSFs of the $\hat{P}^{\pm}$ channels at fixed $\omega=1.76$ (the energy of the chiral roton mode) from TDVP simulations in Fig.~\ref{fig_chiral_roton}. The full dispersion and consistent results from wider cylinders are included in the SI ~\cite{suppl}.
We observe a sharp peak at $(\pi,\pi)$ in the $\hat{P}^+$ channel, while the response vanishes in the $\hat{P}^-$ one.
These are consistent with the earlier ED results in Fig.~\ref{fig_ED_spectrum_SSF} and Tab.~\ref{tab_roton_overlap}, and together support the exactly chiral p-wave nature of the observed singlet roton mode.

\begin{figure}[htp!]
	\centering		
	\includegraphics[width=0.5\textwidth]{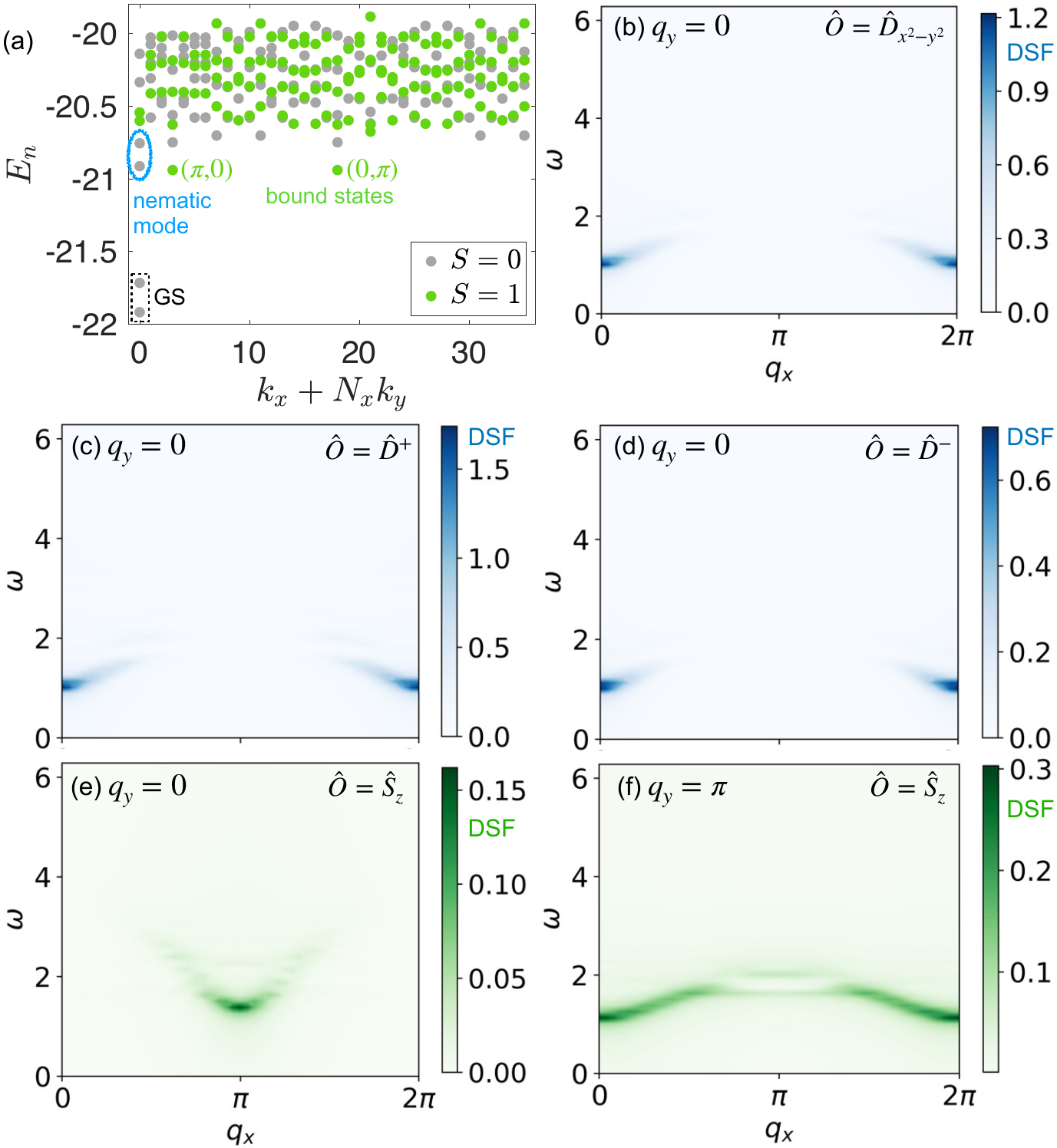}
	\caption{\textbf{Soft collective modes.}  (a) The energy spectrum of $J_2=0.6,J_\chi=0.5$ on the $6\times6$ torus. The TDVP results of different channels  on the $4\times32$ cylinders are shown in panels (b-f). The horizontal/vertical axis is the energy/$q_x$ in each panel with different fixed $q_y$.
	}
	\label{fig_soft_modes}
\end{figure}

\noindent{\textcolor{blue}{\it Nature of the long-wavelength nematic mode.}---} 
Next we shift our focus to the lowest excitation mode in $\mathbf q=0$ sector. We find this mode would become soft by increasing a competing $J_2$ interaction. 
We use $J_2=0.6$, and $J_\chi=0.5$ as an example and show the energy spectrum on the $6\times6$ torus in Fig.\ref{fig_soft_modes} (a).
Notably, in the singlet subspace, the lowest excited states are now in the zero-momentum sector, in contrast to the spectra deep in the CSL phase (c.f. Fig.\ref{fig_ED_spectrum_SSF} (a)), where a higher-energy nematic mode has been found at $\mathbf{q}=0$.
This motivates us to explore the relation between these low-energy states here and the previously studied nematic mode. 
The overlaps between the approximated d-wave nematic mode and the low-energy excited states based on the ED results are shown in Tab.\ref{tab_nematic_overlap}.
Apart from using the $\hat{D}_{x^2-y^2}$ operator to check the orbital spin-2 characteristic, we also try to explore its chirality motivated by the fact that the graviton mode in the FQH system is chiral~\cite{Liou2019_graviton}.
Therefore, we further design the chiral operators:
$\hat{D}^\pm=\hat{D}_{x^2-y^2}\pm\mathrm{i}\hat{D}_{xy}$,
where $\hat{D}_{xy}=\hat{B}_{\mathbf{x},\mathbf{y}}-\hat{B}_{-\mathbf{x},\mathbf{y}}$ is composed of the next-NN bonds.

\begin{table}[t]
	\centering
	\begin{tabular*}{0.46\textwidth}{@{\extracolsep{\fill}} l|c|c|c|c|c|c @{}}
		\toprule
        \hline
        \hline
		{\hspace{1.5em}$|\tilde{\psi}\rangle$}
		& \multicolumn{3}{c|}{$|\tilde{\psi}_0(0,0)\rangle$}
		& \multicolumn{3}{c}{$|\tilde{\psi}_1(0,0)\rangle$} \\
		\hline
		\cmidrule(lr){2-4}\cmidrule(lr){5-7}
		\diagbox[width=3.7em,height=2.1em]{$\langle\psi|$}{\hspace{1em}\raisebox{-1ex}{$\hat{O}$}}
		& {\hspace{-1em}$\hat{D}_{x^2-y^2}$} & {\hspace{-1em}$\hat{D}^+$} & {\hspace{-1em}$\hat{D}^-$}
		& {\hspace{-1em}$\hat{D}_{x^2-y^2}$} & {\hspace{-1em}$\hat{D}^+$} & {\hspace{-1em}$\hat{D}^-$} \\
		\midrule  \hline
		\centering
		$\langle\psi_2(0,0)|$ & {\hspace{-0.45em}0.811\hspace{0.45em}} & {\hspace{-0.45em}0.735\hspace{0.45em}} & {\hspace{-0.45em}0.722\hspace{0.45em}} & {\hspace{-0.45em}0.000\hspace{0.45em}} & {\hspace{-0.45em}0.000\hspace{0.45em}} & {\hspace{-0.45em}0.000\hspace{0.45em}} \\
		\hline
		$\langle\psi_3(0,0)|$ & {\hspace{-0.45em}0.000\hspace{0.45em}} & {\hspace{-0.45em}0.000\hspace{0.45em}} & {\hspace{-0.45em}0.000\hspace{0.45em}} & {\hspace{-0.45em}0.809\hspace{0.45em}} & {\hspace{-0.45em}0.716\hspace{0.45em}} & {\hspace{-0.45em}0.736\hspace{0.45em}} \\
        \hline
        \hline 
		\bottomrule
	\end{tabular*}
	\caption{\textbf{Overlap of the nematic mode.} We show the absolute values of the overlaps ($|\langle\psi|\hat{O}|\tilde{\psi\rangle}|$) from the ED results for different d-wave operators at $J_2=0.6,\ J_\chi=0.5$ on the $6\times6$ torus. 
		$|\tilde{\psi}_{0,1}(0,0)\rangle$ are the two lowest states in the (0,0) momentum sector (the 2-fold degenerate GSs of the CSL) after normalization by imposing $\langle\tilde{\psi}_{0,1}(0,0)|\hat{O}^\dagger\hat{O}|\tilde{\psi}_{0,1}(0,0)\rangle=1$, i.e., the excited state $\hat{O}|\tilde{\psi}_{0,1}(0,0)\rangle$ is renormalized to unit norm. 
		$\langle\psi_{2,3}(0,0)|$ are the two lowest excited states above the two GSs in the (0,0) momentum sector.
	}
	\label{tab_nematic_overlap}
\end{table}

Interestingly, we find that the $\hat{D}_{x^2-y^2}$ operator exhibits a one-to-one (diagonal) selection rule between the two degenerate GSs and the two lowest excited states (circled in blue in Fig.\ref{fig_soft_modes}(a)) with very high overlaps, supporting that these excitations are well captured by a single-mode ansatz. 
In addition, according to the normalized overlaps of the $\hat{D}^\pm$ channels, 
no obvious chirality is observed, while the ratio between their spectral weights is roughly two~\cite{suppl}.
We further show the DSFs of different d-wave operators with fixed $q_y=0$ on the $4\times32$ cylinder in Fig.\ref{fig_soft_modes} (b-d)~\cite{suppl}.
It is consistent with the ED results that the nematic mode (at higher energy at parameters deep in the CSL phase) becomes markedly soft, and
the DSF of the nematic mode is significantly sharp with highly concentrated spectral weight, exhibiting strong nematic instability.
Furthermore, instead of a pure chirality eigenmode, we find that this d-wave nematic mode is elliptically polarized, as the DSF in the $d+\mathrm{i}d$ channel is approximately twice that in the $d-\mathrm{i}d$ channel.
The relative spectral-weight ratio between the $d\pm id$ channels is not universal, and further details are provided in the SI~\cite{suppl}.

\noindent{\textcolor{blue}{\it Spin triplet excitations.}---} 
For the triplet sector deep in the CSL phase ($J_2=0,\ J_\chi=2$), we also show the spin SSF  (using $\hat{S}_z$ for example) in Fig.\ref{fig_ED_spectrum_SSF}(b).
Although no obvious feature can be directly captured from the continuum of the spectrum in Fig.\ref{fig_ED_spectrum_SSF} (a), there are some relative peaks in the SSFs at finite momentum, which are consistent with the bound states observed from the DSFs in Fig.\ref{fig_CSL_spectrum}(a1-a3). 
Note that, in contrast to those of bond operators, there is vanishing weight at $\mathbf{q}\rightarrow0$ as the total spin is conserved. 

The DSFs of the spin excitations are also studied at $J_2=0.6$ and $J_\chi=0.5$ [\ref{fig_soft_modes}(e-f)] and we observe the  $(\pi,0)$/($0,\pi$) triplet mode at the energy minimum (the spectral weight is relatively smaller at $(\pi,0)$ due to the cylinder geometry). The energy gap of the triplet mode and that of the nematic mode are very close and this is consistent with the ED spectrum in Fig.\ref{fig_soft_modes}(a).
This triplet mode is consistent with pronounced spin stripe fluctuations.
Further, this triplet mode is closely connected to the nematic mode, since it also carries nematic fluctuations.

Previous studies have identified finite-momentum triplet modes in CSLs as two-spinon bound states~\cite{Laughlin1989_spin_hamiltonian, Zhu2019_spinon_DSF,drescher2025_dssf_triangle}. 
Here, we emphasize the triplet mode as a spinful (SU(2)-symmetric) analogue of the FQH magnetoroton, a connection that is rarely made explicit in the CSL literature. 
Their coexistence with the chiral p-wave singlet roton further highlights that the latter is a new and distinct type of collective mode rather than a replacement for the magnetoroton-like excitation.

\noindent{\textcolor{blue}{\it Discussions and implications.}---}
In this work, we have identified, for the first time to the best of our knowledge, two distinct spin-singlet collective modes in a CSL.
The chiral p-wave singlet roton mode is unexpected and has no counterpart in FCI or FQH systems.
It does not preclude the simultaneous presence of the  magnetoroton-like triplet mode.
By contrast, the roton minima of the magnetoroton mode in FQH/FCI systems and of the triplet mode in CSLs exhibit neither chirality nor orbital spin.
The $\mathbf{q}=0$ d-wave mode is similar to the graviton mode in FQH/FCI systems in the aspect of the orbital spin-2, suggesting that Haldane's geometric description of FQH states might be further generalized~\cite{Haldane2011_geometrical_description}. 
This mode might be understood as a traceless deformation of the spin-correlation (bond) pattern, in which correlation weight is redistributed among different bond directions while the isotropic component remains unchanged to linear order.
However, in contrast to the chiral graviton, this d-wave mode exhibits only partial helicity selectivity.
Although the chiral p-wave roton mode and the elliptically polarized d-wave mode exhibit different characteristics, the dispersions suggest that they may be connected. 
Whether they belong to the same excitation branch is interesting for further studies.
While these exotic modes are discovered in the spin-singlet sector, breaking the SU(2) symmetry by tuning the relative strengths of different spin components of the Heisenberg interaction does not appear to qualitatively affect the persistence of the channel-selective modes.

Apart from the square-lattice CSL, we have consistently identified the chiral p-wave roton mode, and the $q=0$ d-wave mode with partial helicity selectivity, in a triangular-lattice CSL~\cite{suppl}, suggesting the potential generality of these singlet collective modes. 
We note that, it is interesting for further investigation of singlet collective modes in non-Abelian CSLs~\cite{Yao2007_nonabelian_csl,Greiter2009_nonabelian_csl,Liu2018_CSL_nonabelian,Chen2018_nonabelian_csl,Huang2022_nonabelian_csl}.

Experimentally, detecting a CSL is often more demanding than detecting a fermionic FQH/FCI phase. In FQH/FCI systems, topological order is accompanied by quantized charge transport responses~\cite{Stormer1999_FQH_review}, while in CSLs the corresponding signatures are carried by neutral spin/energy currents and are typically accessed only indirectly~\cite{Haldane1995_spin_currents,Zhang2024_thermal_hall}. 
Therefore, additional experimental probes are highly desirable, and one promising route is to detect the dynamical properties and excitations.
Inelastic neutron scattering is the primary momentum-resolved probe of spin excitations, and it predominantly accesses the $S=1$ channel in an SU(2)-symmetric system with a singlet ground state~\cite{Broholm2020_QSL}. 
However, a broad continuum and even the bound-state-like features in the triplet channel are insufficient for identification.
In this work, the singlet collective modes offer new spectroscopic signatures for future experiments on candidate CSL materials. For example, the $\mathbf{q}=0$ nematic mode could be identified from the (circularly) polarized Raman scattering measurements~\cite{Sandvik1998_B1g_raman,Devereaux2007_ILS,Liang2024_chiral_graviton,koller2025ramancirculardichroism}, where symmetry selection rules allow one to isolate the d-wave channel, and it is also interesting to experimentally explore the (elliptical) polarization of the nematic mode.
A possible route to access the finite-$\mathbf{q}$ chiral p-wave roton mode is momentum- and polarization-resolved (indirect) resonant inelastic X-ray scattering measurements~\cite{Ament2011_RIXS, Sikora2010_Kedge}, where circular-polarization dependence might provide a direct way to distinguish $\hat{P}^\pm$ channels through chiral selection rules.

We close by highlighting the soft collective modes in the CSL are related to interesting quantum phase transitions.
For example, we study the soft modes at $J_2=0.6$ and $J_\chi=0.5$, but the GS phase diagram at fixed $J_\chi=0.5$ larger $J_2$ is still under debate. 
Ref.~\cite{Zhang2024_CSL_square} suggests a direct CSL-stripe transition at $J_2\sim0.75$ and the spin stripe order is ($0,\pi$) or ($\pi,0$), 
However, Ref.~\cite{Yang2024_CSL_square} suggests the possibility of an intermediate nematic spin liquid phase between the CSL and stripe phases, but the identification of the nematic phase still suffers from the finite-size effect.
In fact, the soft collective modes reported in our work are consistent with both scenarios.
For the first one, if the CSL-stripe transition is continuous, one might observe the $\mathbf{q}=0$ nematic mode and ($0,\pi$)/($\pi,0$) triplet mode both become gapless at the critical point, and the stripe phase itself has the co-existing secondary nematic order.
For the second scenario, the nematic spin liquid phase shall be a nematic CSL, and the two-step transition could be triggered by the sequential gap closing of the nematic and triplet modes in the CSL. 
This two-step transition also leads to a unique vestigial scenario~\cite{Nie2017vestigial,Fernandes2019Vestigial,Lu2024_vestigial_gbdw}, in which the vestigial nematic phase inherited from stripe order is topologically ordered.
The mechanism for the nematic CSL discussed here echoes that of FQH nematics~\cite{Regnault2017_fqh_nematic,Yang2020_nematic_fqh,Pu2024_fqh_nematics} and smectic FCIs~\cite{Lu_2024_fqahs,Lu2025_fqah_fqahs}, where the latter phases are triggered by the soft chiral graviton at $\mathbf{q}=0$ and magneto-roton at finite momentum, respectively, and the topological orders remain intact through the symmetry-breaking process.
A more extensive exploration of the parameter space and possibly larger system sizes would be needed to distinguish these possibilities more conclusively~\cite{suppl}, and either interesting scenario may be realized in a broader regime, which we leave for future work.

\begin{acknowledgments}
\noindent{\textcolor{blue}{\it Acknowledgments}---}
H. L. thanks Jie Wang and Xin Shen for helpful discussions, and Xiao-Tian Zhang and Han-Qing Wu for some ground-state benchmark, and Min Long, Zeno Bacciconi, Hernan B. Xavier, Zi Yang Meng, and Marcello Dalmonte for collaborations in studying the FCI excitations, and Zheng Yan for the hospitality at the early stage of this project.
W. Z. thanks S. S. Gong, J. W. Yang, L. Wang for collaborating and fruitful discussion on the previous projects.
H. L. and W. Y. are supported by the National Natural Science Foundation of China (No. 12425406), Research Grant Council of Hong Kong (AoE/P-701/20, HKU SRFS21227S05), and New Cornerstone Science Foundation. We thank Beijing PARATERA Tech Co., Ltd. (https://cloud.paratera.com) for providing HPC resources that supported the research results reported in this paper. 
This work is also supported by the National Key Research and Development Program of China Grant No. 2022YFA1402204 .
\end{acknowledgments}

\subsection{Energy spectra}
The energy spectra are calculated by the ED. For the ED simulations, we use the Lanczos algorithm, working in symmetry-resolved sectors that exploit translation (crystal momentum) and total $S_z$ conservation~\cite{Sandvik2010_ED}.
Part of the ED simulations are carried out using QuSpin package~\cite{QuSpin2019}.
The system size in the ED calculation varies from $N_s=16$ to $36$. 
The Hilbert subspace of the $N_s = 36$ in each momentum sector has $\sim2.52\times10^8$ basis states, which is is close to the practical computational limit of our ED calculations. 

Within the single-mode approximation (SMA), the low-energy excitation could be constructed by applying the local operator to the many-body ground state: $|\psi(\mathbf q)  \rangle = O(\mathbf q) |\psi_0(\mathbf{q}=(0,0)) \rangle$. The comparison of this SMA-inspired state with the ED calculation may validate the operator content of the low-energy excited state.

\subsection{Dynamical Structure Factors}

The DSFs for different local observables are defined as: 
\begin{align}
\mathrm{DSF}^{\hat{O}}(\mathbf{q},\omega)=\sum_{\mathbf{r},t}\frac{e^{\mathrm{i}\omega t-\eta t}
e^{-\mathrm{i}\mathbf{q}\mathbf{r}}}{\sqrt{N_t}}(\langle \hat{O}^\dagger_{\mathbf{r}}(t)\hat{O}_{\mathbf{0}}(0)\rangle-\langle \hat{O}^\dagger_{\mathbf{r}}\rangle \langle \hat{O}_{\mathbf{0}}\rangle)   \label{eq:DSF}
\end{align}
where $\mathbf q$ and $\omega$ denotes the crystal momentum and frequency. 
The time-dependent quantity is $\hat{O}_{\mathbf{r}}(t)=e^{\mathrm{i}\hat{H}t}\hat{O}_{\mathbf{r}}e^{-\mathrm{i}\hat{H}t}$, and $\hat O$ is the local operator.

The dynamical response can be accessed by the tensor network simulations. 
We perform finite and infinite DMRG simulations~\cite{White1992_dmrg} with global $S_z$-conservation implemented.
To calculate DSFs (see Eq. \ref{eq:DSF}),  
we compute the time-dependent correlation function using the TDVP simulations~\cite{Haegeman2011_tdvp}: 
\begin{align}
 \langle 0| \hat{O}^\dagger_{\mathbf{r}}(t)\hat{O}_{\mathbf{0}}|0\rangle 
 = e^{\mathrm{i}E_0t}\langle 0|\hat{O}_{\mathbf{r}}e^{-\mathrm{i}\hat{H}t}\hat{O}_{\mathbf{0}}|0\rangle,
\end{align}
where  $|0\rangle$ is the ground-state wavefunction.
To obtain the DSF of each local operator $\hat{O}$,
In practice, we evolve the state $\psi(t)=e^{-\mathrm{i}\hat{H}t}\hat{O}_{\mathbf{0}}|0\rangle$ step by step with $\Delta t=0.05$, and the correlator is measured after each time step.
For the results in the main text, we evolve totally $N_t=1000$ steps such that the TDVP spectra are obtained on a frequency grid with spacing $\Delta\omega=0.04\pi$. 
In the time-to-frequency Fourier transform, we have used the Lorentzian broadening $\eta=0.05$.
The bond dimension for the TDVP results is up to $\chi=400$, which provides well converged results for the studied systems in this work~\cite{suppl}. 
TeNPy library is used in these matrix-product-state simulations~\cite{Johannes_2024_tenpy}.

As a byproduct, the static structure factor (SSF) can be calculated using the equal-time correlators
$\sum_\mathbf{r}e^{-\mathrm{i}\mathbf{q}\cdot\mathbf{r}}(\langle \hat{O}_{\mathbf{0}}^{\dagger}\hat{O}_{\mathbf{r}}\rangle-\langle\hat{O}_{\mathbf{0}}^{\dagger}\rangle\langle\hat{O}_{\mathbf{r}}\rangle)$. 
The Fourier transformation leads to the SSF defined in the momentum space.


\clearpage

	\onecolumngrid
	 \textbf{The supplementary information} contains:
	I. the ground-state entanglement spectrum;
	II. additional results of the collective modes of the square-lattice CSL;
		III. possible transitions from the CSL to the stripe;
		IV. consistent collective modes in a triangular-lattice CSL.
	
	\section{I. Ground-state entanglement spectrum of the square-lattice CSL}
\begin{figure}[htp!]
		\centering		
		\includegraphics[width=0.53\textwidth]{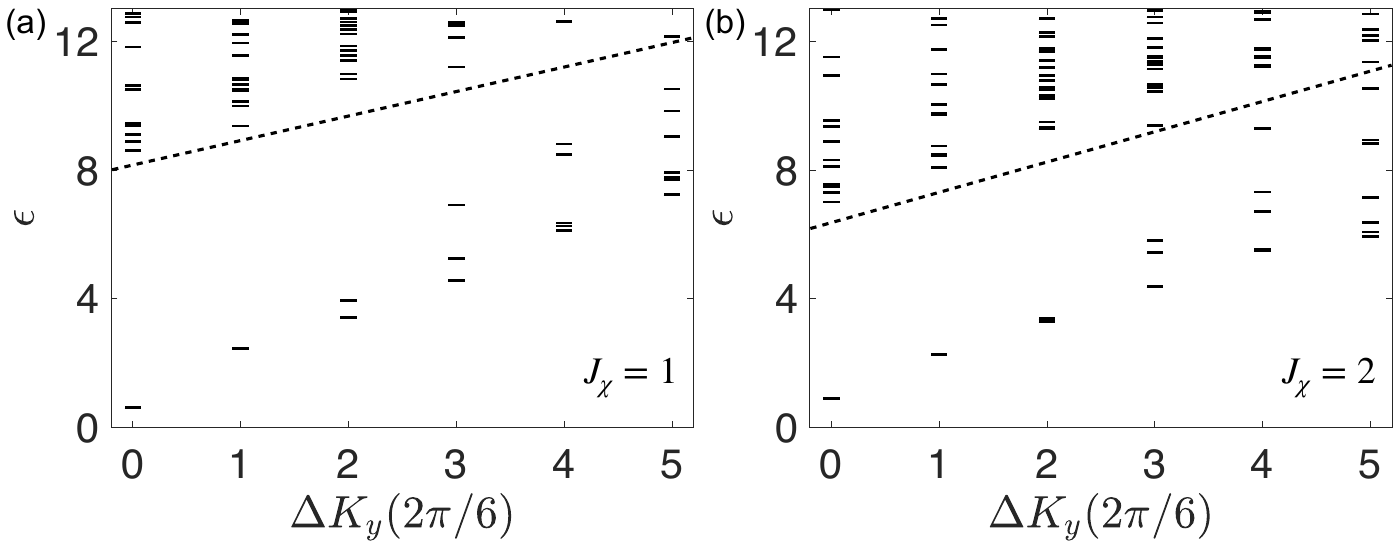}
		\caption{\textbf{Ground-state entanglement spectrum} Here, we show the entanglement spectrum of the CSL at different parameters, demonstrating the  the characteristic edge-mode counting $\{1,1,2,3,5,7,...\}$.
		}
		\label{figs_GS_ES}
	\end{figure}

	Here, we show the entanglement spectrum from the identity sector of the CSL in Fig.~\ref{figs_GS_ES}, which is not only consistent with the edge-mode counting~\cite{LiHaldane2008,Qi2012_ES_edge_spectrum} but also demonstrates the GS chirality. 
	We note that, both the GS chirality and that of the chiral roton mode would be flipped by the sign of $J_\chi$.

	\section{II. Additional results of the collective modes of the square-lattice CSL}
	\subsection{II.A Additional ED calculations }
	
	\begin{figure}[htp!]
		\centering		
		\includegraphics[width=\textwidth]{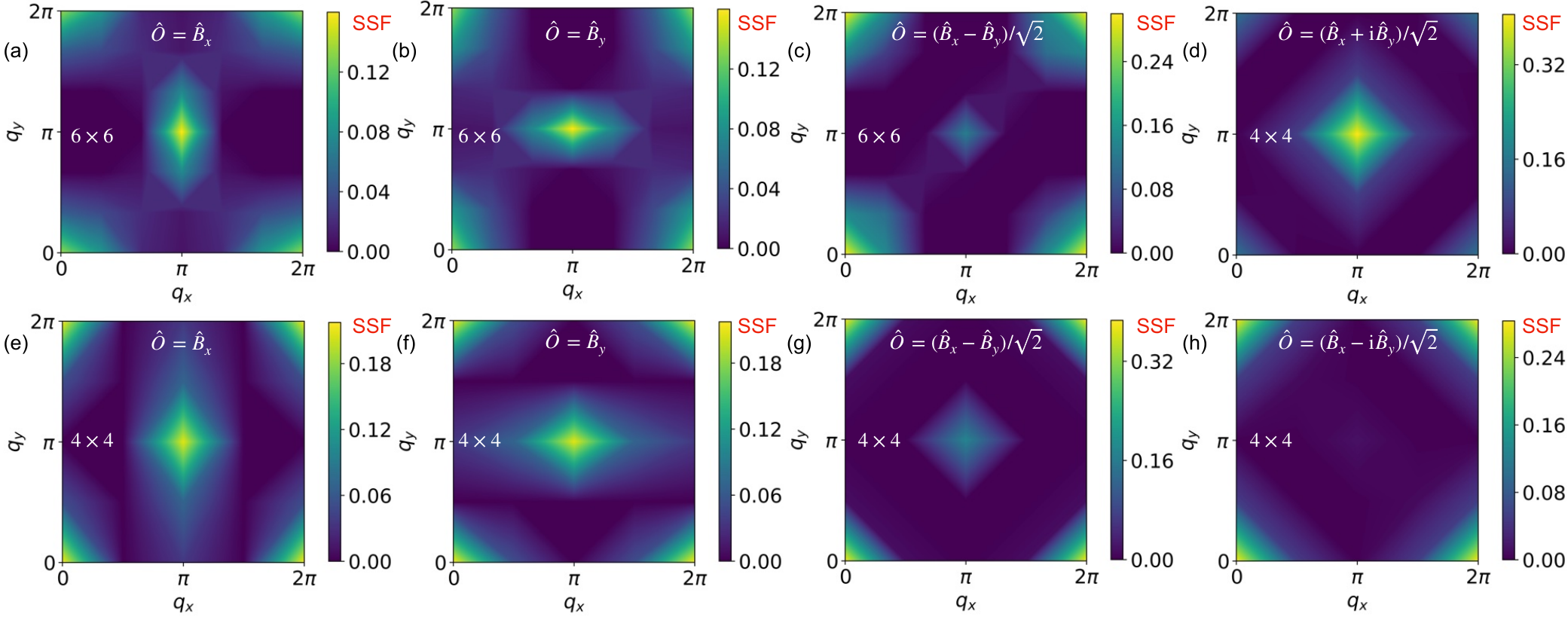}
		\caption{\textbf{Additional results of SSFs at $J_2=0,\ J_\chi=2$.} In panel (a-c), we show the complementary SSFs from the $6\times6$ torus.
			The SSFs of different channels from the $4\times4$ torus are shown in (d-h).
		}
		\label{figs_ED_SSF_Jchi-2}
	\end{figure}
	
\begin{figure}[htp!]
		\centering		
		\includegraphics[width=0.5\textwidth]{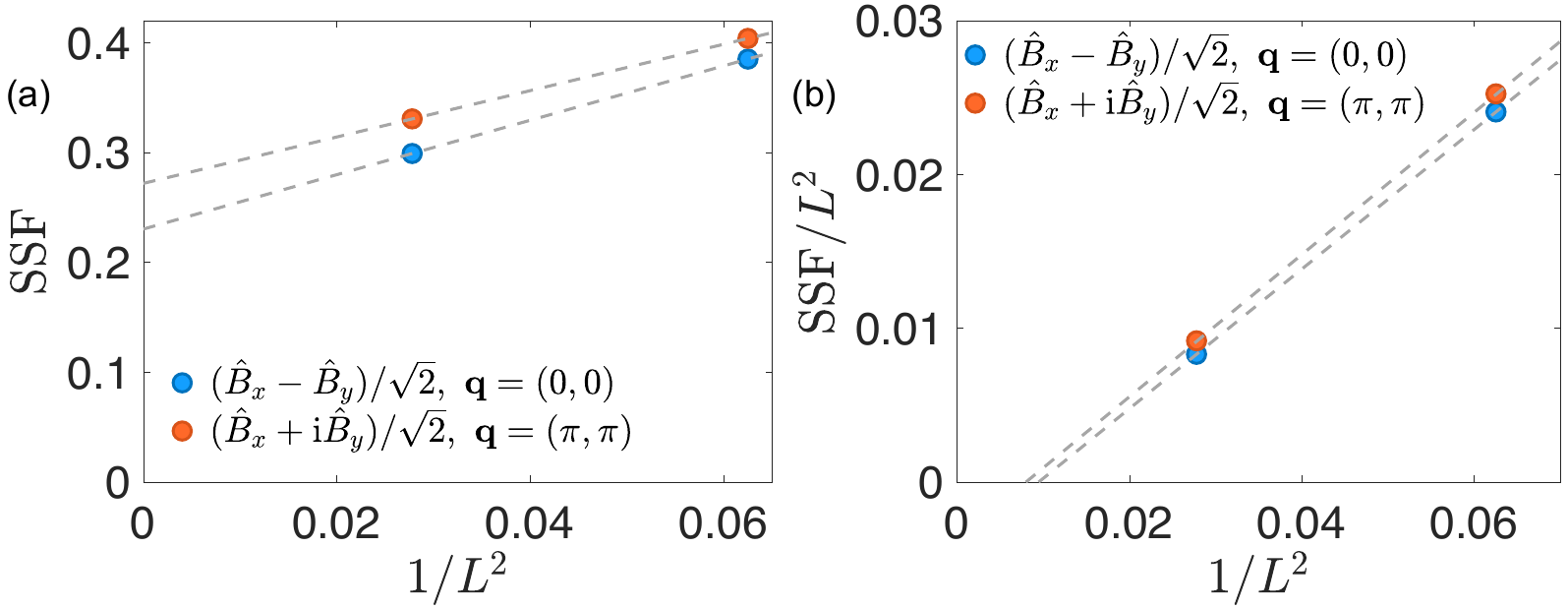}
		\caption{\textbf{Extrapolation of SSFs at $J_2=0,\ J_\chi=2$.} Here, we extrapolate the (a) extensive SSFs and (b) the corresponding intensive values (normalized by lattice sites) from ED simulations on symmetric tori.
		}
		\label{figs_ED_SSF_Jchi-2_extrap}
	\end{figure}

	In the main text, we have shown some ED results of SSFs  at $J_2=0,\ J_\chi=2$ from the $6\times6$ torus in Fig. 3. 
	Here, we further show the complementary SSFs in Fig.\ref{figs_ED_SSF_Jchi-2} (a-c). We note that the SSF peak at $\mathbf{q}=(0,0)$ is much stronger in the d-wave channel than those of other channels.
	Besides, we also show the different SSFs from the $4\times4$ torus in Fig.\ref{figs_ED_SSF_Jchi-2}(b-f), with similar features to those of the $6\times6$ torus.
	Further, we show the extrapolation of the d-wave SSF at $\mathbf{q}=(0,0)$ and the chiral p-wave SSF at $\mathbf{q}=(\pi,\pi)$ in Fig.\ref{figs_ED_SSF_Jchi-2_extrap}(a), and they both extrapolate to finite values in the thermodynamic limit. 
	The robust nematic fluctuations from these systems with exact $C_4$ rotation symmetry suggest that the nematic mode reported in the main text is intrinsic in this CSL phase.
	In addition, the SSFs scale extensively with the system size. Accordingly, normalizing them by the number of lattice sites yields an intensive quantity, which can be used as a measure of the strength of the corresponding order parameter.
	The extrapolations of the intensive quantities are shown in Fig.\ref{figs_ED_SSF_Jchi-2_extrap} (b). The extrapolated zeros support that the peaks in the SSFs only correspond to fluctuations and there are no symmetry-breaking orders.

\begin{figure}[htp!]
		\centering		
		\includegraphics[width=1\textwidth]{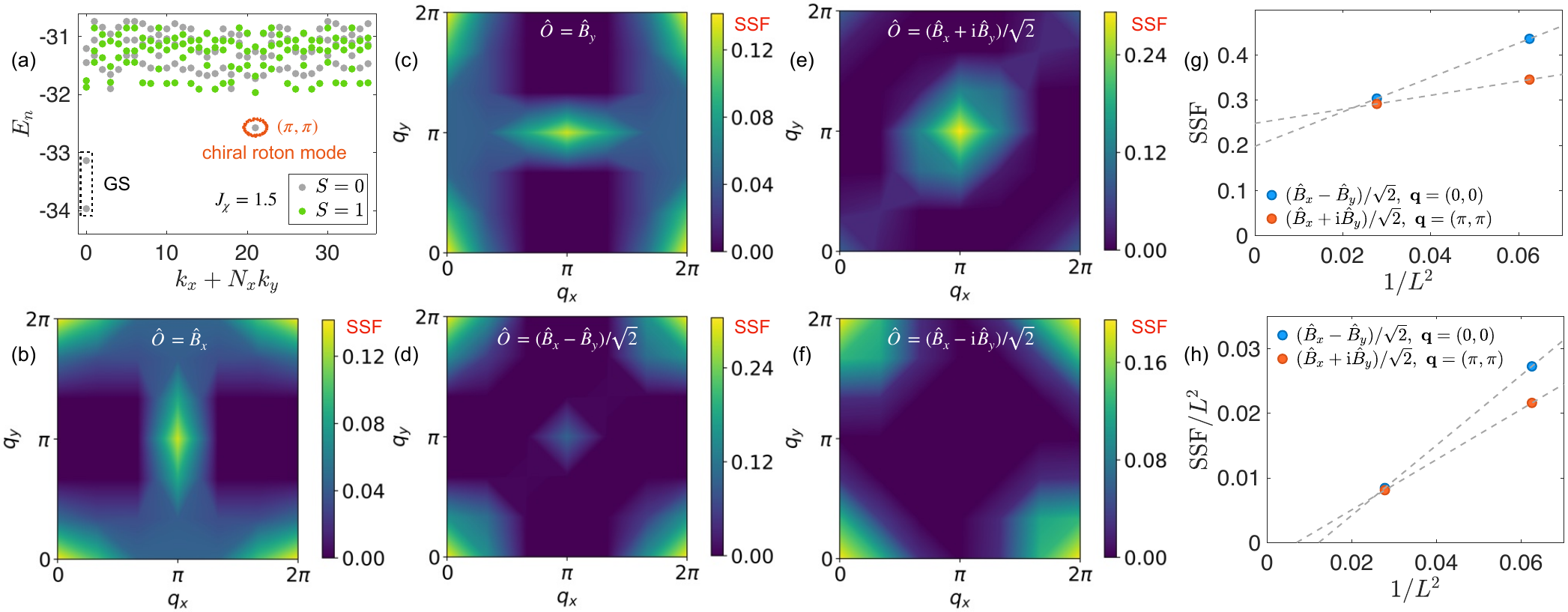}
		\caption{\textbf{ED spectrum and SSFs at $J_2=0,\ J_\chi=1.5$.} The ED spectrum from the $6\times6$ torus is shown in panel (a). The corresponding SSFs of different parameters from the $6\times6$ torus are shown in panels (b-f). 
			The extrapolation of some (g) extensive SSFs and (h) the corresponding intensive values (normalized by lattice sites) from ED simulations on symmetric tori are shown.
		}
		\label{figs_ED_spectra_SSF_Jchi-1.5}
	\end{figure}

	For more examples, we show the ED results of the CSL phase at $J_2=0,\ J_\chi=1.5$.
	In Fig.\ref{figs_ED_spectra_SSF_Jchi-1.5} (a), the low-energy spectrum from the $6\times6$ torus is shown, which is similar to that at $J_\chi=1.5$ and the chiral roton mode is the lowest excitation here.
	We show the SSFs of different operators from the $6\times6$ torus as well, with similar features as reported earlier: strong d-wave nematic fluctuations at $\mathbf{q}=(0,0)$ and chiral p-wave fluctuations at $\mathbf{q}=(\pi,\pi)$.

	We have also extrapolated the peaks of the SSFs as well as those normalized by the lattice sites, as shown in Fig.\ref{figs_ED_spectra_SSF_Jchi-1.5}(g,h), respectively. 
	The results are consistent with our earlier conclusions that the peaks of such SSFs are due to the intrinsic fluctuations instead of the symmetry-breaking orders.
	
	Besides, for the ED spectra, we would like to clarify that the finite-size splitting of the topological ground-state manifold and the excitation energy of the ($\pi,\pi$) roton mode have different physical origins and need not converge on the same scale.
		The former reflects the lifting of the topological ground-state degeneracy on a finite torus and is expected to vanish in the thermodynamic limit, whereas the latter corresponds to the bulk collective excitation that should remain at finite energy.
		Besides, even within the same accessible system size in ED, the splitting of the two topological ground states is parameter dependent. 
		While the splitting is larger when $J_2=0$, it becomes much smaller with finite $J_2$.
		For example, at $J_2=0.6,\ J_\chi=0.5$, as shown in Fig. 5 of the main text and Fig.~\ref{figs_chiral_roton_Jchi0.5_J20.6}, the two ground states are already much closer in energy, and the roton gap is more than 5 times larger than the  energy splitting of the two ground states.
		Besides, we note that, at these different parameters [Fig. 3(a) and Fig. 5(a)], despite the large difference in the splitting of the two topological ground states, the roton gap remains very close, which also illustrates why the topological ground-state splitting should not be directly compared with the excitation energy of the roton as they represent different finite-size energy scales.

	\begin{table}[htp]
		\centering
		\begin{tabular}{l|c|c|c|c}
			\hline 
			Overlaps & \multicolumn{4}{c}{$|\langle\psi_\mathrm{e}(0,0)|\hat{O}|\tilde{\psi}_0(0,0)\rangle|$} 
			\\
			\hline
			\diagbox[width=6em,height=3.55em]{Size:\\ $L_y\times L_x$}{\raisebox{-2.5ex}{\hspace{-1.6em}$\hat{O}$}} & $\hat{B}_{\mathbf{x}}$ & $\hat{B}_{\mathbf{y}}$ & $\hat{B}_{\mathbf{x}}-\hat{B}_{\mathbf{y}}$  & $\hat{B}_{\mathbf{x}}+\hat{B}_{\mathbf{y}}$   \\
			\hline
			\hspace{0.7em}$4\times4$ & 0.59 & 0.59 & 0.80 & 0.00 \\
			\hline        
		\end{tabular}
		\caption{\textbf{Normalized overlap of the $q=0$ mode.} We show the absolute values of the normalized overlaps  from the ED results for different bond operators at $J_2=0,\ J_\chi=2$ on the $4\times4$ torus. 
				$|\tilde{\psi}_{0}(0,0)\rangle$ is the ground state in the (0,0) momentum sector  after normalization by imposing $\langle\tilde{\psi}_{0}(0,0)|\hat{O}^\dagger\hat{O}|\tilde{\psi}_{0}(0,0)\rangle=1$.
				$\langle\psi_\mathrm{e}(0,0)|$ is the target excited state in the (0,0) momentum sector (the 14th state of this sector).
		}
		\label{tabs_nematic_overlap_4x4_Jchi_2_J2_0}
	\end{table}
	
	To further support the intrinsic d-wave nature of the q=0 mode, we show the normalized overlap between the target excited state and different variational states at $J_2=0,\ J_\chi=2$ on the symmetric $4\times4$ torus in Table.~\ref{tabs_nematic_overlap_4x4_Jchi_2_J2_0}.
		The overlaps of the $\hat{B}_x$ and $\hat{B}_x$ operators are the same, and are both much smaller than that of the d-wave operator $\hat{D}_{x^2-y^2}=\hat{B}_x-\hat{B}_y$. 
		Besides, we also show the exactly zero overlap of the uniform operator $\hat{B}_x+\hat{B}_y$ in the s-wave channel.
		These results on the symmetric torus also strongly support that the d-wave nematic nature is intrinsic for the $q=0$ mode.

	\subsection{II.B Robust chiral roton across parameters within the CSL phase}
	
	To further substantiate the chiral roton mode, we show the overlaps between the $(\pi,\pi)$ excited state and the corresponding trial states at different parameters (including $J_\chi=1.5$) in Tab.\ref{tabs_roton_overlap}, which further supports the chiral p-wave nature of this singlet roton mode. We note that, the chiral roton remains the lowest excitation for these considered parameters.

	\begin{table}[htp!]
		\centering
		\begin{tabular}{l|c|c|c|c}
			\toprule
			\hline
			\cmidrule(lr){2-3}\cmidrule(lr){4-5}
			Overlaps & $\hat{B}_x$ & $\hat{B}_y$ 
			& $\hat{B}_x+i\hat{B}_y$  & $\hat{B}_x-i\hat{B}_y$   \\
			\hline
			\midrule
			$4\times4$, $J_\chi=1.3$ & 0.856 & 0.856 & 0.960  & 0.000 \\
			\hline
			$4\times4$, $J_\chi=1.5$ & 0.864 & 0.864 & 0.964  & 0.000 \\
			\hline
			$4\times4$, $J_\chi=1.7$ & 0.863 & 0.863 & 0.964  & 0.000 \\
			\hline
			$4\times4$, $J_\chi=1.9$ & 0.861 & 0.861 & 0.962  & 0.000 \\
			\hline
			$4\times6$, $J_\chi=1.3$ & 0.827 & 0.859 & 0.944  & 0.012 \\
			\hline
			$4\times6$, $J_\chi=1.5$ & 0.818 & 0.870 & 0.944  & 0.072 \\
			\hline
			$4\times6$, $J_\chi=1.7$ & 0.812 & 0.866 & 0.938  & 0.080 \\
			\hline
			$4\times6$, $J_\chi=1.9$ & 0.809 & 0.858 & 0.930  & 0.069 \\
			\bottomrule
		\end{tabular}
		\caption{\textbf{The overlaps $\langle\psi_0(\pi,\pi)|\hat{O}|\tilde{\psi}_0(0,0)\rangle$ of the roton mode}.  Here, $|\psi_0(\mathbf{k})\rangle$ is the lowest-energy state of the $\mathbf{k}$ sector. $|\tilde{\psi}_0(0,0)\rangle$ is normalized such that $\langle\tilde{\psi}_0(0,0)|\hat{O}^\dagger\hat{O}|\tilde{\psi}_0(0,0)\rangle=1$. ED results with fixed $J_2=0$, and different $J_\chi$, $\hat{O}$,  and system sizes are listed.
		}
		\label{tabs_roton_overlap}
	\end{table}
	
\begin{figure}[htp!]
		\centering		
		\includegraphics[width=0.78\textwidth]{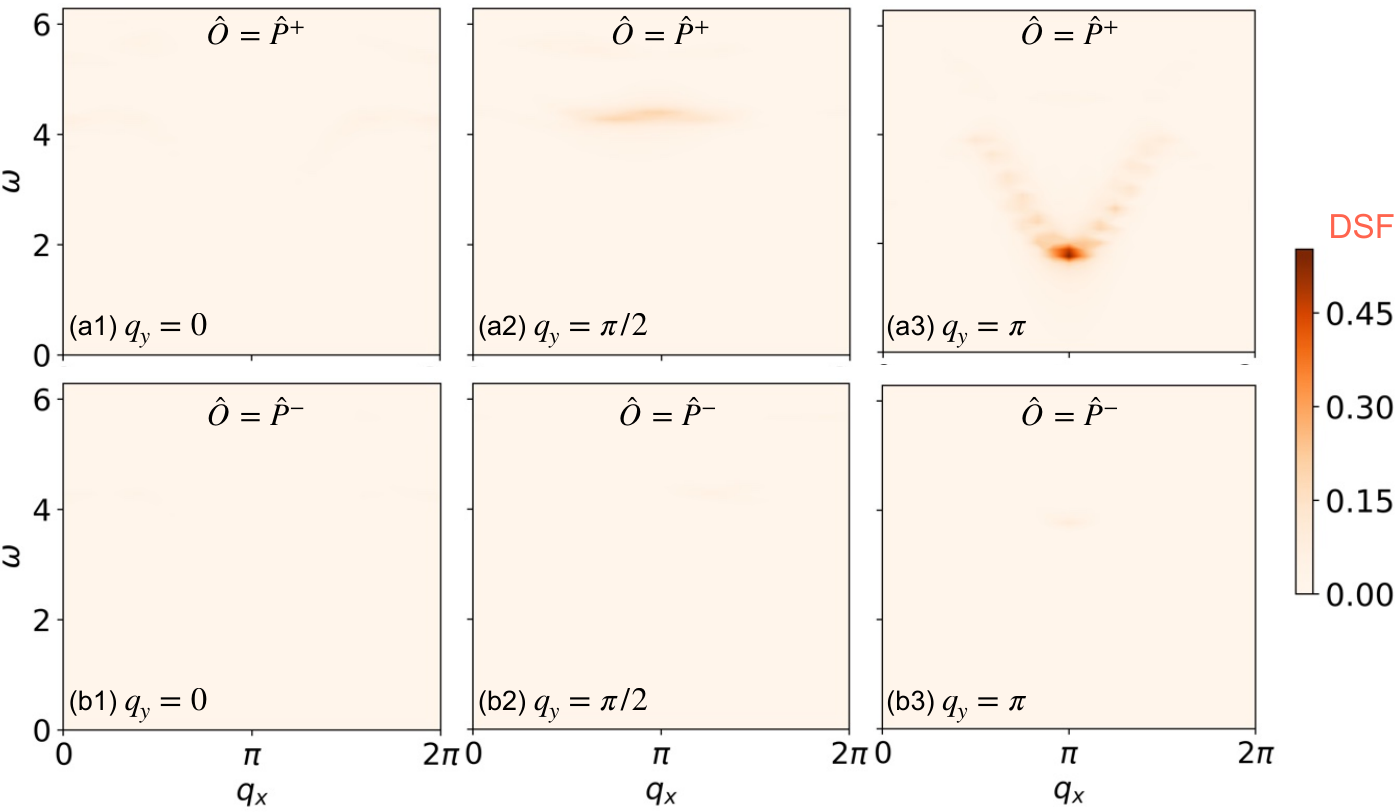}
		\caption{\textbf{DSFs in the chiral p-wave channels.} TDVP results obtained at $J_2=0$ and $J_\chi=2$. The horizontal/vertical axis is the energy/$q_x$ in each panel with different fixed $q_y$.
		}
		\label{figs_Jchi2_chiral_pwave}
	\end{figure}

	In Fig.4 of the main text, we have shown the DSFs in the chiral p-wave channels with fixed $\omega=1.76$ (the energy scale of the roton minimum). Here, as a complement, we show their $\omega$-dependence in Fig.\ref{figs_Jchi2_chiral_pwave}.
	The dispersion of the chiral roton is the same as that from the $\hat{B}_\mu$ operators and the spectral response fades away at other momenta.
	
\begin{figure}[htp!]
		\centering		
		\includegraphics[width=0.78\textwidth]{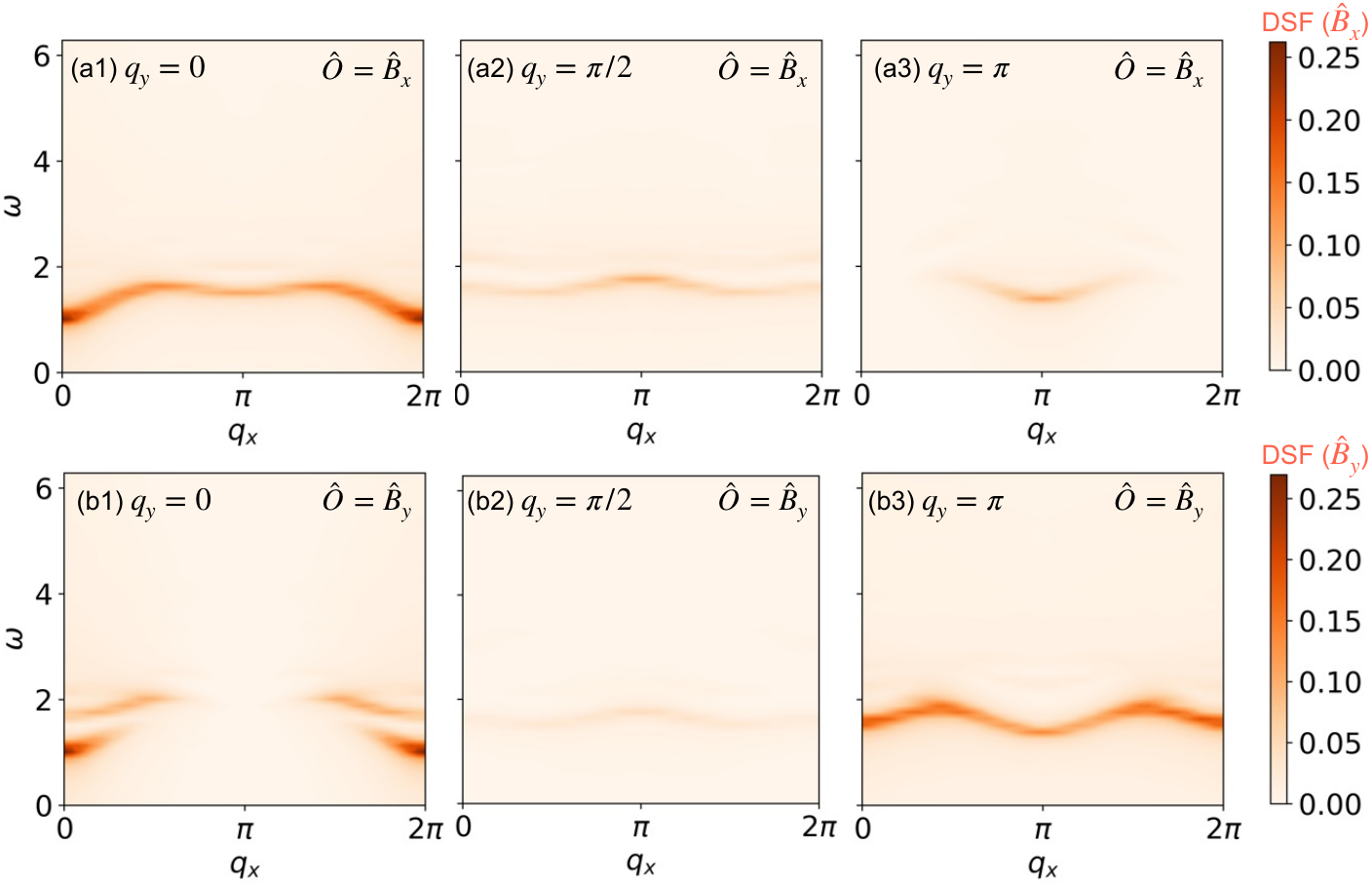}
		\caption{\textbf{Complementary DSFs at $J_2=0.6,\ J_\chi=0.5$.} Here, we show the DSFs of the bond operators $\hat{B}_x$ and $\hat{B}_y$, respectively. 
		}
		\label{figs_DSF_bond_Jchi0.5_J20.6}
	\end{figure}

\begin{figure}[htp!]
		\centering		
		\includegraphics[width=0.53\textwidth]{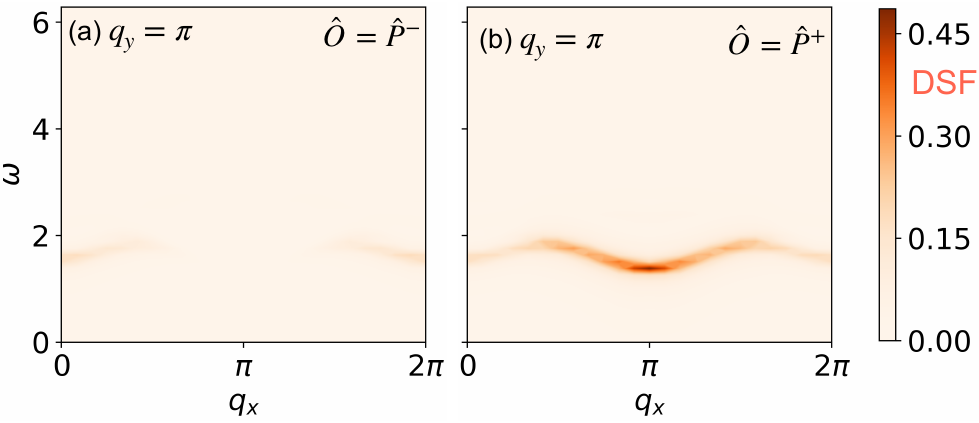}
		\caption{\textbf{Chiral roton mode at $J_2=0.6,\ J_\chi=0.5$.} Here, we show the DSFs in the chiral p-wave channels to demonstrate the low-energy $S=0$ chiral roton mode at ($\pi,\pi$).
		}
		\label{figs_chiral_roton_Jchi0.5_J20.6}
	\end{figure}

	With strong competing $J_2$, we have demonstrated that the nematic mode would be pronouncedly soft and the DSFs in the d-wave channels at $J_2=0.6,\ J_\chi=0.5$ are shown in Fig.5 of the main text for example, which have vanishing response at finite $\mathbf{q}$.
	For complementary information of the singlet excitations near the phase boundaries, we further show the DSFs of the bond operators $\hat{B}_\mu$ at $J_2=0.6,\ J_\chi=0.5$ in Fig.\ref{figs_DSF_bond_Jchi0.5_J20.6}.
	While the lowest excitation is the nematic mode at $\mathbf{q}=0$ with maximum spectral weight, the singlet excitation gap at ($\pi,\pi$) is still very low.
	More importantly, as shown in Fig.\ref{figs_chiral_roton_Jchi0.5_J20.6}, the chiral p-wave roton mode with a sharp spectral peak at ($\pi,\pi$) is still prominent even near the phase boundaries.
	Besides, despite the sharp response in the d-wave channels at $q=0$, the responses of the $q=0$ mode in the $\hat{B_x}$ and $\hat{B_y}$ channels are very close even in this cylinder geometry [Fig.\ref{figs_DSF_bond_Jchi0.5_J20.6}(a1,b1)], further support that the d-wave nematic nature is intrinsic for the $q=0$ mode.
	
	\subsection{II.C Additional information of TDVP simulations}
\begin{figure}[htp!]
		\centering		
		\includegraphics[width=\textwidth]{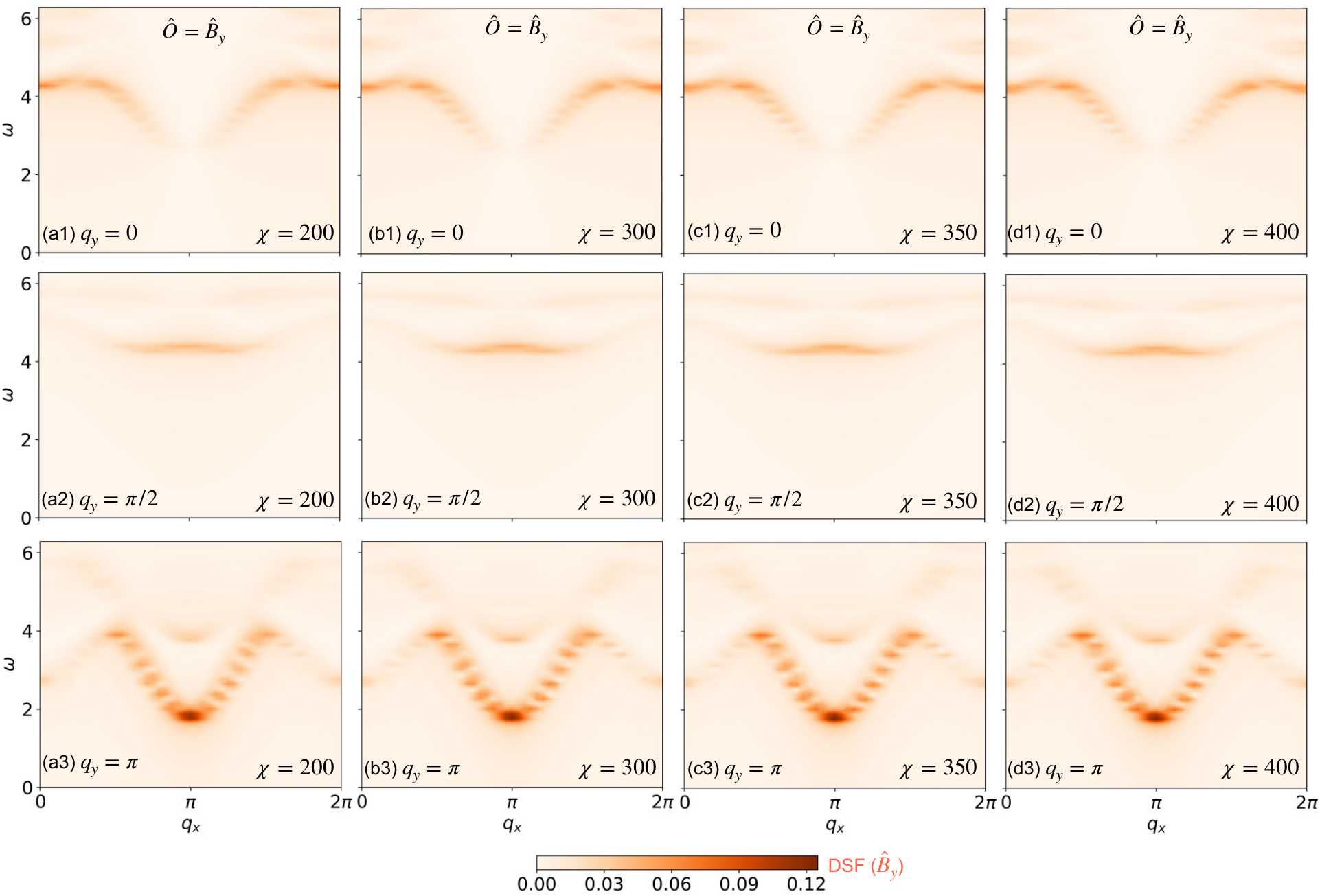}
		\caption{\textbf{TDVP results with different bond dimensions.} Here, we take the DSFs of the $\hat{B}_y$ operator from the $4\times32$ cylinders, as an example to illustrate the convergence with respect to the bond dimension. The model parameters are $J_2=0$ and $J_\chi=2$.
		}
		\label{figs_DSF_By_bond_dimension}
	\end{figure}

	To demonstrate the bond dimension up to $\chi=400$ in the TDVP simulations is well enough for the systems considered in the main text, we show the DSFs obtained from different maximum bond dimensions in the $\hat{B}_y$ channel as an example in Fig.\ref{figs_DSF_By_bond_dimension}, and the spectra are well converged.
	
	\clearpage
	
	\subsection{II.D TDVP results in wider and longer cylinders}
\begin{figure}[htp!]
		\centering		
		\includegraphics[width=\textwidth]{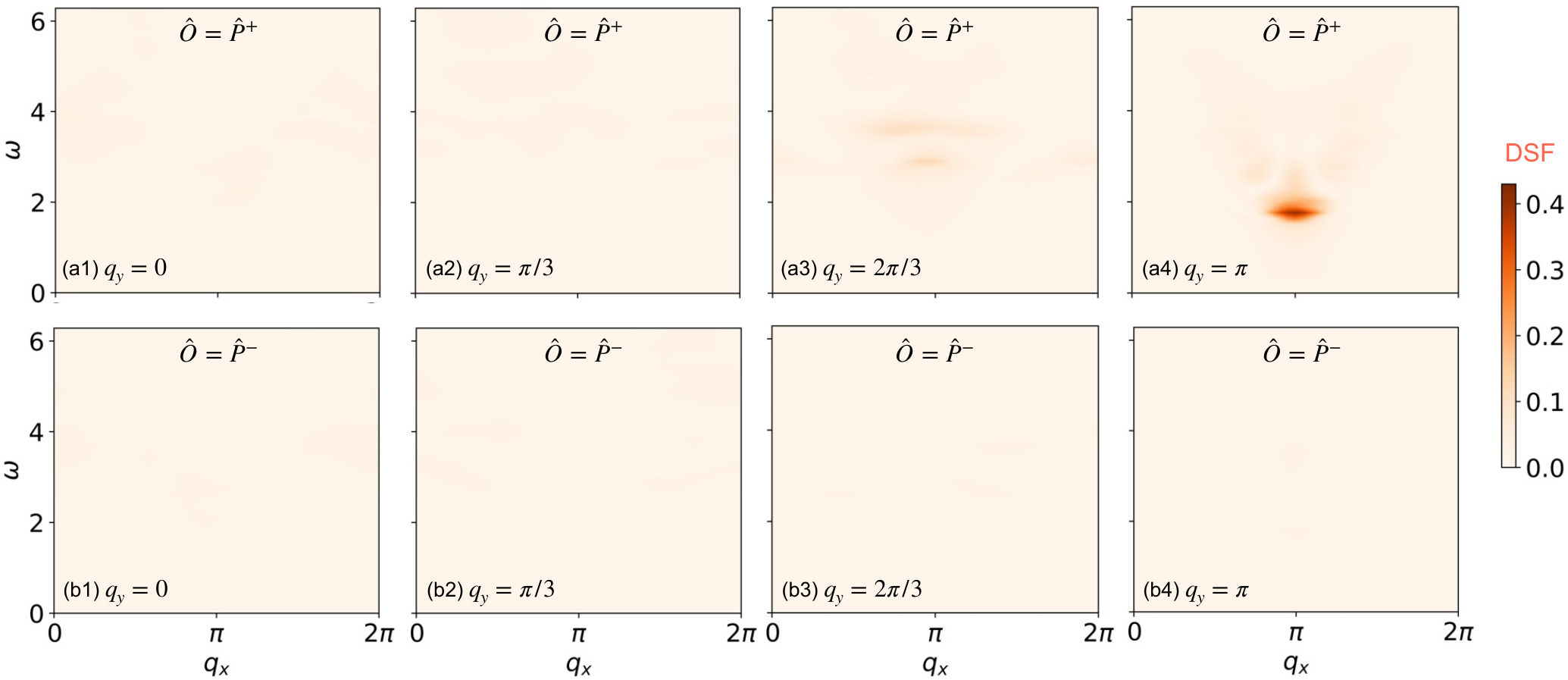}
		\caption{\textbf{Chiral p-wave roton from the $6\times24$ cylinder}. These spectra are from  TDVP simulations at $J_2=0$ and $J_\chi=2$, with bond dimension $\chi=400$. 
				The chiral roton is at $\mathbf{q}=(\pi,\pi)$ and $\omega=1.76$.
		}
		\label{figs_W6L24_chiral_roton_spectrum}
	\end{figure}

	To study the chiral roton mode beyond the $L_y=4$ cylinders, we have performed TDVP simulations on the $L_y\times L_x=6\times24$ cylinder.
		The results at $J_2=0,\ J_\chi=2$ are shown in Fig.~\ref{figs_W6L24_chiral_roton_spectrum}.
		We consistently find the sharp chiral p-wave roton mode at $(\pi,\pi)$.
		The mode remains strongly selected by the same $\hat{P}^+$ channel,
		and its excitation energy is essentially unchanged from that for the $L_y=4$ cylinder, within the numerical frequency resolution.
		Besides, the scales of roton gap in these TDVP results are very close to those of the ED results at this parameter, suggesting that the finite-size dependence is very small for the robust singlet ($\pi,\pi$) roton mode.
		These results from different system sizes, with consistent gap value and chiral p-wave nature, further support the robustness of the chiral roton mode.

\begin{figure}[htp!]
		\centering		
		\includegraphics[width=0.8\textwidth]{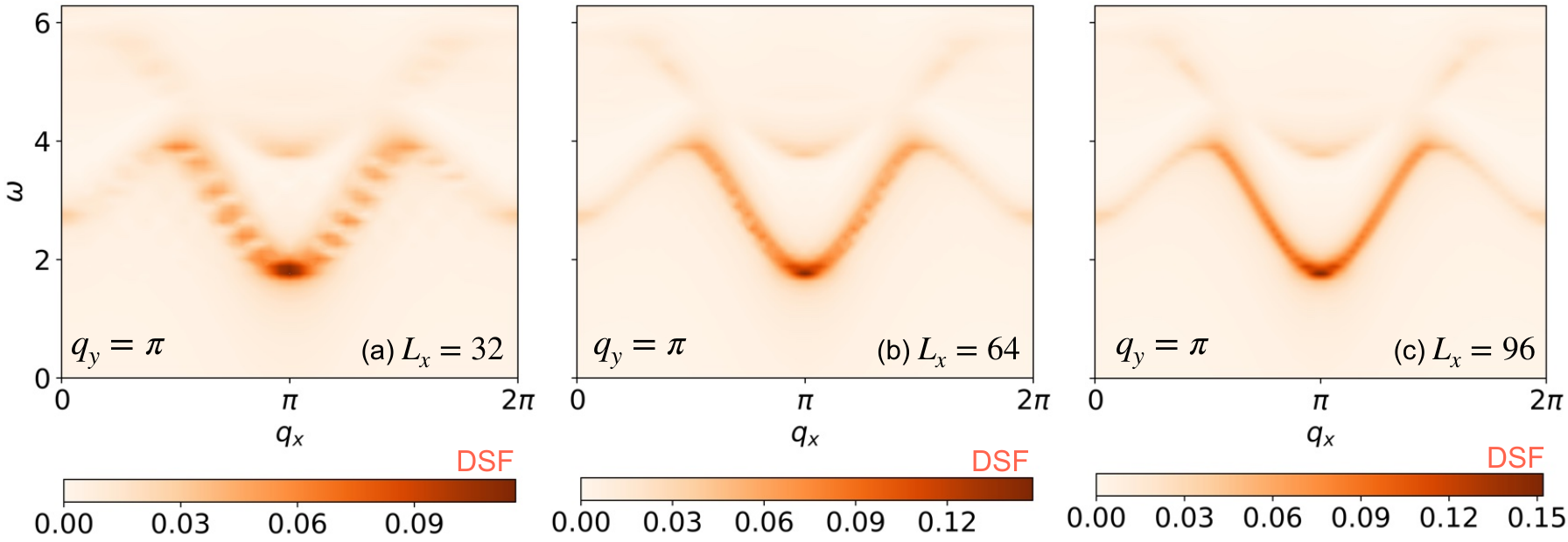}
		\caption{\textbf{Dynamical structure factors with different $L_x$}.
				We take the $\hat{B}_y$ operator for example to compare the spectral functions from the $L_y=4$ cylinders with different $L_x$= (a) 32, (b) 64, and (c) 96. These results are from TDVP simulations with bond dimension $\chi=200$.
		}
		\label{figs_By_Jchi2_different_Lx}
	\end{figure}

	Besides, we note that there exist seemingly discrete behaviors of the spectral weights in such as Fig.2 a2, b3, and c3 of the main text, and this is indeed due to the the finite cylinder length $L_x$ that produces a discrete set of accessible $q_x$ points.
		We take the $\hat{O}=\hat{B}_y$ spectrum with $q_y=\pi$ at the same parameter $J_2=0,\ J_\chi=2$ for example. The spectra of the $L_y=4$ cylinders with different $L_x$ are shown in Fig.~\ref{figs_By_Jchi2_different_Lx}.
		At $L_x=32$, the dispersive excitations look discrete due to the finite number of data points. At $L_x=64$, the discrete behavior basically disappears, and the spectrum gets more smooth at $L_x=96$.
		Despite these, the essential features of the spectrum are already well captured by the results in shorter cylinders.

	\subsection{II.E Choice of local probe for the chiral roton mode}
	In this work, the local probe of the chiral roton mode is based on the NN bond operators. Here, we examine whether other local singlet operator can reveal the finite-momentum mode.
		We perform the SMA using the next-NN bond operators for comparison. We start from the parameters $J_2=0,\ J_\chi=2$, where Table I of the main text shows very high overlaps between the target excited state at $(\pi,\pi)$ and the variational states made of NN bond operators.
		The overlaps of the variational states using next-NN bond operators ($\hat{B}_{\mathbf{x},\mathbf{y}}$ and $\hat{B}_{\mathbf{x},-\mathbf{y}}$) are shown in Table.~\ref{tabs_roton_overlap_NNN_J2_0_Jchi_2}, and the results turn out to be zero, showing that the next-NN operators do not reveal the same finite-momentum mode.
		Further, we consider another parameter at $J_\chi=0.5$ and finite $J_2=0.6$, and again compare the overlaps between different variational states and the target roton mode at ($\pi,\pi$).
		The results are shown in Table.~\ref{tabs_roton_overlap_J2_0.6_Jchi_0.5}.
		The NN bond operators couple strongly to the chiral roton mode, while the next-NN bond operators still show zero overlaps.

	\begin{table}[htp!]
		\centering
		\begin{tabular}{l|c|c|c|c}
			\toprule
			\hline 
			Overlaps & \multicolumn{4}{c}{$|\langle\psi_0(\pi,\pi)|\hat{O}|\tilde{\psi}_0(0,0)\rangle|$} 
			\\
			\hline
			\diagbox[width=6em,height=3.6em]{Size:\\ $L_y\times L_x$}{\raisebox{-2.5ex}{\hspace{-1.6em}$\hat{O}$}} & $\hat{B}_{\mathbf{x},\mathbf{y}}$ & $\hat{B}_{\mathbf{x},\mathbf{-y}}$ & $\hat{B}_{\mathbf{x},\mathbf{y}}+i\hat{B}_{\mathbf{x},\mathbf{-y}}$  & $\hat{B}_{\mathbf{x},\mathbf{y}}-i\hat{B}_{\mathbf{x},-\mathbf{y}}$   \\
			\hline
			\midrule
			\hspace{0.7em}$4\times4$ & 0.00 & 0.00 & 0.00 & 0.00 \\
			\hline 
			\bottomrule
		\end{tabular}
		\caption{\textbf{SMA using next-NN bond operators}. The results are calculated at $J_2=0,\ J_\chi=2$, and the overlaps of the variational states using next-NN bond operators and the target excited state at $(\pi,\pi)$ are zero. 
		}
		\label{tabs_roton_overlap_NNN_J2_0_Jchi_2}
	\end{table}

	\begin{table}[!htp]
		\centering
		\begin{tabular}{l|c|c|c|c|c|c|c|c}
			\toprule
			\hline
			Overlaps & \multicolumn{8}{c}{$|\langle\psi_0(\pi,\pi)|\hat{O}|\tilde{\psi}_0(0,0)\rangle|$} \\
			\hline
			\multirow{2}{*}{
				\diagbox[width=6em,height=2.45em]
				{$L_y\times L_x$}
				{\raisebox{-1.7ex}{\hspace{2.6em}$\hat{O}$}}
			}
			& \multicolumn{4}{c|}{NN bond operators}
			& \multicolumn{4}{c}{Next-NN bond operators}
			\\
			\cline{2-9}
			&
			$\hat{B}_{\mathbf{x}}$
			&
			$\hat{B}_{\mathbf{y}}$
			&
			$\hat{B}_{\mathbf{x}}+i\hat{B}_{\mathbf{y}}$
			&
			$\hat{B}_{\mathbf{x}}-i\hat{B}_{\mathbf{y}}$
			&
			$\hat{B}_{\mathbf{x},\mathbf{y}}$
			&
			$\hat{B}_{\mathbf{x},\mathbf{-y}}$
			&
			$\hat{B}_{\mathbf{x},\mathbf{y}}
			+i\hat{B}_{\mathbf{x},\mathbf{-y}}$
			&
			$\hat{B}_{\mathbf{x},\mathbf{y}}
			-i\hat{B}_{\mathbf{x},\mathbf{-y}}$
			\\
			\midrule
			\hline
			\hspace{0.7em}$4\times4$ & 0.88 & 0.88 & 0.97 & 0.00 & 0.00 & 0.00 & 0.00 & 0.00 \\ 
			\hline
			\bottomrule
		\end{tabular}
		\caption{\textbf{SMA at $J_\chi=0.5,\ J_2=0.6$.} The overlaps of the variational states using both the NN and next-NN bond operators are calculated. While the NN bond operators well reveal the chiral roton mode, the next-NN bond operators show zero overlaps.
		}
		\label{tabs_roton_overlap_J2_0.6_Jchi_0.5}
	\end{table}
	
	\subsection{II.F Partial helicity
			selectivity of the $q=0$ d-wave mode}
	
	In the main text, we have introduced that the $q=0$ d-wave mode is elliptically polarized as the polarization ratio of the spectral weights in the $\hat{D}^\pm$ channels is not so large from the TDVP results. 
		For the ED results in Tab. II, we have shown the normalized overlap:
		$\frac{|\langle \psi_\mathrm{excited\ state}|\hat{O}|\psi_\mathrm{ground\ state} \rangle|}{\sqrt{\langle \psi_\mathrm{ground\ state}|\hat{O}^\dagger\hat{O}|\psi_\mathrm{ground\ state} \rangle}},$
		such that the chirality dependence is less clear than the spectral weights from TDVP results. Therefore, we further show the spectral weights $|\langle \psi_\mathrm{excited\ state}|\hat{O}|\psi_\mathrm{ground\ state} \rangle|^2$ of the two $\hat{D}^\pm$ channels in Table.\ref{tabs_dwave_mode_spectral_weights_J2_0.6_Jchi_0.5}, which are obtained from the same ED calculations for the normalized overlaps in Table II of the main text.
		The chirality dependence of the $\hat{D}^\pm$ channels in ED results is also clear from the spectral weights, and the relative difference is consistent with the TDVP results.
		Next, we provide more explanation about the partial helicity
		selectivity.
	
	\begin{table}[htp!]
		\centering
		\begin{tabular}{l|c|c|c|c|c|c}
			\toprule
			\hline
			Spectral weight & \multicolumn{3}{c|}{$|\psi_0(0,0)\rangle$} & \multicolumn{3}{c}{$|\psi_1(0,0)\rangle $}\\
			\hline
			\hspace{2.2em}$\hat{O}$ &
			$\hat{D}_{x^2-y^2}$ &
			$\hat{D}^+$ &
			$\hat{D}^-$ &
			$\hat{D}_{x^2-y^2}$ &
			$\hat{D}^+$ & 
			$\hat{D}^-$ \\
			\midrule
			\hline
			\hspace{1.2em} $\langle\psi_2(0,0)|$ & 1.43 & 2.04 & 0.94 & 0.00 & 0.00 & 0.00 \\  
			\midrule
			\hline
			\hspace{1.2em} $\langle\psi_3(0,0)|$ & 0.00 & 0.00 & 0.00 & 1.56 & 2.06 & 1.13 \\  
			\bottomrule
			\hline
		\end{tabular}
		\caption{\textbf{The spectral weights of the d-wave mode}. The spectral weights ($\left|\langle\psi_\mathrm{excited\ state}|\hat{O}|\psi_\mathrm{ground\ state}\rangle\right|^2$) of different d-wave operators are computed at $J_2=0.6,\ J_\chi=0.5$ on the $6\times6$ torus. $|\psi_{0,1}(0,0)\rangle$ are the two lowest states in the (0,0) momentum sector (the 2-fold degenerate GSs of the CSL).
				$\langle\psi_{2,3}(0,0)|$ are the two lowest excited states above the two GSs in the (0,0) momentum sector.
		}
		\label{tabs_dwave_mode_spectral_weights_J2_0.6_Jchi_0.5}
	\end{table}

		Let us briefly review the chiral graviton in FQH systems.
		For the continuum Laughlin FQH states, the two helicity sectors \(l=+2\) and \(l=-2\) are distinguished by continuous rotational symmetry. For the model Hamiltonians (such as the $V_0$ pseudopotential for the bosonic 1/2 state), the response in the opposite graviton helicity can vanish identically~\cite{Golkar_2016,Liou2019_graviton}. 
		For more generic interactions, such as the Coulomb interaction, the opposite-helicity response need not vanish exactly, 
		although it is generally strongly suppressed compared with the dominant helicity channel. 
		In the studied square-lattice CSL here, by contrast, the two d-wave channels show much weaker helicity selectivity.
		
		An important difference from the continuum system is the discrete lattice rotation symmetry, which can affect the identification of opposite chiral channels.
		On a lattice with ($C_n$) rotational symmetry, the crystal angular momentum is defined modulo ($n$), so that angular-momentum channels differing by ($n$) are equivalent.
		Therefore, for the square lattice with discrete $C_4$ symmetry,  $l=\pm2$ are the same, and the $C_4$ symmetry alone cannot distinguish $\pm2$.
		However, in the $C_4$ symmetric system, two d-wave channels $D_{x^2-y^2}$ and $D_{xy}$ belong to the same rotational sector, and can hybridize.
		Thus, it is natural in practice to consider the more general pair of chiral operators:
		$$D^{\pm}_\alpha=D_{x^2-y^2}\pm i\alpha D_{xy},$$ where $\alpha$ controls the relative normalization of the two d-wave components.
		Note that, these two channels can also hybridize.
		Then we can analyze the spectral properties.
		For the ground state $|0\rangle$ and the target excited state $|e_n\rangle$, we can define: 
		$$\langle e_n|D_{x^2-y^2}|0\rangle=a_1e^{i\phi_1},\ \langle e_n|D_{xy}|0\rangle=a_2e^{i\phi_2}. $$
		The relative magnitude is $a_2/a_1$ and the relative phase is $\delta=\phi_2-\phi_1$.
		So that for the designed $D^{\pm}_\alpha$ operators, the spectral weight and relative ratio are:
		$$I_\pm=a_1^2+\alpha^2a_2^2\mp2\alpha a_1a_2\sin{\delta},\ r=\frac{I_+}{I_-}=\frac{a_1^2+\alpha^2a_2^2-2\alpha a_1a_2\sin{\delta}}{a_1^2+\alpha^2a_2^2+2\alpha a_1a_2\sin{\delta}}.$$
		Hence, the ratio $r$ would depend on microscopic parameters that affect the relative magnitude $a_2/a_1$ and phase $\delta$. 
		In addition, as long as $a_1,a_2,\sin{\delta}\neq0$, the ratio $r$ would also be $\alpha$-dependent (relative normalization).
		
		For this square-lattice model, there is an additional symmetry constraint, which is important for the spectral ratio.
		Although the time-reversal symmetry ($\mathcal{T}$) and the mirror symmetry ($\mathcal{M}$) are both broken, the combined $\mathcal{MT}$ symmetry is still preserved.
		Besides, since the studied $q=0$ $D_{x^2-y^2}$ and $D_{xy}$ operators have opposite mirror parities, while they are both $\mathcal{T}$-even, the relative phase is required to be $\delta=\pm\frac{\pi}{2}$, and $\sin{\delta}=\pm1$. Our ED results are consistent with this constraint.
		Thus, the spectral ratio could  be simplified as $r=\frac{(a_1+\alpha a_2)^2}{(a_1-\alpha a_2)^2}$, which depends on $\alpha$ and the relative magnitude $a_2/a_1$.
		
\begin{figure}[htp!]
		\centering		
		\includegraphics[width=0.35\textwidth]{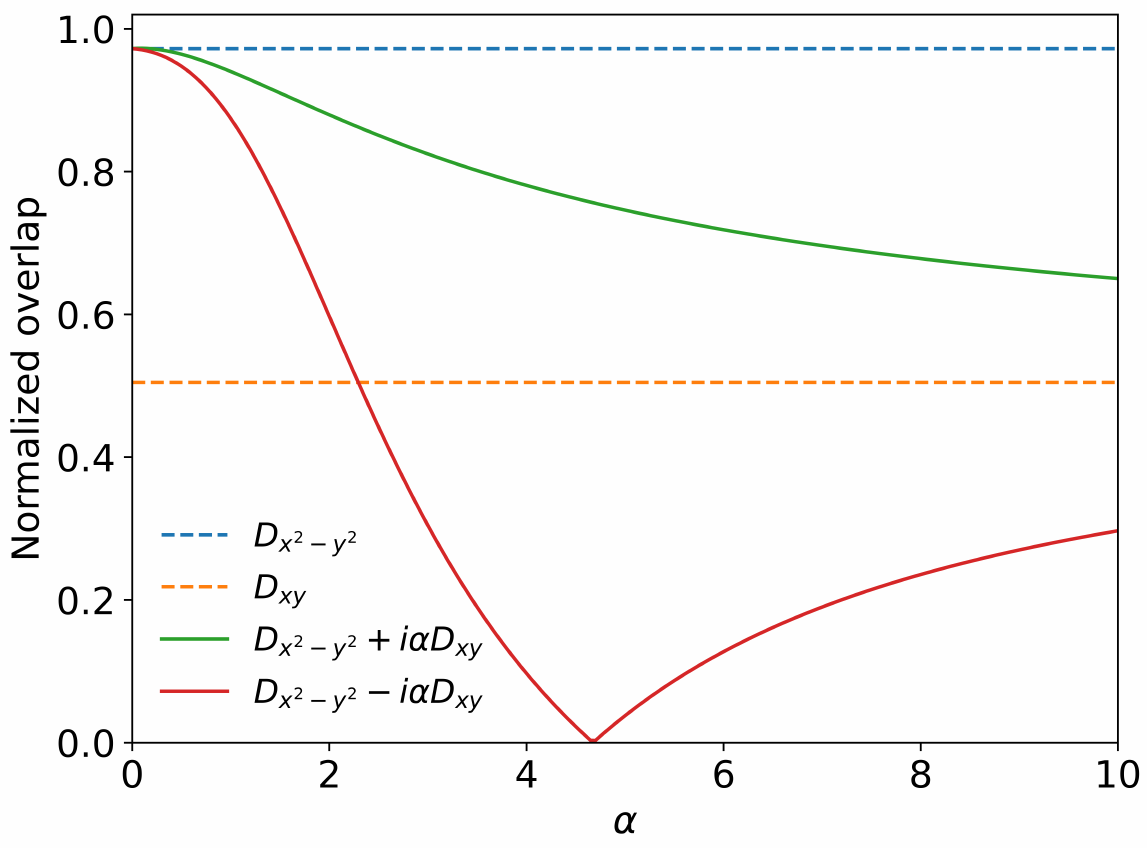}
		\caption{\textbf{The relative-normalization dependence of the $D^{\pm}_\alpha$ operators}. We calculate the normalized overlaps $|\langle e_n|\hat{O}|\tilde{0}\rangle|$ of the d-wave mode at $J_2=0.6,\ J_\chi=0.5$ when tuning the relative-normalization $\alpha$ on the $4\times4$ torus.  The results of $D_{x^2-y^2}$ and $D_{xy}$ operators are also shown for comparison.
		}
		\label{figs_overlap_dwave_alpha}
	\end{figure}

		There is an important subtlety concerning the meaning of polarization on the $C_4$ symmetric square lattice.
		Because of the constrained relative phase $\delta=\pm\frac{\pi}{2}$, and the simplified spectral ratio $r=\frac{(a_1+\alpha a_2)^2}{(a_1-\alpha a_2)^2}$, one can mathematically choose a particular value of $\alpha\rightarrow \frac{a_1}{a_2}$ such that  one of the two chiral matrix elements can become very small or even zero.
		For example, at $J_2=0.6,\ J_\chi=0.5$,
		we have checked that $\sin{\delta}=-1, \frac{a_1}{a_2}\sim4.7$ ($\sin{\delta}$ can be flipped by the sign of $J_\chi$).
		If $\alpha\rightarrow \frac{a_1}{a_2}=4.7$, the ratio can be extremely large.
		However, such a circularly polarized response reflects fine tuning rather than an intrinsic property of the mode. In particular, the strong polarization observed in certain \(\hat{D}^{\pm}_{\alpha}\) channels for specific choices of \(\alpha\) should not by itself be interpreted as evidence that the mode is intrinsically circularly polarized. To characterize the intrinsic nature of the \(q=0\) singlet mode, one should also examine the normalized overlap of the chiral operators with the corresponding eigenstates, ($|\langle e_n|\hat{D}^{\pm}_{\alpha}|\tilde{0}\rangle|$).
		As shown in Fig.~\ref{figs_overlap_dwave_alpha}, although the spectral ratio between the two $D_\alpha^\pm$ operators could be fine-tuned to be very large in a particular basis when $\alpha\rightarrow \frac{a_1}{a_2}\sim4.7$, the normalized overlap of the $D_\alpha^+$ channel also decreases substantially in the meantime. 
		Therefore, this is far from a physically motivated or desired approximation, especially compared to the $D_{x^2-y^2}$ operator alone. 
		The choice $\alpha=1$ used in this manuscript corresponds to a natural equal-weight combination. It provides a simple fixed convention without fine-tuning the probe to a particular eigenstate, while maintaining a substantial normalized overlap with the target mode.

		We have also checked the polarization ratio at model parameters with $J_2=0$. 
		At $J_\chi=1.5$ for example,  $\frac{a_1}{a_2}\sim3.6$.
		Although one can again fine-tune $\alpha\rightarrow\frac{a_1}{a_2}$ to strongly suppress one of the chiral channels,
		the normalized overlap of $D_{\alpha=\frac{a_1}{a_2}}^+$ is only $0.45$, which is substantially reduced compared with the $0.81$ overlap for $D_{x^2-y^2}$ alone.
		At the convention $\alpha=1$, the ratio is $r=3.18$.

		Therefore, the $C_4$ symmetry itself does not distinguish the $l=\pm2$ angular momentum, so we cannot identify the chirality or polarization directly from the $D_{x^2-y^2}$ operator alone. 
		By constructing the $D_\alpha^\pm$ operators within relatively reasonable choice of $\alpha$, the difference and polarization ratio of the spectral weights in the two channels are relatively small.

	\subsection{II.G The effect of the spin SU(2) symmetry}
	
		In this work, we study the CSL with spin SU(2) symmetry. While the topology order would not change without this spin symmetry, it is also interesting to study the importance of this symmetry for the collective modes discovered in this work.
		To address this question, we consider changing the Hamiltonian to:
		\begin{equation}
			\hat{H}=J_1\sum_{\langle i,j\rangle}
			(\hat{\mathbf{S}}_i^x\cdot\hat{\mathbf{S}}_j^x
			+\hat{\mathbf{S}}_i^y\cdot\hat{\mathbf{S}}_j^y
			+\delta\hat{\mathbf{S}}_i^z\cdot\hat{\mathbf{S}}_j^z)
			+J_2\sum_{\langle\langle i,j\rangle\rangle}\hat{\mathbf{S}}_i\cdot\hat{\mathbf{S}}_j
			+J_\chi\sum_{\triangle}\hat{\mathbf{S}}_i\cdot(\hat{\mathbf{S}}_j\times\hat{\mathbf{S}}_k).
		\end{equation}
		We introduce spin anisotropy through the relative strengths of the transverse and longitudinal components of the Heisenberg interaction, which provides a controlled and simple representative way of breaking SU(2) symmetry.
		We start from $J_\chi=2$, $J_2=0$, and $\delta=1$ (the SU(2) point), and gradually increase $\delta$.
		At large-$\delta$ limit, the ground state should be adiabatically connected to the classical Ising order. 
		However, we will focus on the parameter regime within the CSL phase and start from the chiral p-wave mode at ($\pi,\pi$).
	
	\begin{figure}[htp!]
		\centering		
		\includegraphics[width=0.3\textwidth]{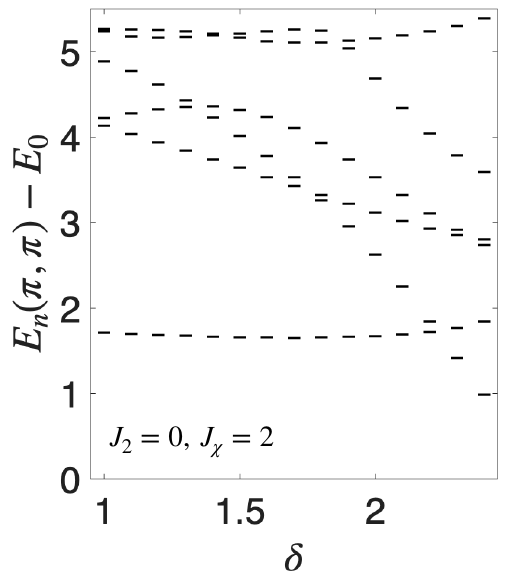}
		\caption{\textbf{Energy spectra in the $(\pi,\pi)$ momentum sector within the CSL phase}. 
				The low-energy states (subtracted by the ground state energy)  when tuning spin anisotropy $\delta$ are calculated on the $4\times4$ torus by ED. In this momentum sector, the chiral roton mode  remains the lowest excitation of the CSL phase until $\delta\sim2.2$.
		}
		\label{figs_Jchi_2_tune_J1z_gap_pipi}
	\end{figure}

	When tuning $\delta$, the low-energy spectra in the $(\pi,\pi)$ momentum sector within the CSL phase are shown in Fig.~\ref{figs_Jchi_2_tune_J1z_gap_pipi}. 
		At $\delta=1$, the lowest excited state is the chiral roton mode, and it remains the lowest excited state of this sector until $\delta\sim2.2$, although the SU(2) symmetry is broken.
		We further calculate the normalized overlaps and the spectral weights of different p-wave operators in Table.~\ref{tabs_chiral_roton_tune_spin_anisotropy}. 
		Apart from the spin-singlet operators, we also consider their transverse and longitudinal components, respectively.
		At large $\delta=2$ (away from the SU(2) symmetric point), the results are roughly similar to those at $\delta=1$, and the nature of the chiral roton is still well captured by the original p+ip operator. 
		Note that, at $\delta=1$, the spectral weights of the singlet, transverse, and longitudinal operators satisfy the exact ratio $9:4:1$ (values rounded in the table).
		The deviation from the relation is small  at $\delta=2$.
		Besides, the normalized overlaps of the original $\hat{P}^+$ operator are also larger than those of the transverse and longitudinal operators.
	
	\begin{table}[htp!]
		\centering
		\begin{tabular}{l|c|c|c|c|c|c}
			\toprule
			\hline
			Overlap (spectral weight) & \multicolumn{6}{c|}{$|\langle\psi_0(\pi,\pi)|\hat{O}|\tilde{\psi}_0(0,0)\rangle|$ \ $\left(|\langle\psi_0(\pi,\pi)|\hat{O}|\psi_0(0,0)\rangle|^2\right)$} \\
			\hline
			\hspace{4.2em}$\hat{O}$ &
			$\hat{P}^+$ &
			$\hat{P}^+_{xy}$ &
			$\hat{P}^+_{zz}$ &
			$\hat{P}^-$ &
			$\hat{P}^-_{xy}$ & 
			$\hat{P}^-_{zz}$    \\
			\midrule
			\hline
			\hspace{3.2em} $\delta=1$ & 0.96  (1.28) & 0.92 (0.57) & 0.83  (0.14)& 0.00 (0.00) & 0.00 (0.00) & 0.00 (0.00) \\  
			\midrule
			\hline
			\hspace{3.2em} $\delta=1.5$ & 0.96 (1.24) & 0.93 (0.55)& 0.83 (0.14)& 0.00 (0.00)& 0.00 (0.00)& 0.00 (0.00) \\  
			\midrule
			\hline
			\hspace{3.2em} $\delta=2$ & 0.96 (1.16)& 0.94 (0.52)& 0.81 (0.12)& 0.00 (0.00)& 0.00 (0.00)& 0.00 (0.00) \\  
			\hline
			\bottomrule
		\end{tabular}
		\caption{\textbf{SMA of the chiral p-wave roton when tuning $\delta$}. 
				We calculate the normalized overlaps $\left(|\langle\psi_0(\pi,\pi)|\hat{O}|\tilde{\psi}_0(0,0)\rangle|\right)$ and the spectral weights $\left(|\langle\psi_0(\pi,\pi)|\hat{O}|\psi_0(0,0)\rangle|^2\right)$ of different operators on the $4\times4 $ torus by ED.
				Apart from $P^\pm$ operators themselves, we also consider their transverse ($xy$) and longitudinal ($zz$) components, respectively. For example, $P^\pm_{xy}(\mathbf{r})=\left(S_x(\mathbf{r})S_x(\mathbf{r}+\mathbf{x})+S_y(\mathbf{r})S_y(\mathbf{r}+\mathbf{x})\right)\pm i\left(S_x(\mathbf{r})S_x(\mathbf{r}+\mathbf{y})+S_y(\mathbf{r})S_y(\mathbf{r}+\mathbf{y})\right)$, and $P^\pm_{zz}(\mathbf{r})=S_z(\mathbf{r})S_z(\mathbf{r}+\mathbf{x})\pm iS_z(\mathbf{r})S_z(\mathbf{r}+\mathbf{y})$.
		}
		\label{tabs_chiral_roton_tune_spin_anisotropy}
	\end{table}
	
	\begin{figure}[htp!]
		\centering		
		\includegraphics[width=\textwidth]{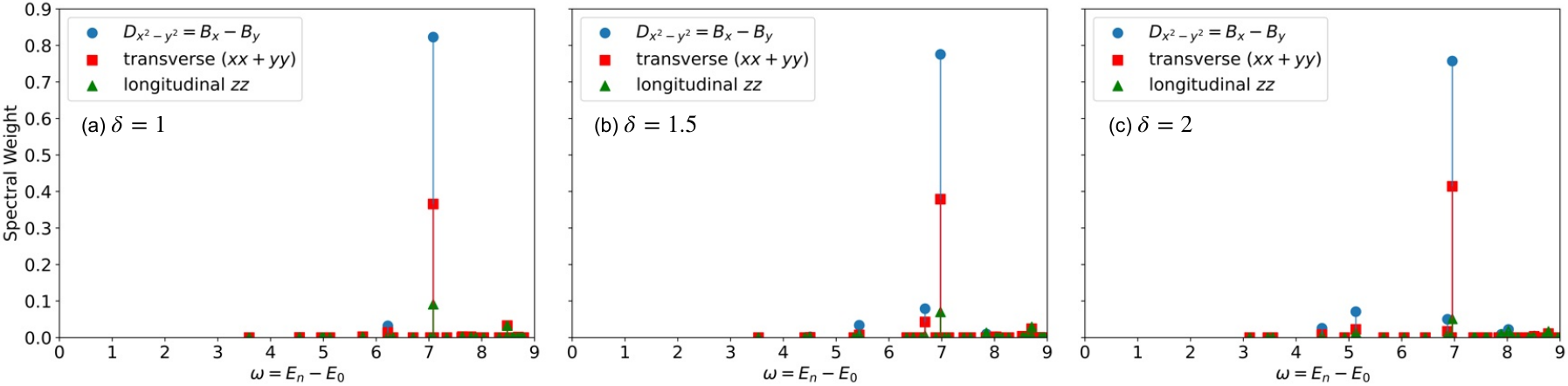}
		\caption{\textbf{The $q=0$ d-wave mode when tuning $\delta$}. We calculate the spectral weights $|\langle\psi_n(0,0)|\hat{O}|\psi_0(0,0)\rangle|^2$ of different d-wave operators (labelled by different colored markers) on the $4\times4$ torus by ED. Different parameters are considered: (a) $\delta=1$, (b) $\delta=1.5$, and (c) $\delta=2$.}
		\label{figs_dwave_spectral_weight_tune_delta}
	\end{figure}
	
		Next, we turn to study the $q=0$ d-wave mode when tuning $\delta$. The spectral weights of different d-wave operators at different $\delta$ are shown in Fig.~\ref{figs_dwave_spectral_weight_tune_delta}.
		As we have discussed the chirality of the d-wave mode in earlier parts of the reply, we focus on the $D_{x^2-y^2}$ operator and its transverse and longitudinal components.
		At large $\delta=2$, the $q=0$ d-wave mode is well captured by the original operator  $D_{x^2-y^2}$, while there is  some internal deviation from the exact proportion of different components at the SU(2) symmetric point.
		
		According to the results above, breaking the SU(2) symmetry by tuning the relative strengths of different spin components of the Heisenberg interaction does not appear to qualitatively affect the persistence of the channel-selective modes. 
		Nevertheless, we cannot exclude the possibility that these modes may become substantially weakened in more strongly spin-anisotropic regimes.

	
	\section{III. Possible transitions from the CSL to the stripe}
	
	In the main text, we have observed the softened d-wave mode and triplet mode.
		Whether these modes soften simultaneously or sequentially is important for distinguishing a direct CSL–stripe transition~\cite{Zhang2024_CSL_square} from an intermediate nematic CSL phase~\cite{Yang2024_CSL_square}.
	
	\begin{figure}[htp!]
		\centering		
		\includegraphics[width=0.8\textwidth]{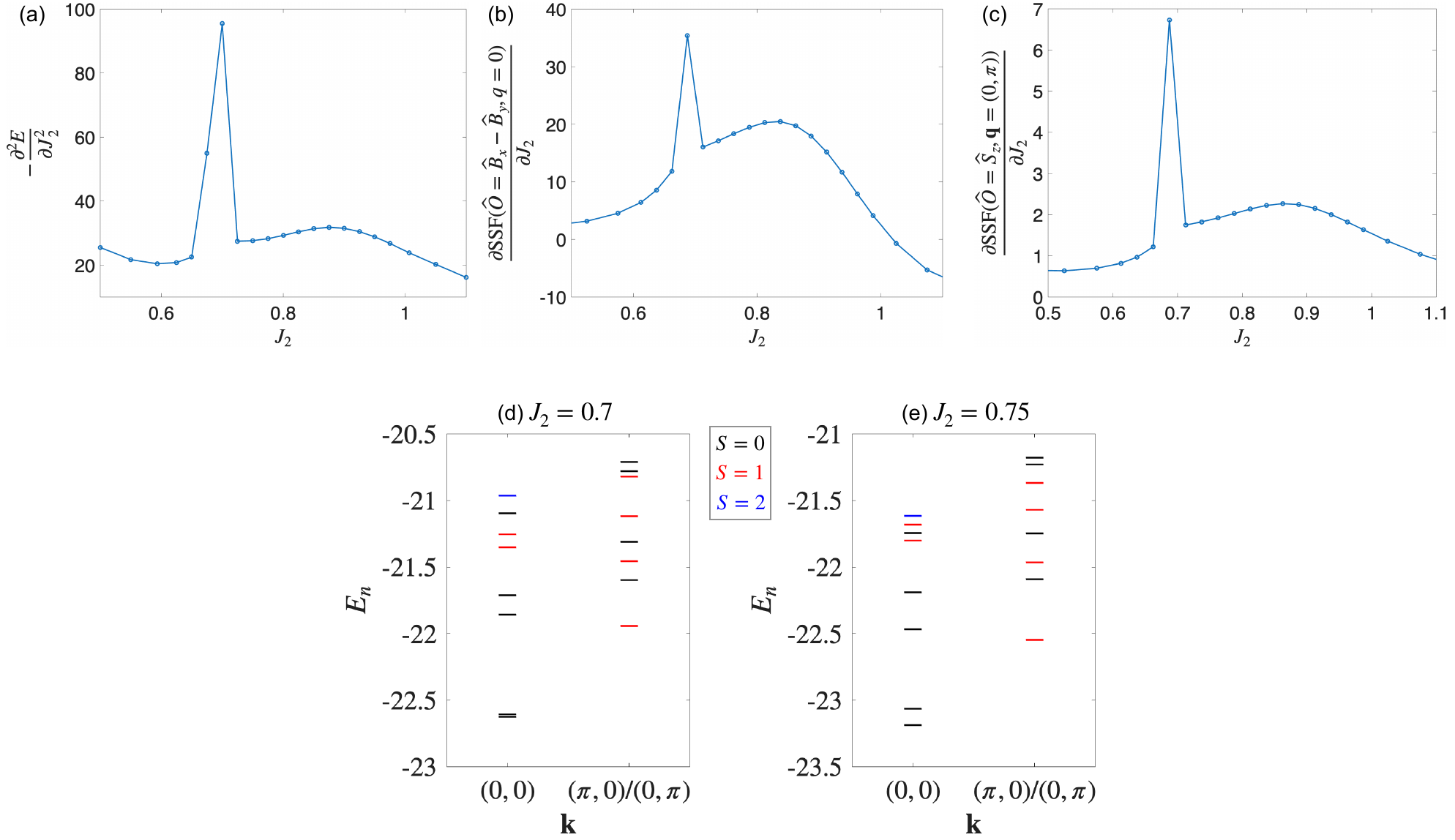}
		\caption{\textbf{Phase transitions at larger $J_2$}. We fix $J_\chi=0.5$ and tune $J_2$ to study the possible phase transitions on the $6\times6$ torus by ED simulations. (a) The second derivative of ground-state energy. 
				(b) The first derivative of the static structure factor (SSF) of the bond nematic operator $\hat{B}_x-\hat{B}_y$.
				(c) The first derivative of the spin SSF.
				We also show the energy spectra at (d) $J_2=0.7$ and (e) $J_2=0.75$ for the $(0,0)$ and degenerate $(\pi,0)/(0,\pi)$  sectors. The black, red, and blue levels are from the $S=0,1,2$ sectors, respectively.
		}
		\label{figs_transition_Jchi_0d5_tune_J2}
	\end{figure}
	
	We try to study this question by symmetric ED simulations on the $6\times6$ torus with fixed $J_\chi=0.5$ and tuning $J_2$.
		First, starting from the CSL at relatively smaller $J_2$, 
		the second-derivative of ground-state energy, the first derivative of the static structure factors (SSF) of the bond nematic operator ($\hat{B}_x-\hat{B}_y$) and the spin operator ($\hat{S}_z$), are shown in Fig.~\ref{figs_transition_Jchi_0d5_tune_J2} (a-c), respectively. 
		All these results show two peaks at $J_2\sim0.7$ and $J_2\sim0.86$, suggesting that there are possibly two phase transitions within the parameter range, an intermediate phase, and sequential softening of the collective modes.
		This is also consistent with the ground-state phase diagram in Ref. \cite{Yang2024_CSL_square}. 
		
		If the $q=0$ d-wave singlet mode softens faster than the triplet mode and there is indeed an intermediate nematic CSL phase, one should expect a gapped spectrum with 4-fold degenerate ground states (additional 2-fold due to rotation symmetry breaking from $C_4$ to $C_2$) at $k=0$.
		However, as shown from the many-body spectra at $J_2=0.7$ (close to first possible transition point) and $J_2=0.75$ (deep in the possible nematic CSL phase), the energies of the triplet mode in the (degenerate) $(\pi,0)$ and $(0,\pi)$ momentum sectors are even lower than that of the third state in the $(0,0)$ momentum sector, suggesting that there is not a stable many-body gap (for this system size) for the sought nematic CSL phase.
		
		Therefore, current finite-size data do not allow us to determine whether the two modes soften simultaneously or sequentially in the thermodynamic limit, as different finite-size diagnostics suggest different interpretations.
		Whether the collective modes soften simultaneously or sequentially is still an open question, and either scenario may be realized in a broader parameter regime.  We leave these for future investigations.
	\\
	\\
	\\
	
	
	\section{IV. Consistent collective modes in a triangular-lattice CSL}

\begin{figure}[htp!]
		\centering		
		\includegraphics[width=\textwidth]{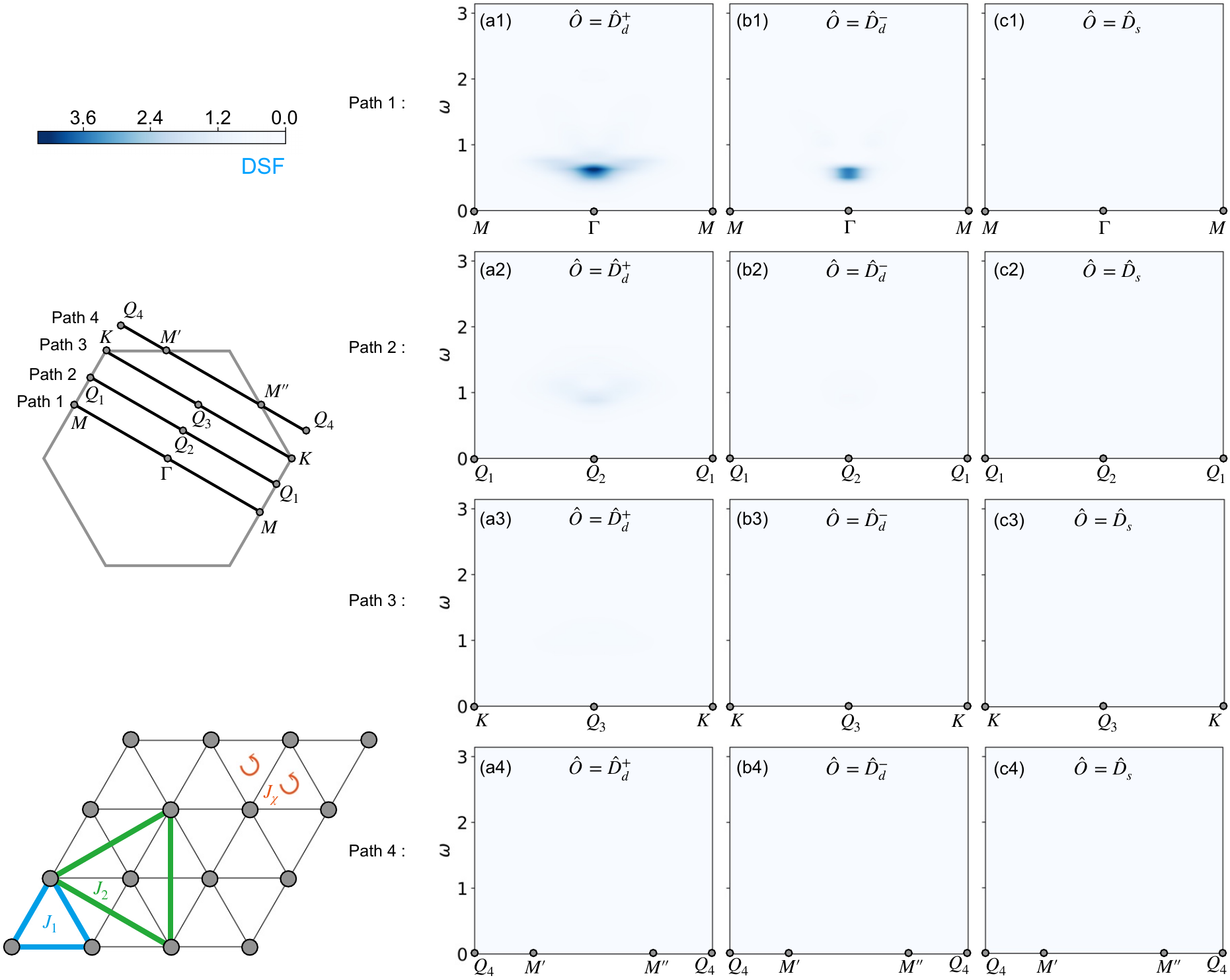}
		\caption{\textbf{$q=0$ d-wave mode of the triangular-lattice CSL}.
				We show the dynamical structure factors in different channels with the same shared color bar, and the energy scale is fixed by the Heisenberg coupling strength $J_1=1$.
				The four different paths and notations of the momentum points in the BZ are also illustrated on the left side.
				The lower-left corner is the schematic of the lattice model.
				These results are from TDVP simulations on the $L_y\times L_x=6\times66$ cylinder with bond dimension $\chi=300$.
				(a1-a4) and (b1-b4) panels include the d-wave $\hat{D}^\pm_d$ channels with opposite chiralities, and (c1-c4) panels show the  s-wave channel.
				The $q=0$ d-wave mode is consistently found, and it is still partially chiral in the sense that the relative ratio between the spectral weights in the $l=\pm2$ channels is $\sim1.23$.
		}
		\label{figs_triangle_dwave_DSF_J2_0.11_Jchi_0.2}
	\end{figure}

		To explore the generality of these singlet collective modes,
		we perform additional calculations to study the antiferromagnetic $J_1-J_2-J_\chi$ model on triangular lattice:
		\begin{equation}
			H=J_1\sum_{\langle i,j\rangle}\hat{\mathbf{S}}_i\cdot\hat{\mathbf{S}}_j+J_2\sum_{\langle\langle i,j\rangle\rangle}\hat{\mathbf{S}}_i\cdot\hat{\mathbf{S}}_j
			+J_\chi\sum_{\triangle}\hat{\mathbf{S}}_i\cdot(\hat{\mathbf{S}}_j\times\hat{\mathbf{S}}_k),
		\end{equation}
		where $J_1$ (fixed $J_1=1$ as the energy scale) and $J_2$ are the NN and next-NN Heisenberg interactions. $J_\chi$ refers to the three-spin scalar chiral coupling, and $\triangle$ includes all triangles with the same chirality, as shown in the lower-left corner of Fig.~\ref{figs_triangle_dwave_DSF_J2_0.11_Jchi_0.2}.
		The ground-state phase diagram has been well established~\cite{Wietek2017_CSL_triangular,Gong2017_QSL_triangle}, with a robust CSL phase.
		To study the collective modes here, we focus on the parameter ($J_2=0.11,\ J_\chi=0.2$), which is almost at the centre of the CSL phase regime.

	We notice the dynamical dimer structure factor of the CSL phase in this model has been studied in a recent work by Drescher et al. ~\cite{drescher2026dynamicaldimerstructurefactor}. 
		However, they considered the rotationally symmetric equal-weight NN bond operator (s-wave channel in our convention): $\hat{D}_s(\mathbf{r})=\sum_{\mu}\hat{S}_\mathbf{r}\cdot\hat{S}_{\mathbf{r+\mu}}$, with $\mu$ denoting the 6 NN vectors.
		Therefore, whether the $q=0$ d-wave mode and the chiral roton mode exist in such a CSL, were not analyzed there.
		
		To address this question, we will first explore the existence of $q=0$ d-wave mode, as the existence of graviton-like mode in CSL is not simply expected from the FQH/FCI studies, and the $q=0$ d-wave mode in the square-lattice CSL has only partial chirality, in contrast to the FQH/FCI literature.
		Therefore, we design the chiral d-wave operators by the NN bond operators:  $\hat{D}^{\pm}_d(\mathbf{r})=\sum_{\mu}e^{\pm i2\theta_\mu}\hat{S}_\mathbf{r}\cdot\hat{S}_{\mathbf{r+\mu}}$, corresponding to the different $l=\pm2$ channels.
		The DSFs along different momentum paths are shown in Fig.\ref{figs_triangle_dwave_DSF_J2_0.11_Jchi_0.2}.
		Notably and consistently, we find the d-wave mode with a sharp spectral peak at $q=0$ in the $\hat{D}^+_d$ channel.
		Moreover, in contrast to the graviton modes in FQH/FCI states, the response in the $\hat{D}^-_d$ channel is not vanishing, and the relative ratio is only $\sim1.23$.
		The DSFs in the  s-wave channel with relatively vanishing response are also shown for comparison.
		These results further support the partial helicity selectivity of the $q=0$ d-wave mode in CSL.
	
	\begin{figure}[htp!]
		\centering		
		\includegraphics[width=\textwidth]{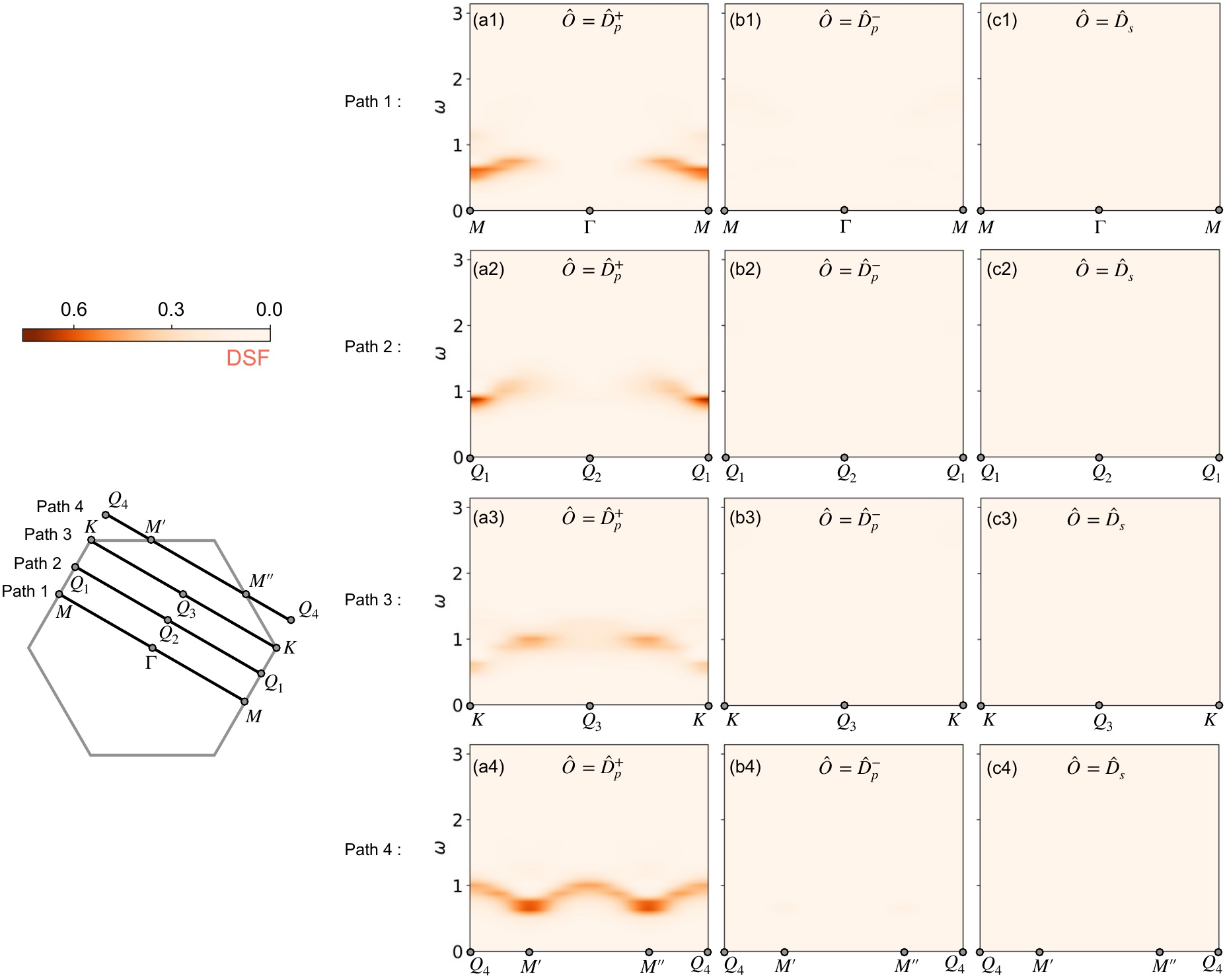}
		\par\medskip
		\hfill
		\includegraphics[width=0.9\textwidth]{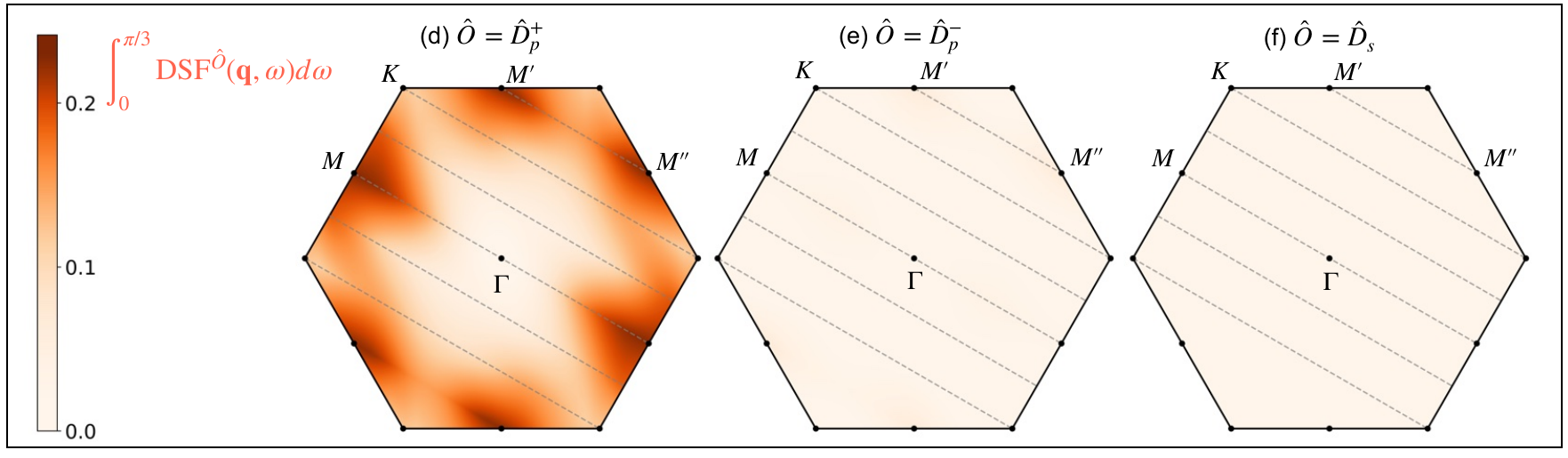}
		\caption{\textbf{Chiral p-wave roton mode of the triangular-lattice CSL}.
				We show the dynamical structure factors (DSFs) of different channels in panels (a1-a4, b1-b4, c1-c4), with the same shared color bar,
				and the energy scale is fixed by the Heisenberg coupling strength $J_1=1$. 
				The four different paths and notations of the momentum points in the BZ are also illustrated on the left side.
				These results are from TDVP simulations on the $L_y\times L_x=6\times66$ cylinder with bond dimension $\chi=300$.
				(a1-a4) and (b1-b4) panels include the p-wave $\hat{D}^\pm_p$ channels with opposite chiralities, and (c1-c4) panels show the  s-wave channel.
				The chiral p-wave roton mode is consistently found.
				To better demonstrate the momentum dependence of such a mode, we integrate the low-energy DSFs ($0<\omega<\pi/3$), and the results in different channels are shown in panels (d-f), with the dashed grey lines denote the real momentum cuts of the $L_y=6$ system.
		}
		\label{figs_triangle_pwave_DSF_J2_0.11_Jchi_0.2}
	\end{figure}

	Furthermore, to explore the chiral roton mode, we define the different chiral p-wave operators as:
		$\hat{D}^{\pm}_p(\mathbf{r})=\sum_{\mu}e^{\pm i\theta_\mu}\hat{S}_\mathbf{r}\cdot\hat{S}_{\mathbf{r+\mu}}$,
		corresponding to the $l=\pm1$ channels.
		The DSFs in different channels are shown in Fig.~\ref{figs_triangle_pwave_DSF_J2_0.11_Jchi_0.2} (a1-a4,b1-b4,c1-c4), respectively.
		Consistently, in the $\hat{D}_p^+$ channel, we find the low-energy chiral roton mode, with dispersive roton minima at the $M$ points.
		In contrast, the responses in the $\hat{D}_p^-$ and $\hat{D}_s$ channels are vanishing.
		To better demonstrate the momentum dependence of the chiral roton mode, we integrate the DSFs at low energy ($0<\omega<\pi/3$), and the results of different channels are shown in Fig.~\ref{figs_triangle_pwave_DSF_J2_0.11_Jchi_0.2} (d-f), respectively. 
		Consistently, in the $\hat{D}_p^+$ channel, the chiral p-wave roton mode has concentrated spectral weights at (around) $M$ points.
		We note that, from the results in Ref.~\cite{drescher2026dynamicaldimerstructurefactor}, the low-energy spectral weights at $M$ points are vanishing in the  s-wave channel, even compared with the higher-energy continuum. 
		This also supports the channel selectivity of the chiral p-wave roton mode we identify here.
		
		To summarize, we believe these new results in a different triangular-lattice model are consistent with what we discovered on the square lattice model: the $q=0$ d-wave mode with partial chirality, and the chiral p-wave roton mode, which further supports the potential generality of the conclusions.
		Based on these fruitful results, more extensive works could be motivated in future studies. 
	\clearpage
	\bibliographystyle{apsrev4-2}

\begin{thebibliography}{78}%
		\makeatletter
		\providecommand \@ifxundefined [1]{%
			\@ifx{#1\undefined}
		}%
		\providecommand \@ifnum [1]{%
			\ifnum #1\expandafter \@firstoftwo
			\else \expandafter \@secondoftwo
			\fi
		}%
		\providecommand \@ifx [1]{%
			\ifx #1\expandafter \@firstoftwo
			\else \expandafter \@secondoftwo
			\fi
		}%
		\providecommand \natexlab [1]{#1}%
		\providecommand \enquote  [1]{``#1''}%
		\providecommand \bibnamefont  [1]{#1}%
		\providecommand \bibfnamefont [1]{#1}%
		\providecommand \citenamefont [1]{#1}%
		\providecommand \href@noop [0]{\@secondoftwo}%
		\providecommand \href [0]{\begingroup \@sanitize@url \@href}%
		\providecommand \@href[1]{\@@startlink{#1}\@@href}%
		\providecommand \@@href[1]{\endgroup#1\@@endlink}%
		\providecommand \@sanitize@url [0]{\catcode `\\12\catcode `\$12\catcode
			`\&12\catcode `\#12\catcode `\^12\catcode `\_12\catcode `\%12\relax}%
		\providecommand \@@startlink[1]{}%
		\providecommand \@@endlink[0]{}%
		\providecommand \url  [0]{\begingroup\@sanitize@url \@url }%
		\providecommand \@url [1]{\endgroup\@href {#1}{\urlprefix }}%
		\providecommand \urlprefix  [0]{URL }%
		\providecommand \Eprint [0]{\href }%
		\providecommand \doibase [0]{https://doi.org/}%
		\providecommand \selectlanguage [0]{\@gobble}%
		\providecommand \bibinfo  [0]{\@secondoftwo}%
		\providecommand \bibfield  [0]{\@secondoftwo}%
		\providecommand \translation [1]{[#1]}%
		\providecommand \BibitemOpen [0]{}%
		\providecommand \bibitemStop [0]{}%
		\providecommand \bibitemNoStop [0]{.\EOS\space}%
		\providecommand \EOS [0]{\spacefactor3000\relax}%
		\providecommand \BibitemShut  [1]{\csname bibitem#1\endcsname}%
		\let\auto@bib@innerbib\@empty
		\bibitem [{\citenamefont {Savary}\ and\ \citenamefont
			{Balents}(2016)}]{Savary2017_QSL}%
		\BibitemOpen
		\bibfield  {author} {\bibinfo {author} {\bibfnamefont {L.}~\bibnamefont
				{Savary}}\ and\ \bibinfo {author} {\bibfnamefont {L.}~\bibnamefont
				{Balents}},\ }\href {https://doi.org/10.1088/0034-4885/80/1/016502}
		{\bibfield  {journal} {\bibinfo  {journal} {Reports on Progress in Physics}\
			}\textbf {\bibinfo {volume} {80}},\ \bibinfo {pages} {016502} (\bibinfo
			{year} {2016})}\BibitemShut {NoStop}%
		\bibitem [{\citenamefont {Zhou}\ \emph {et~al.}(2017)\citenamefont {Zhou},
			\citenamefont {Kanoda},\ and\ \citenamefont {Ng}}]{Zhou2017_QSL}%
		\BibitemOpen
		\bibfield  {author} {\bibinfo {author} {\bibfnamefont {Y.}~\bibnamefont
				{Zhou}}, \bibinfo {author} {\bibfnamefont {K.}~\bibnamefont {Kanoda}},\ and\
			\bibinfo {author} {\bibfnamefont {T.-K.}\ \bibnamefont {Ng}},\ }\href
		{https://doi.org/10.1103/RevModPhys.89.025003} {\bibfield  {journal}
			{\bibinfo  {journal} {Rev. Mod. Phys.}\ }\textbf {\bibinfo {volume} {89}},\
			\bibinfo {pages} {025003} (\bibinfo {year} {2017})}\BibitemShut {NoStop}%
		\bibitem [{\citenamefont {Broholm}\ \emph {et~al.}(2020)\citenamefont
			{Broholm}, \citenamefont {Cava}, \citenamefont {Kivelson}, \citenamefont
			{Nocera}, \citenamefont {Norman},\ and\ \citenamefont
			{Senthil}}]{Broholm2020_QSL}%
		\BibitemOpen
		\bibfield  {author} {\bibinfo {author} {\bibfnamefont {C.}~\bibnamefont
				{Broholm}}, \bibinfo {author} {\bibfnamefont {R.~J.}\ \bibnamefont {Cava}},
			\bibinfo {author} {\bibfnamefont {S.~A.}\ \bibnamefont {Kivelson}}, \bibinfo
			{author} {\bibfnamefont {D.~G.}\ \bibnamefont {Nocera}}, \bibinfo {author}
			{\bibfnamefont {M.~R.}\ \bibnamefont {Norman}},\ and\ \bibinfo {author}
			{\bibfnamefont {T.}~\bibnamefont {Senthil}},\ }\href
		{https://doi.org/10.1126/science.aay0668} {\bibfield  {journal} {\bibinfo
				{journal} {Science}\ }\textbf {\bibinfo {volume} {367}},\ \bibinfo {pages}
			{eaay0668} (\bibinfo {year} {2020})},\ \Eprint
		{https://arxiv.org/abs/https://www.science.org/doi/pdf/10.1126/science.aay0668}
		{https://www.science.org/doi/pdf/10.1126/science.aay0668} \BibitemShut
		{NoStop}%
		\bibitem [{\citenamefont {Kalmeyer}\ and\ \citenamefont
			{Laughlin}(1987)}]{Kalmeyer1987_RVB_FQH}%
		\BibitemOpen
		\bibfield  {author} {\bibinfo {author} {\bibfnamefont {V.}~\bibnamefont
				{Kalmeyer}}\ and\ \bibinfo {author} {\bibfnamefont {R.~B.}\ \bibnamefont
				{Laughlin}},\ }\href {https://doi.org/10.1103/PhysRevLett.59.2095} {\bibfield
			{journal} {\bibinfo  {journal} {Phys. Rev. Lett.}\ }\textbf {\bibinfo
				{volume} {59}},\ \bibinfo {pages} {2095} (\bibinfo {year}
			{1987})}\BibitemShut {NoStop}%
		\bibitem [{\citenamefont {Wen}\ \emph {et~al.}(1989)\citenamefont {Wen},
			\citenamefont {Wilczek},\ and\ \citenamefont
			{Zee}}]{Wen1989_chiral_spin_states}%
		\BibitemOpen
		\bibfield  {author} {\bibinfo {author} {\bibfnamefont {X.~G.}\ \bibnamefont
				{Wen}}, \bibinfo {author} {\bibfnamefont {F.}~\bibnamefont {Wilczek}},\ and\
			\bibinfo {author} {\bibfnamefont {A.}~\bibnamefont {Zee}},\ }\href
		{https://doi.org/10.1103/PhysRevB.39.11413} {\bibfield  {journal} {\bibinfo
				{journal} {Phys. Rev. B}\ }\textbf {\bibinfo {volume} {39}},\ \bibinfo
			{pages} {11413} (\bibinfo {year} {1989})}\BibitemShut {NoStop}%
		\bibitem [{\citenamefont {Stormer}\ \emph {et~al.}(1999)\citenamefont
			{Stormer}, \citenamefont {Tsui},\ and\ \citenamefont
			{Gossard}}]{Stormer1999_FQH_review}%
		\BibitemOpen
		\bibfield  {author} {\bibinfo {author} {\bibfnamefont {H.~L.}\ \bibnamefont
				{Stormer}}, \bibinfo {author} {\bibfnamefont {D.~C.}\ \bibnamefont {Tsui}},\
			and\ \bibinfo {author} {\bibfnamefont {A.~C.}\ \bibnamefont {Gossard}},\
		}\href {https://doi.org/10.1103/RevModPhys.71.S298} {\bibfield  {journal}
			{\bibinfo  {journal} {Rev. Mod. Phys.}\ }\textbf {\bibinfo {volume} {71}},\
			\bibinfo {pages} {S298} (\bibinfo {year} {1999})}\BibitemShut {NoStop}%
		\bibitem [{\citenamefont {Yang}\ \emph {et~al.}(1993)\citenamefont {Yang},
			\citenamefont {Warman},\ and\ \citenamefont {Girvin}}]{Yang1993_spin_liquid}%
		\BibitemOpen
		\bibfield  {author} {\bibinfo {author} {\bibfnamefont {K.}~\bibnamefont
				{Yang}}, \bibinfo {author} {\bibfnamefont {L.~K.}\ \bibnamefont {Warman}},\
			and\ \bibinfo {author} {\bibfnamefont {S.~M.}\ \bibnamefont {Girvin}},\
		}\href {https://doi.org/10.1103/PhysRevLett.70.2641} {\bibfield  {journal}
			{\bibinfo  {journal} {Phys. Rev. Lett.}\ }\textbf {\bibinfo {volume} {70}},\
			\bibinfo {pages} {2641} (\bibinfo {year} {1993})}\BibitemShut {NoStop}%
		\bibitem [{\citenamefont {Haldane}\ and\ \citenamefont
			{Arovas}(1995)}]{Haldane1995_spin_currents}%
		\BibitemOpen
		\bibfield  {author} {\bibinfo {author} {\bibfnamefont {F.~D.~M.}\
				\bibnamefont {Haldane}}\ and\ \bibinfo {author} {\bibfnamefont {D.~P.}\
				\bibnamefont {Arovas}},\ }\href {https://doi.org/10.1103/PhysRevB.52.4223}
		{\bibfield  {journal} {\bibinfo  {journal} {Phys. Rev. B}\ }\textbf {\bibinfo
				{volume} {52}},\ \bibinfo {pages} {4223} (\bibinfo {year}
			{1995})}\BibitemShut {NoStop}%
		\bibitem [{\citenamefont {Nielsen}\ \emph {et~al.}(2013)\citenamefont
			{Nielsen}, \citenamefont {Sierra},\ and\ \citenamefont
			{Cirac}}]{nielsen2013}%
		\BibitemOpen
		\bibfield  {author} {\bibinfo {author} {\bibfnamefont {A.~E.~B.}\
				\bibnamefont {Nielsen}}, \bibinfo {author} {\bibfnamefont {G.}~\bibnamefont
				{Sierra}},\ and\ \bibinfo {author} {\bibfnamefont {J.~I.}\ \bibnamefont
				{Cirac}},\ }\href {https://doi.org/10.1038/ncomms3864} {\bibfield  {journal}
			{\bibinfo  {journal} {Nat. Commun.}\ }\textbf {\bibinfo {volume} {4}},\
			\bibinfo {pages} {1} (\bibinfo {year} {2013})}\BibitemShut {NoStop}%
		\bibitem [{\citenamefont {He}\ \emph {et~al.}(2014)\citenamefont {He},
			\citenamefont {Sheng},\ and\ \citenamefont {Chen}}]{He2014_CSL}%
		\BibitemOpen
		\bibfield  {author} {\bibinfo {author} {\bibfnamefont {Y.-C.}\ \bibnamefont
				{He}}, \bibinfo {author} {\bibfnamefont {D.~N.}\ \bibnamefont {Sheng}},\ and\
			\bibinfo {author} {\bibfnamefont {Y.}~\bibnamefont {Chen}},\ }\href
		{https://doi.org/10.1103/PhysRevLett.112.137202} {\bibfield  {journal}
			{\bibinfo  {journal} {Phys. Rev. Lett.}\ }\textbf {\bibinfo {volume} {112}},\
			\bibinfo {pages} {137202} (\bibinfo {year} {2014})}\BibitemShut {NoStop}%
		\bibitem [{\citenamefont {Gong}\ \emph {et~al.}(2014)\citenamefont {Gong},
			\citenamefont {Zhu},\ and\ \citenamefont {Sheng}}]{Gong2014_CSL}%
		\BibitemOpen
		\bibfield  {author} {\bibinfo {author} {\bibfnamefont {S.-S.}\ \bibnamefont
				{Gong}}, \bibinfo {author} {\bibfnamefont {W.}~\bibnamefont {Zhu}},\ and\
			\bibinfo {author} {\bibfnamefont {D.~N.}\ \bibnamefont {Sheng}},\ }\href
		{https://doi.org/10.1038/srep06317} {\bibfield  {journal} {\bibinfo
				{journal} {Scientific Reports}\ }\textbf {\bibinfo {volume} {4}},\ \bibinfo
			{pages} {6317} (\bibinfo {year} {2014})}\BibitemShut {NoStop}%
		\bibitem [{\citenamefont {Bauer}\ \emph {et~al.}(2014)\citenamefont {Bauer},
			\citenamefont {Cincio}, \citenamefont {Keller}, \citenamefont {Dolfi},
			\citenamefont {Vidal}, \citenamefont {Trebst},\ and\ \citenamefont
			{Ludwig}}]{bauer2014}%
		\BibitemOpen
		\bibfield  {author} {\bibinfo {author} {\bibfnamefont {B.}~\bibnamefont
				{Bauer}}, \bibinfo {author} {\bibfnamefont {L.}~\bibnamefont {Cincio}},
			\bibinfo {author} {\bibfnamefont {B.~P.}\ \bibnamefont {Keller}}, \bibinfo
			{author} {\bibfnamefont {M.}~\bibnamefont {Dolfi}}, \bibinfo {author}
			{\bibfnamefont {G.}~\bibnamefont {Vidal}}, \bibinfo {author} {\bibfnamefont
				{S.}~\bibnamefont {Trebst}},\ and\ \bibinfo {author} {\bibfnamefont
				{A.~W.~W.}\ \bibnamefont {Ludwig}},\ }\href
		{https://doi.org/10.1038/ncomms6137} {\bibfield  {journal} {\bibinfo
				{journal} {Nat. Commun.}\ }\textbf {\bibinfo {volume} {5}},\ \bibinfo {pages}
			{1} (\bibinfo {year} {2014})}\BibitemShut {NoStop}%
		\bibitem [{\citenamefont {Wietek}\ and\ \citenamefont
			{L\"auchli}(2017)}]{Wietek2017_CSL_triangular}%
		\BibitemOpen
		\bibfield  {author} {\bibinfo {author} {\bibfnamefont {A.}~\bibnamefont
				{Wietek}}\ and\ \bibinfo {author} {\bibfnamefont {A.~M.}\ \bibnamefont
				{L\"auchli}},\ }\href {https://doi.org/10.1103/PhysRevB.95.035141} {\bibfield
			{journal} {\bibinfo  {journal} {Phys. Rev. B}\ }\textbf {\bibinfo {volume}
				{95}},\ \bibinfo {pages} {035141} (\bibinfo {year} {2017})}\BibitemShut
		{NoStop}%
		\bibitem [{\citenamefont {Gong}\ \emph {et~al.}(2017)\citenamefont {Gong},
			\citenamefont {Zhu}, \citenamefont {Zhu}, \citenamefont {Sheng},\ and\
			\citenamefont {Yang}}]{Gong2017_QSL_triangle}%
		\BibitemOpen
		\bibfield  {author} {\bibinfo {author} {\bibfnamefont {S.-S.}\ \bibnamefont
				{Gong}}, \bibinfo {author} {\bibfnamefont {W.}~\bibnamefont {Zhu}}, \bibinfo
			{author} {\bibfnamefont {J.-X.}\ \bibnamefont {Zhu}}, \bibinfo {author}
			{\bibfnamefont {D.~N.}\ \bibnamefont {Sheng}},\ and\ \bibinfo {author}
			{\bibfnamefont {K.}~\bibnamefont {Yang}},\ }\href
		{https://doi.org/10.1103/PhysRevB.96.075116} {\bibfield  {journal} {\bibinfo
				{journal} {Phys. Rev. B}\ }\textbf {\bibinfo {volume} {96}},\ \bibinfo
			{pages} {075116} (\bibinfo {year} {2017})}\BibitemShut {NoStop}%
		\bibitem [{\citenamefont {Hu}\ \emph {et~al.}(2015)\citenamefont {Hu},
			\citenamefont {Zhu}, \citenamefont {Zhang}, \citenamefont {Gong},
			\citenamefont {Becca},\ and\ \citenamefont {Sheng}}]{Hu_2015}%
		\BibitemOpen
		\bibfield  {author} {\bibinfo {author} {\bibfnamefont {W.-J.}\ \bibnamefont
				{Hu}}, \bibinfo {author} {\bibfnamefont {W.}~\bibnamefont {Zhu}}, \bibinfo
			{author} {\bibfnamefont {Y.}~\bibnamefont {Zhang}}, \bibinfo {author}
			{\bibfnamefont {S.}~\bibnamefont {Gong}}, \bibinfo {author} {\bibfnamefont
				{F.}~\bibnamefont {Becca}},\ and\ \bibinfo {author} {\bibfnamefont {D.~N.}\
				\bibnamefont {Sheng}},\ }\bibfield  {journal} {\bibinfo  {journal} {Physical
				Review B}\ }\textbf {\bibinfo {volume} {91}},\ \href
		{https://doi.org/10.1103/physrevb.91.041124} {10.1103/physrevb.91.041124}
		(\bibinfo {year} {2015})\BibitemShut {NoStop}%
		\bibitem [{\citenamefont {Laughlin}(1983)}]{Laughlin1983_FQH}%
		\BibitemOpen
		\bibfield  {author} {\bibinfo {author} {\bibfnamefont {R.~B.}\ \bibnamefont
				{Laughlin}},\ }\href {https://doi.org/10.1103/PhysRevLett.50.1395} {\bibfield
			{journal} {\bibinfo  {journal} {Phys. Rev. Lett.}\ }\textbf {\bibinfo
				{volume} {50}},\ \bibinfo {pages} {1395} (\bibinfo {year}
			{1983})}\BibitemShut {NoStop}%
		\bibitem [{\citenamefont {Laughlin}(1989)}]{Laughlin1989_spin_hamiltonian}%
		\BibitemOpen
		\bibfield  {author} {\bibinfo {author} {\bibfnamefont {R.}~\bibnamefont
				{Laughlin}},\ }\href
		{https://doi.org/https://doi.org/10.1016/0003-4916(89)90339-4} {\bibfield
			{journal} {\bibinfo  {journal} {Annals of Physics}\ }\textbf {\bibinfo
				{volume} {191}},\ \bibinfo {pages} {163} (\bibinfo {year}
			{1989})}\BibitemShut {NoStop}%
		\bibitem [{\citenamefont {Balatsky}(1991)}]{Balatsky1991_csl_spin_rigidity}%
		\BibitemOpen
		\bibfield  {author} {\bibinfo {author} {\bibfnamefont {A.}~\bibnamefont
				{Balatsky}},\ }\href {https://doi.org/10.1103/PhysRevLett.66.814} {\bibfield
			{journal} {\bibinfo  {journal} {Phys. Rev. Lett.}\ }\textbf {\bibinfo
				{volume} {66}},\ \bibinfo {pages} {814} (\bibinfo {year} {1991})}\BibitemShut
		{NoStop}%
		\bibitem [{\citenamefont {Zhu}\ \emph {et~al.}(2019)\citenamefont {Zhu},
			\citenamefont {shu Gong},\ and\ \citenamefont {Sheng}}]{Zhu2019_spinon_DSF}%
		\BibitemOpen
		\bibfield  {author} {\bibinfo {author} {\bibfnamefont {W.}~\bibnamefont
				{Zhu}}, \bibinfo {author} {\bibfnamefont {S.}~\bibnamefont {shu Gong}},\ and\
			\bibinfo {author} {\bibfnamefont {D.~N.}\ \bibnamefont {Sheng}},\ }\href
		{https://doi.org/10.1073/pnas.1807840116} {\bibfield  {journal} {\bibinfo
				{journal} {Proceedings of the National Academy of Sciences}\ }\textbf
			{\bibinfo {volume} {116}},\ \bibinfo {pages} {5437} (\bibinfo {year}
			{2019})},\ \Eprint
		{https://arxiv.org/abs/https://www.pnas.org/doi/pdf/10.1073/pnas.1807840116}
		{https://www.pnas.org/doi/pdf/10.1073/pnas.1807840116} \BibitemShut {NoStop}%
		\bibitem [{\citenamefont {Pichler}\ \emph {et~al.}(2024)\citenamefont
			{Pichler}, \citenamefont {Kadow}, \citenamefont {Kuhlenkamp},\ and\
			\citenamefont {Knap}}]{Pichler2024_CSL_DSSF}%
		\BibitemOpen
		\bibfield  {author} {\bibinfo {author} {\bibfnamefont {F.}~\bibnamefont
				{Pichler}}, \bibinfo {author} {\bibfnamefont {W.}~\bibnamefont {Kadow}},
			\bibinfo {author} {\bibfnamefont {C.}~\bibnamefont {Kuhlenkamp}},\ and\
			\bibinfo {author} {\bibfnamefont {M.}~\bibnamefont {Knap}},\ }\href
		{https://doi.org/10.1103/PhysRevB.110.045116} {\bibfield  {journal} {\bibinfo
				{journal} {Phys. Rev. B}\ }\textbf {\bibinfo {volume} {110}},\ \bibinfo
			{pages} {045116} (\bibinfo {year} {2024})}\BibitemShut {NoStop}%
		\bibitem [{\citenamefont {Drescher}\ \emph {et~al.}(2025)\citenamefont
			{Drescher}, \citenamefont {Vanderstraeten}, \citenamefont {Moessner},\ and\
			\citenamefont {Pollmann}}]{drescher2025_dssf_triangle}%
		\BibitemOpen
		\bibfield  {author} {\bibinfo {author} {\bibfnamefont {M.}~\bibnamefont
				{Drescher}}, \bibinfo {author} {\bibfnamefont {L.}~\bibnamefont
				{Vanderstraeten}}, \bibinfo {author} {\bibfnamefont {R.}~\bibnamefont
				{Moessner}},\ and\ \bibinfo {author} {\bibfnamefont {F.}~\bibnamefont
				{Pollmann}},\ }\href {https://arxiv.org/abs/2508.17292} {\bibinfo {title}
			{Spectral functions of an extended antiferromagnetic $s=1/2$ heisenberg model
				on the triangular lattice}} (\bibinfo {year} {2025}),\ \Eprint
		{https://arxiv.org/abs/2508.17292} {arXiv:2508.17292 [cond-mat.str-el]}
		\BibitemShut {NoStop}%
		\bibitem [{\citenamefont {Girvin}\ \emph {et~al.}(1985)\citenamefont {Girvin},
			\citenamefont {MacDonald},\ and\ \citenamefont {Platzman}}]{Girvin1985}%
		\BibitemOpen
		\bibfield  {author} {\bibinfo {author} {\bibfnamefont {S.~M.}\ \bibnamefont
				{Girvin}}, \bibinfo {author} {\bibfnamefont {A.~H.}\ \bibnamefont
				{MacDonald}},\ and\ \bibinfo {author} {\bibfnamefont {P.~M.}\ \bibnamefont
				{Platzman}},\ }\href {https://doi.org/10.1103/PhysRevLett.54.581} {\bibfield
			{journal} {\bibinfo  {journal} {Phys. Rev. Lett.}\ }\textbf {\bibinfo
				{volume} {54}},\ \bibinfo {pages} {581} (\bibinfo {year} {1985})}\BibitemShut
		{NoStop}%
		\bibitem [{\citenamefont {Girvin}\ \emph {et~al.}(1986)\citenamefont {Girvin},
			\citenamefont {MacDonald},\ and\ \citenamefont {Platzman}}]{GMP1986}%
		\BibitemOpen
		\bibfield  {author} {\bibinfo {author} {\bibfnamefont {S.~M.}\ \bibnamefont
				{Girvin}}, \bibinfo {author} {\bibfnamefont {A.~H.}\ \bibnamefont
				{MacDonald}},\ and\ \bibinfo {author} {\bibfnamefont {P.~M.}\ \bibnamefont
				{Platzman}},\ }\href {https://doi.org/10.1103/PhysRevB.33.2481} {\bibfield
			{journal} {\bibinfo  {journal} {Phys. Rev. B}\ }\textbf {\bibinfo {volume}
				{33}},\ \bibinfo {pages} {2481} (\bibinfo {year} {1986})}\BibitemShut
		{NoStop}%
		\bibitem [{\citenamefont {Yang}(2013)}]{BoYang2013}%
		\BibitemOpen
		\bibfield  {author} {\bibinfo {author} {\bibfnamefont {B.}~\bibnamefont
				{Yang}},\ }\href {https://doi.org/10.1103/PhysRevB.87.245132} {\bibfield
			{journal} {\bibinfo  {journal} {Phys. Rev. B}\ }\textbf {\bibinfo {volume}
				{87}},\ \bibinfo {pages} {245132} (\bibinfo {year} {2013})}\BibitemShut
		{NoStop}%
		\bibitem [{\citenamefont {Haldane}(2011{\natexlab{a}})}]{Haldane_2011}%
		\BibitemOpen
		\bibfield  {author} {\bibinfo {author} {\bibfnamefont {F.~D.~M.}\
				\bibnamefont {Haldane}},\ }\bibfield  {journal} {\bibinfo  {journal}
			{Physical Review Letters}\ }\textbf {\bibinfo {volume} {107}},\ \href
		{https://doi.org/10.1103/physrevlett.107.116801}
		{10.1103/physrevlett.107.116801} (\bibinfo {year}
		{2011}{\natexlab{a}})\BibitemShut {NoStop}%
		\bibitem [{\citenamefont {Yang}\ \emph {et~al.}(2012)\citenamefont {Yang},
			\citenamefont {Hu}, \citenamefont {Papi\ifmmode~\acute{c}\else \'{c}\fi{}},\
			and\ \citenamefont {Haldane}}]{Yang2012_collective_modes}%
		\BibitemOpen
		\bibfield  {author} {\bibinfo {author} {\bibfnamefont {B.}~\bibnamefont
				{Yang}}, \bibinfo {author} {\bibfnamefont {Z.-X.}\ \bibnamefont {Hu}},
			\bibinfo {author} {\bibfnamefont {Z.}~\bibnamefont
				{Papi\ifmmode~\acute{c}\else \'{c}\fi{}}},\ and\ \bibinfo {author}
			{\bibfnamefont {F.~D.~M.}\ \bibnamefont {Haldane}},\ }\href
		{https://doi.org/10.1103/PhysRevLett.108.256807} {\bibfield  {journal}
			{\bibinfo  {journal} {Phys. Rev. Lett.}\ }\textbf {\bibinfo {volume} {108}},\
			\bibinfo {pages} {256807} (\bibinfo {year} {2012})}\BibitemShut {NoStop}%
		\bibitem [{\citenamefont {Golkar}\ \emph {et~al.}(2016)\citenamefont {Golkar},
			\citenamefont {Nguyen},\ and\ \citenamefont {Son}}]{Golkar_2016}%
		\BibitemOpen
		\bibfield  {author} {\bibinfo {author} {\bibfnamefont {S.}~\bibnamefont
				{Golkar}}, \bibinfo {author} {\bibfnamefont {D.~X.}\ \bibnamefont {Nguyen}},\
			and\ \bibinfo {author} {\bibfnamefont {D.~T.}\ \bibnamefont {Son}},\
		}\bibfield  {journal} {\bibinfo  {journal} {Journal of High Energy Physics}\
		}\textbf {\bibinfo {volume} {2016}},\ \href
		{https://doi.org/10.1007/jhep01(2016)021} {10.1007/jhep01(2016)021} (\bibinfo
		{year} {2016})\BibitemShut {NoStop}%
		\bibitem [{\citenamefont {Liu}\ \emph {et~al.}(2018{\natexlab{a}})\citenamefont
			{Liu}, \citenamefont {Gromov},\ and\ \citenamefont
			{Papi\ifmmode~\acute{c}\else \'{c}\fi{}}}]{ZhaoLiu2018}%
		\BibitemOpen
		\bibfield  {author} {\bibinfo {author} {\bibfnamefont {Z.}~\bibnamefont
				{Liu}}, \bibinfo {author} {\bibfnamefont {A.}~\bibnamefont {Gromov}},\ and\
			\bibinfo {author} {\bibfnamefont {Z.}~\bibnamefont
				{Papi\ifmmode~\acute{c}\else \'{c}\fi{}}},\ }\href
		{https://doi.org/10.1103/PhysRevB.98.155140} {\bibfield  {journal} {\bibinfo
				{journal} {Phys. Rev. B}\ }\textbf {\bibinfo {volume} {98}},\ \bibinfo
			{pages} {155140} (\bibinfo {year} {2018}{\natexlab{a}})}\BibitemShut
		{NoStop}%
		\bibitem [{\citenamefont {Liou}\ \emph {et~al.}(2019)\citenamefont {Liou},
			\citenamefont {Haldane}, \citenamefont {Yang},\ and\ \citenamefont
			{Rezayi}}]{Liou2019_graviton}%
		\BibitemOpen
		\bibfield  {author} {\bibinfo {author} {\bibfnamefont {S.-F.}\ \bibnamefont
				{Liou}}, \bibinfo {author} {\bibfnamefont {F.~D.~M.}\ \bibnamefont
				{Haldane}}, \bibinfo {author} {\bibfnamefont {K.}~\bibnamefont {Yang}},\ and\
			\bibinfo {author} {\bibfnamefont {E.~H.}\ \bibnamefont {Rezayi}},\ }\href
		{https://doi.org/10.1103/PhysRevLett.123.146801} {\bibfield  {journal}
			{\bibinfo  {journal} {Phys. Rev. Lett.}\ }\textbf {\bibinfo {volume} {123}},\
			\bibinfo {pages} {146801} (\bibinfo {year} {2019})}\BibitemShut {NoStop}%
		\bibitem [{\citenamefont {Kumar}\ and\ \citenamefont
			{Haldane}(2022)}]{Kumar2022_neutral_excitations_FQH}%
		\BibitemOpen
		\bibfield  {author} {\bibinfo {author} {\bibfnamefont {P.}~\bibnamefont
				{Kumar}}\ and\ \bibinfo {author} {\bibfnamefont {F.~D.~M.}\ \bibnamefont
				{Haldane}},\ }\href {https://doi.org/10.1103/PhysRevB.106.075116} {\bibfield
			{journal} {\bibinfo  {journal} {Phys. Rev. B}\ }\textbf {\bibinfo {volume}
				{106}},\ \bibinfo {pages} {075116} (\bibinfo {year} {2022})}\BibitemShut
		{NoStop}%
		\bibitem [{\citenamefont {Liu}\ \emph {et~al.}(2024)\citenamefont {Liu},
			\citenamefont {Zhao},\ and\ \citenamefont
			{Xiang}}]{Liu2024_geometric_excitations_fqh}%
		\BibitemOpen
		\bibfield  {author} {\bibinfo {author} {\bibfnamefont {Y.}~\bibnamefont
				{Liu}}, \bibinfo {author} {\bibfnamefont {T.}~\bibnamefont {Zhao}},\ and\
			\bibinfo {author} {\bibfnamefont {T.}~\bibnamefont {Xiang}},\ }\href
		{https://doi.org/10.1103/PhysRevB.110.195137} {\bibfield  {journal} {\bibinfo
				{journal} {Phys. Rev. B}\ }\textbf {\bibinfo {volume} {110}},\ \bibinfo
			{pages} {195137} (\bibinfo {year} {2024})}\BibitemShut {NoStop}%
		\bibitem [{\citenamefont {Kukushkin}\ \emph {et~al.}(2009)\citenamefont
			{Kukushkin}, \citenamefont {Smet}, \citenamefont {Scarola}, \citenamefont
			{Umansky},\ and\ \citenamefont {von Klitzing}}]{Igor2009_dispersion_FQH}%
		\BibitemOpen
		\bibfield  {author} {\bibinfo {author} {\bibfnamefont {I.~V.}\ \bibnamefont
				{Kukushkin}}, \bibinfo {author} {\bibfnamefont {J.~H.}\ \bibnamefont {Smet}},
			\bibinfo {author} {\bibfnamefont {V.~W.}\ \bibnamefont {Scarola}}, \bibinfo
			{author} {\bibfnamefont {V.}~\bibnamefont {Umansky}},\ and\ \bibinfo {author}
			{\bibfnamefont {K.}~\bibnamefont {von Klitzing}},\ }\href
		{https://doi.org/10.1126/science.1171472} {\bibfield  {journal} {\bibinfo
				{journal} {Science}\ }\textbf {\bibinfo {volume} {324}},\ \bibinfo {pages}
			{1044} (\bibinfo {year} {2009})},\ \Eprint
		{https://arxiv.org/abs/https://www.science.org/doi/pdf/10.1126/science.1171472}
		{https://www.science.org/doi/pdf/10.1126/science.1171472} \BibitemShut
		{NoStop}%
		\bibitem [{\citenamefont {Liang}\ \emph {et~al.}(2024)\citenamefont {Liang},
			\citenamefont {Liu}, \citenamefont {Yang}, \citenamefont {Huang},
			\citenamefont {Wurstbauer}, \citenamefont {Dean}, \citenamefont {West},
			\citenamefont {Pfeiffer}, \citenamefont {Du},\ and\ \citenamefont
			{Pinczuk}}]{Liang2024_chiral_graviton}%
		\BibitemOpen
		\bibfield  {author} {\bibinfo {author} {\bibfnamefont {J.}~\bibnamefont
				{Liang}}, \bibinfo {author} {\bibfnamefont {Z.}~\bibnamefont {Liu}}, \bibinfo
			{author} {\bibfnamefont {Z.}~\bibnamefont {Yang}}, \bibinfo {author}
			{\bibfnamefont {Y.}~\bibnamefont {Huang}}, \bibinfo {author} {\bibfnamefont
				{U.}~\bibnamefont {Wurstbauer}}, \bibinfo {author} {\bibfnamefont {C.~R.}\
				\bibnamefont {Dean}}, \bibinfo {author} {\bibfnamefont {K.~W.}\ \bibnamefont
				{West}}, \bibinfo {author} {\bibfnamefont {L.~N.}\ \bibnamefont {Pfeiffer}},
			\bibinfo {author} {\bibfnamefont {L.}~\bibnamefont {Du}},\ and\ \bibinfo
			{author} {\bibfnamefont {A.}~\bibnamefont {Pinczuk}},\ }\href
		{https://doi.org/10.1038/s41586-024-07201-w} {\bibfield  {journal} {\bibinfo
				{journal} {Nature}\ }\textbf {\bibinfo {volume} {628}},\ \bibinfo {pages}
			{78} (\bibinfo {year} {2024})}\BibitemShut {NoStop}%
		\bibitem [{\citenamefont {Repellin}\ \emph {et~al.}(2014)\citenamefont
			{Repellin}, \citenamefont {Neupert}, \citenamefont
			{Papi\ifmmode~\acute{c}\else \'{c}\fi{}},\ and\ \citenamefont
			{Regnault}}]{Repellin2014_SMA_FCI}%
		\BibitemOpen
		\bibfield  {author} {\bibinfo {author} {\bibfnamefont {C.}~\bibnamefont
				{Repellin}}, \bibinfo {author} {\bibfnamefont {T.}~\bibnamefont {Neupert}},
			\bibinfo {author} {\bibfnamefont {Z.}~\bibnamefont
				{Papi\ifmmode~\acute{c}\else \'{c}\fi{}}},\ and\ \bibinfo {author}
			{\bibfnamefont {N.}~\bibnamefont {Regnault}},\ }\href
		{https://doi.org/10.1103/PhysRevB.90.045114} {\bibfield  {journal} {\bibinfo
				{journal} {Phys. Rev. B}\ }\textbf {\bibinfo {volume} {90}},\ \bibinfo
			{pages} {045114} (\bibinfo {year} {2014})}\BibitemShut {NoStop}%
		\bibitem [{\citenamefont {Lu}\ \emph {et~al.}(2024{\natexlab{a}})\citenamefont
			{Lu}, \citenamefont {Chen}, \citenamefont {Wu}, \citenamefont {Sun},\ and\
			\citenamefont {Meng}}]{Lu2024_fqah_neutral_excitations}%
		\BibitemOpen
		\bibfield  {author} {\bibinfo {author} {\bibfnamefont {H.}~\bibnamefont
				{Lu}}, \bibinfo {author} {\bibfnamefont {B.-B.}\ \bibnamefont {Chen}},
			\bibinfo {author} {\bibfnamefont {H.-Q.}\ \bibnamefont {Wu}}, \bibinfo
			{author} {\bibfnamefont {K.}~\bibnamefont {Sun}},\ and\ \bibinfo {author}
			{\bibfnamefont {Z.~Y.}\ \bibnamefont {Meng}},\ }\href
		{https://doi.org/10.1103/PhysRevLett.132.236502} {\bibfield  {journal}
			{\bibinfo  {journal} {Phys. Rev. Lett.}\ }\textbf {\bibinfo {volume} {132}},\
			\bibinfo {pages} {236502} (\bibinfo {year} {2024}{\natexlab{a}})}\BibitemShut
		{NoStop}%
		\bibitem [{\citenamefont {Kousa}\ \emph {et~al.}(2025)\citenamefont {Kousa},
			\citenamefont {Morales-Durán}, \citenamefont {Wolf}, \citenamefont
			{Khalaf},\ and\ \citenamefont {MacDonald}}]{Bishoy2025_magneotoroton}%
		\BibitemOpen
		\bibfield  {author} {\bibinfo {author} {\bibfnamefont {B.~M.}\ \bibnamefont
				{Kousa}}, \bibinfo {author} {\bibfnamefont {N.}~\bibnamefont
				{Morales-Durán}}, \bibinfo {author} {\bibfnamefont {T.~M.~R.}\ \bibnamefont
				{Wolf}}, \bibinfo {author} {\bibfnamefont {E.}~\bibnamefont {Khalaf}},\ and\
			\bibinfo {author} {\bibfnamefont {A.~H.}\ \bibnamefont {MacDonald}},\ }\href
		{https://arxiv.org/abs/2502.17574} {\bibinfo {title} {Theory of magnetoroton
				bands in moir\'e materials}} (\bibinfo {year} {2025}),\ \Eprint
		{https://arxiv.org/abs/2502.17574} {arXiv:2502.17574 [cond-mat.mes-hall]}
		\BibitemShut {NoStop}%
		\bibitem [{\citenamefont {Long}\ \emph
			{et~al.}(2026{\natexlab{a}})\citenamefont {Long}, \citenamefont {Lu},
			\citenamefont {Wu},\ and\ \citenamefont {Meng}}]{Long2025_spectra_FCI}%
		\BibitemOpen
		\bibfield  {author} {\bibinfo {author} {\bibfnamefont {M.}~\bibnamefont
				{Long}}, \bibinfo {author} {\bibfnamefont {H.}~\bibnamefont {Lu}}, \bibinfo
			{author} {\bibfnamefont {H.-Q.}\ \bibnamefont {Wu}},\ and\ \bibinfo {author}
			{\bibfnamefont {Z.~Y.}\ \bibnamefont {Meng}},\ }\href
		{https://doi.org/10.1103/bjrf-b8s9} {\bibfield  {journal} {\bibinfo
				{journal} {Phys. Rev. B}\ }\textbf {\bibinfo {volume} {113}},\ \bibinfo
			{pages} {L041108} (\bibinfo {year} {2026}{\natexlab{a}})}\BibitemShut
		{NoStop}%
		\bibitem [{\citenamefont {Wang}\ \emph {et~al.}(2025)\citenamefont {Wang},
			\citenamefont {Huxford}, \citenamefont {Nguyen}, \citenamefont {Ji},
			\citenamefont {Kim},\ and\ \citenamefont
			{Yang}}]{Wang2025_geometric_excitations_moire}%
		\BibitemOpen
		\bibfield  {author} {\bibinfo {author} {\bibfnamefont {Y.}~\bibnamefont
				{Wang}}, \bibinfo {author} {\bibfnamefont {J.}~\bibnamefont {Huxford}},
			\bibinfo {author} {\bibfnamefont {D.~X.}\ \bibnamefont {Nguyen}}, \bibinfo
			{author} {\bibfnamefont {G.}~\bibnamefont {Ji}}, \bibinfo {author}
			{\bibfnamefont {Y.~B.}\ \bibnamefont {Kim}},\ and\ \bibinfo {author}
			{\bibfnamefont {B.}~\bibnamefont {Yang}},\ }\href
		{https://arxiv.org/abs/2502.02640} {\bibinfo {title} {Dynamics and lifetime
				of geometric excitations in moir\'e systems}} (\bibinfo {year} {2025}),\
		\Eprint {https://arxiv.org/abs/2502.02640} {arXiv:2502.02640
			[cond-mat.str-el]} \BibitemShut {NoStop}%
		\bibitem [{\citenamefont {Shen}\ \emph {et~al.}(2025)\citenamefont {Shen},
			\citenamefont {Wang}, \citenamefont {Hu}, \citenamefont {Guo}, \citenamefont
			{Yao}, \citenamefont {Wang}, \citenamefont {Duan},\ and\ \citenamefont
			{Xu}}]{shen2025magnetorotonsmoirefractionalchern}%
		\BibitemOpen
		\bibfield  {author} {\bibinfo {author} {\bibfnamefont {X.}~\bibnamefont
				{Shen}}, \bibinfo {author} {\bibfnamefont {C.}~\bibnamefont {Wang}}, \bibinfo
			{author} {\bibfnamefont {X.}~\bibnamefont {Hu}}, \bibinfo {author}
			{\bibfnamefont {R.}~\bibnamefont {Guo}}, \bibinfo {author} {\bibfnamefont
				{H.}~\bibnamefont {Yao}}, \bibinfo {author} {\bibfnamefont {C.}~\bibnamefont
				{Wang}}, \bibinfo {author} {\bibfnamefont {W.}~\bibnamefont {Duan}},\ and\
			\bibinfo {author} {\bibfnamefont {Y.}~\bibnamefont {Xu}},\ }\href
		{https://arxiv.org/abs/2412.01211} {\bibinfo {title} {Magnetorotons in
				moir\'e fractional chern insulators}} (\bibinfo {year} {2025}),\ \Eprint
		{https://arxiv.org/abs/2412.01211} {arXiv:2412.01211 [cond-mat.str-el]}
		\BibitemShut {NoStop}%
		\bibitem [{\citenamefont {Xavier}\ \emph {et~al.}(2025)\citenamefont {Xavier},
			\citenamefont {Bacciconi}, \citenamefont {Chanda}, \citenamefont {Son},\ and\
			\citenamefont {Dalmonte}}]{Xavier2025_graviton}%
		\BibitemOpen
		\bibfield  {author} {\bibinfo {author} {\bibfnamefont {H.~B.}\ \bibnamefont
				{Xavier}}, \bibinfo {author} {\bibfnamefont {Z.}~\bibnamefont {Bacciconi}},
			\bibinfo {author} {\bibfnamefont {T.}~\bibnamefont {Chanda}}, \bibinfo
			{author} {\bibfnamefont {D.~T.}\ \bibnamefont {Son}},\ and\ \bibinfo {author}
			{\bibfnamefont {M.}~\bibnamefont {Dalmonte}},\ }\href
		{https://doi.org/10.1103/1636-kl65} {\bibfield  {journal} {\bibinfo
				{journal} {Phys. Rev. Lett.}\ }\textbf {\bibinfo {volume} {135}},\ \bibinfo
			{pages} {196501} (\bibinfo {year} {2025})}\BibitemShut {NoStop}%
		\bibitem [{\citenamefont {Long}\ \emph
			{et~al.}(2026{\natexlab{b}})\citenamefont {Long}, \citenamefont {Bacciconi},
			\citenamefont {Lu}, \citenamefont {Xavier}, \citenamefont {Meng},\ and\
			\citenamefont {Dalmonte}}]{Long2026_graviton_FCI}%
		\BibitemOpen
		\bibfield  {author} {\bibinfo {author} {\bibfnamefont {M.}~\bibnamefont
				{Long}}, \bibinfo {author} {\bibfnamefont {Z.}~\bibnamefont {Bacciconi}},
			\bibinfo {author} {\bibfnamefont {H.}~\bibnamefont {Lu}}, \bibinfo {author}
			{\bibfnamefont {H.~B.}\ \bibnamefont {Xavier}}, \bibinfo {author}
			{\bibfnamefont {Z.~Y.}\ \bibnamefont {Meng}},\ and\ \bibinfo {author}
			{\bibfnamefont {M.}~\bibnamefont {Dalmonte}},\ }\href
		{https://arxiv.org/abs/2601.05196} {\bibinfo {title} {Chiral graviton modes
				in fermionic fractional chern insulators}} (\bibinfo {year}
		{2026}{\natexlab{b}}),\ \Eprint {https://arxiv.org/abs/2601.05196}
		{arXiv:2601.05196 [cond-mat.str-el]} \BibitemShut {NoStop}%
		\bibitem [{\citenamefont {Samajdar}\ \emph {et~al.}(2019)\citenamefont
			{Samajdar}, \citenamefont {Scheurer}, \citenamefont {Chatterjee},
			\citenamefont {Guo}, \citenamefont {Xu},\ and\ \citenamefont
			{Sachdev}}]{Samajdar2019_thermal_hall_Square}%
		\BibitemOpen
		\bibfield  {author} {\bibinfo {author} {\bibfnamefont {R.}~\bibnamefont
				{Samajdar}}, \bibinfo {author} {\bibfnamefont {M.~S.}\ \bibnamefont
				{Scheurer}}, \bibinfo {author} {\bibfnamefont {S.}~\bibnamefont
				{Chatterjee}}, \bibinfo {author} {\bibfnamefont {H.}~\bibnamefont {Guo}},
			\bibinfo {author} {\bibfnamefont {C.}~\bibnamefont {Xu}},\ and\ \bibinfo
			{author} {\bibfnamefont {S.}~\bibnamefont {Sachdev}},\ }\href
		{https://doi.org/10.1038/s41567-019-0669-3} {\bibfield  {journal} {\bibinfo
				{journal} {Nature Physics}\ }\textbf {\bibinfo {volume} {15}},\ \bibinfo
			{pages} {1290} (\bibinfo {year} {2019})}\BibitemShut {NoStop}%
		\bibitem [{\citenamefont {Zhang}\ \emph
			{et~al.}(2024{\natexlab{a}})\citenamefont {Zhang}, \citenamefont {Huang},
			\citenamefont {Wu}, \citenamefont {Sheng},\ and\ \citenamefont
			{Gong}}]{Zhang2024_CSL_square}%
		\BibitemOpen
		\bibfield  {author} {\bibinfo {author} {\bibfnamefont {X.-T.}\ \bibnamefont
				{Zhang}}, \bibinfo {author} {\bibfnamefont {Y.}~\bibnamefont {Huang}},
			\bibinfo {author} {\bibfnamefont {H.-Q.}\ \bibnamefont {Wu}}, \bibinfo
			{author} {\bibfnamefont {D.~N.}\ \bibnamefont {Sheng}},\ and\ \bibinfo
			{author} {\bibfnamefont {S.-S.}\ \bibnamefont {Gong}},\ }\href
		{https://doi.org/10.1103/PhysRevB.109.125146} {\bibfield  {journal} {\bibinfo
				{journal} {Phys. Rev. B}\ }\textbf {\bibinfo {volume} {109}},\ \bibinfo
			{pages} {125146} (\bibinfo {year} {2024}{\natexlab{a}})}\BibitemShut
		{NoStop}%
		\bibitem [{\citenamefont {Yang}\ \emph {et~al.}(2024)\citenamefont {Yang},
			\citenamefont {Liu},\ and\ \citenamefont {Wang}}]{Yang2024_CSL_square}%
		\BibitemOpen
		\bibfield  {author} {\bibinfo {author} {\bibfnamefont {J.}~\bibnamefont
				{Yang}}, \bibinfo {author} {\bibfnamefont {Z.}~\bibnamefont {Liu}},\ and\
			\bibinfo {author} {\bibfnamefont {L.}~\bibnamefont {Wang}},\ }\href
		{https://doi.org/10.1103/PhysRevB.110.224404} {\bibfield  {journal} {\bibinfo
				{journal} {Phys. Rev. B}\ }\textbf {\bibinfo {volume} {110}},\ \bibinfo
			{pages} {224404} (\bibinfo {year} {2024})}\BibitemShut {NoStop}%
		\bibitem [{\citenamefont {Jin}\ \emph {et~al.}(2025)\citenamefont {Jin},
			\citenamefont {Tu},\ and\ \citenamefont {Zhang}}]{Jin2025_dirac_csl_square}%
		\BibitemOpen
		\bibfield  {author} {\bibinfo {author} {\bibfnamefont {H.-K.}\ \bibnamefont
				{Jin}}, \bibinfo {author} {\bibfnamefont {H.-H.}\ \bibnamefont {Tu}},\ and\
			\bibinfo {author} {\bibfnamefont {Y.-H.}\ \bibnamefont {Zhang}},\ }\href
		{https://doi.org/10.1103/4yrt-nsth} {\bibfield  {journal} {\bibinfo
				{journal} {Phys. Rev. B}\ }\textbf {\bibinfo {volume} {112}},\ \bibinfo
			{pages} {035159} (\bibinfo {year} {2025})}\BibitemShut {NoStop}%
		\bibitem [{sup()}]{suppl}%
		\BibitemOpen
		\href@noop {} {\bibinfo  {journal} {Additional results in the Supplemental
				Information}\ }\BibitemShut {NoStop}%
		\bibitem [{\citenamefont {Halimeh}\ and\ \citenamefont
			{Punk}(2016)}]{Halimeh2016_dsf_spin_csl}%
		\BibitemOpen
		\bibfield  {journal} {  }\bibfield  {author} {\bibinfo {author} {\bibfnamefont
				{J.~C.}\ \bibnamefont {Halimeh}}\ and\ \bibinfo {author} {\bibfnamefont
				{M.}~\bibnamefont {Punk}},\ }\href
		{https://doi.org/10.1103/PhysRevB.94.104413} {\bibfield  {journal} {\bibinfo
				{journal} {Phys. Rev. B}\ }\textbf {\bibinfo {volume} {94}},\ \bibinfo
			{pages} {104413} (\bibinfo {year} {2016})}\BibitemShut {NoStop}%
		\bibitem [{\citenamefont {Willsher}\ and\ \citenamefont
			{Knolle}(2025)}]{Josef2025_spin_liquid_DSSF}%
		\BibitemOpen
		\bibfield  {author} {\bibinfo {author} {\bibfnamefont {J.}~\bibnamefont
				{Willsher}}\ and\ \bibinfo {author} {\bibfnamefont {J.}~\bibnamefont
				{Knolle}},\ }\href {https://arxiv.org/abs/2503.13831} {\bibinfo {title}
			{Dynamics and stability of u(1) spin liquids beyond mean-field theory:
				Triangular-lattice $j_1$-$j_2$ heisenberg model}} (\bibinfo {year} {2025}),\
		\Eprint {https://arxiv.org/abs/2503.13831} {arXiv:2503.13831
			[cond-mat.str-el]} \BibitemShut {NoStop}%
		\bibitem [{\citenamefont {Yang}(2025)}]{Yang2025_geometric_fluctuation_FQH}%
		\BibitemOpen
		\bibfield  {author} {\bibinfo {author} {\bibfnamefont {B.}~\bibnamefont
				{Yang}},\ }\href {https://doi.org/10.1103/rxy9-4dr8} {\bibfield  {journal}
			{\bibinfo  {journal} {Phys. Rev. B}\ }\textbf {\bibinfo {volume} {112}},\
			\bibinfo {pages} {075132} (\bibinfo {year} {2025})}\BibitemShut {NoStop}%
		\bibitem [{ove()}]{overlap_size}%
		\BibitemOpen
		\href@noop {} {\bibinfo  {journal} {For cross-sector overlap estimates, the
				$6\times6$ cluster is not included due to the larger Hilbert-space
				dimension}\ }\BibitemShut {NoStop}%
		\bibitem [{\citenamefont
			{Haldane}(2011{\natexlab{b}})}]{Haldane2011_geometrical_description}%
		\BibitemOpen
		\bibfield  {journal} {  }\bibfield  {author} {\bibinfo {author} {\bibfnamefont
				{F.~D.~M.}\ \bibnamefont {Haldane}},\ }\href
		{https://doi.org/10.1103/PhysRevLett.107.116801} {\bibfield  {journal}
			{\bibinfo  {journal} {Phys. Rev. Lett.}\ }\textbf {\bibinfo {volume} {107}},\
			\bibinfo {pages} {116801} (\bibinfo {year} {2011}{\natexlab{b}})}\BibitemShut
		{NoStop}%
		\bibitem [{\citenamefont {Yao}\ and\ \citenamefont
			{Kivelson}(2007)}]{Yao2007_nonabelian_csl}%
		\BibitemOpen
		\bibfield  {author} {\bibinfo {author} {\bibfnamefont {H.}~\bibnamefont
				{Yao}}\ and\ \bibinfo {author} {\bibfnamefont {S.~A.}\ \bibnamefont
				{Kivelson}},\ }\href {https://doi.org/10.1103/PhysRevLett.99.247203}
		{\bibfield  {journal} {\bibinfo  {journal} {Phys. Rev. Lett.}\ }\textbf
			{\bibinfo {volume} {99}},\ \bibinfo {pages} {247203} (\bibinfo {year}
			{2007})}\BibitemShut {NoStop}%
		\bibitem [{\citenamefont {Greiter}\ and\ \citenamefont
			{Thomale}(2009)}]{Greiter2009_nonabelian_csl}%
		\BibitemOpen
		\bibfield  {author} {\bibinfo {author} {\bibfnamefont {M.}~\bibnamefont
				{Greiter}}\ and\ \bibinfo {author} {\bibfnamefont {R.}~\bibnamefont
				{Thomale}},\ }\href {https://doi.org/10.1103/PhysRevLett.102.207203}
		{\bibfield  {journal} {\bibinfo  {journal} {Phys. Rev. Lett.}\ }\textbf
			{\bibinfo {volume} {102}},\ \bibinfo {pages} {207203} (\bibinfo {year}
			{2009})}\BibitemShut {NoStop}%
		\bibitem [{\citenamefont {Liu}\ \emph {et~al.}(2018{\natexlab{b}})\citenamefont
			{Liu}, \citenamefont {Tu}, \citenamefont {Wu}, \citenamefont {He},
			\citenamefont {Liu}, \citenamefont {Zhou},\ and\ \citenamefont
			{Ng}}]{Liu2018_CSL_nonabelian}%
		\BibitemOpen
		\bibfield  {author} {\bibinfo {author} {\bibfnamefont {Z.-X.}\ \bibnamefont
				{Liu}}, \bibinfo {author} {\bibfnamefont {H.-H.}\ \bibnamefont {Tu}},
			\bibinfo {author} {\bibfnamefont {Y.-H.}\ \bibnamefont {Wu}}, \bibinfo
			{author} {\bibfnamefont {R.-Q.}\ \bibnamefont {He}}, \bibinfo {author}
			{\bibfnamefont {X.-J.}\ \bibnamefont {Liu}}, \bibinfo {author} {\bibfnamefont
				{Y.}~\bibnamefont {Zhou}},\ and\ \bibinfo {author} {\bibfnamefont {T.-K.}\
				\bibnamefont {Ng}},\ }\href {https://doi.org/10.1103/PhysRevB.97.195158}
		{\bibfield  {journal} {\bibinfo  {journal} {Phys. Rev. B}\ }\textbf {\bibinfo
				{volume} {97}},\ \bibinfo {pages} {195158} (\bibinfo {year}
			{2018}{\natexlab{b}})}\BibitemShut {NoStop}%
		\bibitem [{\citenamefont {Chen}\ \emph {et~al.}(2018)\citenamefont {Chen},
			\citenamefont {Vanderstraeten}, \citenamefont {Capponi},\ and\ \citenamefont
			{Poilblanc}}]{Chen2018_nonabelian_csl}%
		\BibitemOpen
		\bibfield  {author} {\bibinfo {author} {\bibfnamefont {J.-Y.}\ \bibnamefont
				{Chen}}, \bibinfo {author} {\bibfnamefont {L.}~\bibnamefont
				{Vanderstraeten}}, \bibinfo {author} {\bibfnamefont {S.}~\bibnamefont
				{Capponi}},\ and\ \bibinfo {author} {\bibfnamefont {D.}~\bibnamefont
				{Poilblanc}},\ }\href {https://doi.org/10.1103/PhysRevB.98.184409} {\bibfield
			{journal} {\bibinfo  {journal} {Phys. Rev. B}\ }\textbf {\bibinfo {volume}
				{98}},\ \bibinfo {pages} {184409} (\bibinfo {year} {2018})}\BibitemShut
		{NoStop}%
		\bibitem [{\citenamefont {Huang}\ \emph {et~al.}(2022)\citenamefont {Huang},
			\citenamefont {Zhu}, \citenamefont {Gong}, \citenamefont {Jiang},\ and\
			\citenamefont {Sheng}}]{Huang2022_nonabelian_csl}%
		\BibitemOpen
		\bibfield  {author} {\bibinfo {author} {\bibfnamefont {Y.}~\bibnamefont
				{Huang}}, \bibinfo {author} {\bibfnamefont {W.}~\bibnamefont {Zhu}}, \bibinfo
			{author} {\bibfnamefont {S.-S.}\ \bibnamefont {Gong}}, \bibinfo {author}
			{\bibfnamefont {H.-C.}\ \bibnamefont {Jiang}},\ and\ \bibinfo {author}
			{\bibfnamefont {D.~N.}\ \bibnamefont {Sheng}},\ }\href
		{https://doi.org/10.1103/PhysRevB.105.155104} {\bibfield  {journal} {\bibinfo
				{journal} {Phys. Rev. B}\ }\textbf {\bibinfo {volume} {105}},\ \bibinfo
			{pages} {155104} (\bibinfo {year} {2022})}\BibitemShut {NoStop}%
		\bibitem [{\citenamefont {Zhang}\ \emph
			{et~al.}(2024{\natexlab{b}})\citenamefont {Zhang}, \citenamefont {Gao},\ and\
			\citenamefont {Chen}}]{Zhang2024_thermal_hall}%
		\BibitemOpen
		\bibfield  {author} {\bibinfo {author} {\bibfnamefont {X.-T.}\ \bibnamefont
				{Zhang}}, \bibinfo {author} {\bibfnamefont {Y.~H.}\ \bibnamefont {Gao}},\
			and\ \bibinfo {author} {\bibfnamefont {G.}~\bibnamefont {Chen}},\ }\href
		{https://doi.org/https://doi.org/10.1016/j.physrep.2024.03.004} {\bibfield
			{journal} {\bibinfo  {journal} {Physics Reports}\ }\textbf {\bibinfo {volume}
				{1070}},\ \bibinfo {pages} {1} (\bibinfo {year} {2024}{\natexlab{b}})},\
		\bibinfo {note} {thermal Hall effects in quantum magnets}\BibitemShut
		{NoStop}%
		\bibitem [{\citenamefont {Sandvik}\ \emph {et~al.}(1998)\citenamefont
			{Sandvik}, \citenamefont {Capponi}, \citenamefont {Poilblanc},\ and\
			\citenamefont {Dagotto}}]{Sandvik1998_B1g_raman}%
		\BibitemOpen
		\bibfield  {author} {\bibinfo {author} {\bibfnamefont {A.~W.}\ \bibnamefont
				{Sandvik}}, \bibinfo {author} {\bibfnamefont {S.}~\bibnamefont {Capponi}},
			\bibinfo {author} {\bibfnamefont {D.}~\bibnamefont {Poilblanc}},\ and\
			\bibinfo {author} {\bibfnamefont {E.}~\bibnamefont {Dagotto}},\ }\href
		{https://doi.org/10.1103/PhysRevB.57.8478} {\bibfield  {journal} {\bibinfo
				{journal} {Phys. Rev. B}\ }\textbf {\bibinfo {volume} {57}},\ \bibinfo
			{pages} {8478} (\bibinfo {year} {1998})}\BibitemShut {NoStop}%
		\bibitem [{\citenamefont {Devereaux}\ and\ \citenamefont
			{Hackl}(2007)}]{Devereaux2007_ILS}%
		\BibitemOpen
		\bibfield  {author} {\bibinfo {author} {\bibfnamefont {T.~P.}\ \bibnamefont
				{Devereaux}}\ and\ \bibinfo {author} {\bibfnamefont {R.}~\bibnamefont
				{Hackl}},\ }\href {https://doi.org/10.1103/RevModPhys.79.175} {\bibfield
			{journal} {\bibinfo  {journal} {Rev. Mod. Phys.}\ }\textbf {\bibinfo {volume}
				{79}},\ \bibinfo {pages} {175} (\bibinfo {year} {2007})}\BibitemShut
		{NoStop}%
		\bibitem [{\citenamefont {Koller}\ \emph {et~al.}(2025)\citenamefont {Koller},
			\citenamefont {Leeb}, \citenamefont {Perkins},\ and\ \citenamefont
			{Knolle}}]{koller2025ramancirculardichroism}%
		\BibitemOpen
		\bibfield  {author} {\bibinfo {author} {\bibfnamefont {E.}~\bibnamefont
				{Koller}}, \bibinfo {author} {\bibfnamefont {V.}~\bibnamefont {Leeb}},
			\bibinfo {author} {\bibfnamefont {N.~B.}\ \bibnamefont {Perkins}},\ and\
			\bibinfo {author} {\bibfnamefont {J.}~\bibnamefont {Knolle}},\ }\href
		{https://arxiv.org/abs/2503.14091} {\bibinfo {title} {Raman circular
				dichroism and quantum geometry of chiral quantum spin liquids}} (\bibinfo
		{year} {2025}),\ \Eprint {https://arxiv.org/abs/2503.14091} {arXiv:2503.14091
			[cond-mat.str-el]} \BibitemShut {NoStop}%
		\bibitem [{\citenamefont {Ament}\ \emph {et~al.}(2011)\citenamefont {Ament},
			\citenamefont {van Veenendaal}, \citenamefont {Devereaux}, \citenamefont
			{Hill},\ and\ \citenamefont {van~den Brink}}]{Ament2011_RIXS}%
		\BibitemOpen
		\bibfield  {author} {\bibinfo {author} {\bibfnamefont {L.~J.~P.}\
				\bibnamefont {Ament}}, \bibinfo {author} {\bibfnamefont {M.}~\bibnamefont
				{van Veenendaal}}, \bibinfo {author} {\bibfnamefont {T.~P.}\ \bibnamefont
				{Devereaux}}, \bibinfo {author} {\bibfnamefont {J.~P.}\ \bibnamefont
				{Hill}},\ and\ \bibinfo {author} {\bibfnamefont {J.}~\bibnamefont {van~den
					Brink}},\ }\href {https://doi.org/10.1103/RevModPhys.83.705} {\bibfield
			{journal} {\bibinfo  {journal} {Rev. Mod. Phys.}\ }\textbf {\bibinfo {volume}
				{83}},\ \bibinfo {pages} {705} (\bibinfo {year} {2011})}\BibitemShut
		{NoStop}%
		\bibitem [{\citenamefont {Sikora}\ \emph {et~al.}(2010)\citenamefont {Sikora},
			\citenamefont {Juhin}, \citenamefont {Weng}, \citenamefont {Sainctavit},
			\citenamefont {Detlefs}, \citenamefont {de~Groot},\ and\ \citenamefont
			{Glatzel}}]{Sikora2010_Kedge}%
		\BibitemOpen
		\bibfield  {author} {\bibinfo {author} {\bibfnamefont {M.}~\bibnamefont
				{Sikora}}, \bibinfo {author} {\bibfnamefont {A.}~\bibnamefont {Juhin}},
			\bibinfo {author} {\bibfnamefont {T.-C.}\ \bibnamefont {Weng}}, \bibinfo
			{author} {\bibfnamefont {P.}~\bibnamefont {Sainctavit}}, \bibinfo {author}
			{\bibfnamefont {C.}~\bibnamefont {Detlefs}}, \bibinfo {author} {\bibfnamefont
				{F.}~\bibnamefont {de~Groot}},\ and\ \bibinfo {author} {\bibfnamefont
				{P.}~\bibnamefont {Glatzel}},\ }\href
		{https://doi.org/10.1103/PhysRevLett.105.037202} {\bibfield  {journal}
			{\bibinfo  {journal} {Phys. Rev. Lett.}\ }\textbf {\bibinfo {volume} {105}},\
			\bibinfo {pages} {037202} (\bibinfo {year} {2010})}\BibitemShut {NoStop}%
		\bibitem [{\citenamefont {Nie}\ \emph {et~al.}(2017)\citenamefont {Nie},
			\citenamefont {Maharaj}, \citenamefont {Fradkin},\ and\ \citenamefont
			{Kivelson}}]{Nie2017vestigial}%
		\BibitemOpen
		\bibfield  {author} {\bibinfo {author} {\bibfnamefont {L.}~\bibnamefont
				{Nie}}, \bibinfo {author} {\bibfnamefont {A.~V.}\ \bibnamefont {Maharaj}},
			\bibinfo {author} {\bibfnamefont {E.}~\bibnamefont {Fradkin}},\ and\ \bibinfo
			{author} {\bibfnamefont {S.~A.}\ \bibnamefont {Kivelson}},\ }\href
		{https://doi.org/10.1103/PhysRevB.96.085142} {\bibfield  {journal} {\bibinfo
				{journal} {Phys. Rev. B}\ }\textbf {\bibinfo {volume} {96}},\ \bibinfo
			{pages} {085142} (\bibinfo {year} {2017})}\BibitemShut {NoStop}%
		\bibitem [{\citenamefont {Fernandes}\ \emph {et~al.}(2019)\citenamefont
			{Fernandes}, \citenamefont {Orth},\ and\ \citenamefont
			{Schmalian}}]{Fernandes2019Vestigial}%
		\BibitemOpen
		\bibfield  {author} {\bibinfo {author} {\bibfnamefont {R.~M.}\ \bibnamefont
				{Fernandes}}, \bibinfo {author} {\bibfnamefont {P.~P.}\ \bibnamefont
				{Orth}},\ and\ \bibinfo {author} {\bibfnamefont {J.}~\bibnamefont
				{Schmalian}},\ }\href
		{https://doi.org/https://doi.org/10.1146/annurev-conmatphys-031218-013200}
		{\bibfield  {journal} {\bibinfo  {journal} {Annual Review of Condensed Matter
					Physics}\ }\textbf {\bibinfo {volume} {10}},\ \bibinfo {pages} {133}
			(\bibinfo {year} {2019})}\BibitemShut {NoStop}%
		\bibitem [{\citenamefont {Lu}\ \emph {et~al.}(2026)\citenamefont {Lu},
			\citenamefont {Wu}, \citenamefont {Chen},\ and\ \citenamefont
			{Meng}}]{Lu2024_vestigial_gbdw}%
		\BibitemOpen
		\bibfield  {author} {\bibinfo {author} {\bibfnamefont {H.}~\bibnamefont
				{Lu}}, \bibinfo {author} {\bibfnamefont {H.-Q.}\ \bibnamefont {Wu}}, \bibinfo
			{author} {\bibfnamefont {B.-B.}\ \bibnamefont {Chen}},\ and\ \bibinfo
			{author} {\bibfnamefont {Z.~Y.}\ \bibnamefont {Meng}},\ }\href
		{https://doi.org/10.1103/1bhm-9pk4} {\bibfield  {journal} {\bibinfo
				{journal} {Phys. Rev. B}\ }\textbf {\bibinfo {volume} {113}},\ \bibinfo
			{pages} {035141} (\bibinfo {year} {2026})}\BibitemShut {NoStop}%
		\bibitem [{\citenamefont {Regnault}\ \emph {et~al.}(2017)\citenamefont
			{Regnault}, \citenamefont {Maciejko}, \citenamefont {Kivelson},\ and\
			\citenamefont {Sondhi}}]{Regnault2017_fqh_nematic}%
		\BibitemOpen
		\bibfield  {author} {\bibinfo {author} {\bibfnamefont {N.}~\bibnamefont
				{Regnault}}, \bibinfo {author} {\bibfnamefont {J.}~\bibnamefont {Maciejko}},
			\bibinfo {author} {\bibfnamefont {S.~A.}\ \bibnamefont {Kivelson}},\ and\
			\bibinfo {author} {\bibfnamefont {S.~L.}\ \bibnamefont {Sondhi}},\ }\href
		{https://doi.org/10.1103/PhysRevB.96.035150} {\bibfield  {journal} {\bibinfo
				{journal} {Phys. Rev. B}\ }\textbf {\bibinfo {volume} {96}},\ \bibinfo
			{pages} {035150} (\bibinfo {year} {2017})}\BibitemShut {NoStop}%
		\bibitem [{\citenamefont {Yang}(2020)}]{Yang2020_nematic_fqh}%
		\BibitemOpen
		\bibfield  {author} {\bibinfo {author} {\bibfnamefont {B.}~\bibnamefont
				{Yang}},\ }\href {https://doi.org/10.1103/PhysRevResearch.2.033362}
		{\bibfield  {journal} {\bibinfo  {journal} {Phys. Rev. Res.}\ }\textbf
			{\bibinfo {volume} {2}},\ \bibinfo {pages} {033362} (\bibinfo {year}
			{2020})}\BibitemShut {NoStop}%
		\bibitem [{\citenamefont {Pu}\ \emph {et~al.}(2024)\citenamefont {Pu},
			\citenamefont {Balram}, \citenamefont {Taylor}, \citenamefont {Fradkin},\
			and\ \citenamefont {Papi\ifmmode~\acute{c}\else
				\'{c}\fi{}}}]{Pu2024_fqh_nematics}%
		\BibitemOpen
		\bibfield  {author} {\bibinfo {author} {\bibfnamefont {S.}~\bibnamefont
				{Pu}}, \bibinfo {author} {\bibfnamefont {A.~C.}\ \bibnamefont {Balram}},
			\bibinfo {author} {\bibfnamefont {J.}~\bibnamefont {Taylor}}, \bibinfo
			{author} {\bibfnamefont {E.}~\bibnamefont {Fradkin}},\ and\ \bibinfo {author}
			{\bibfnamefont {Z.}~\bibnamefont {Papi\ifmmode~\acute{c}\else \'{c}\fi{}}},\
		}\href {https://doi.org/10.1103/PhysRevLett.132.236503} {\bibfield  {journal}
			{\bibinfo  {journal} {Phys. Rev. Lett.}\ }\textbf {\bibinfo {volume} {132}},\
			\bibinfo {pages} {236503} (\bibinfo {year} {2024})}\BibitemShut {NoStop}%
		\bibitem [{\citenamefont {Lu}\ \emph {et~al.}(2024{\natexlab{b}})\citenamefont
			{Lu}, \citenamefont {Wu}, \citenamefont {Chen}, \citenamefont {Sun},\ and\
			\citenamefont {Yang~Meng}}]{Lu_2024_fqahs}%
		\BibitemOpen
		\bibfield  {author} {\bibinfo {author} {\bibfnamefont {H.}~\bibnamefont
				{Lu}}, \bibinfo {author} {\bibfnamefont {H.-Q.}\ \bibnamefont {Wu}}, \bibinfo
			{author} {\bibfnamefont {B.-B.}\ \bibnamefont {Chen}}, \bibinfo {author}
			{\bibfnamefont {K.}~\bibnamefont {Sun}},\ and\ \bibinfo {author}
			{\bibfnamefont {Z.}~\bibnamefont {Yang~Meng}},\ }\href
		{https://doi.org/10.1088/1361-6633/ad7640} {\bibfield  {journal} {\bibinfo
				{journal} {Reports on Progress in Physics}\ }\textbf {\bibinfo {volume}
				{87}},\ \bibinfo {pages} {108003} (\bibinfo {year}
			{2024}{\natexlab{b}})}\BibitemShut {NoStop}%
		\bibitem [{\citenamefont {Lu}\ \emph {et~al.}(2025)\citenamefont {Lu},
			\citenamefont {Wu}, \citenamefont {Chen},\ and\ \citenamefont
			{Meng}}]{Lu2025_fqah_fqahs}%
		\BibitemOpen
		\bibfield  {author} {\bibinfo {author} {\bibfnamefont {H.}~\bibnamefont
				{Lu}}, \bibinfo {author} {\bibfnamefont {H.-Q.}\ \bibnamefont {Wu}}, \bibinfo
			{author} {\bibfnamefont {B.-B.}\ \bibnamefont {Chen}},\ and\ \bibinfo
			{author} {\bibfnamefont {Z.~Y.}\ \bibnamefont {Meng}},\ }\bibfield  {journal}
		{\bibinfo  {journal} {Newton}\ }\href
		{https://doi.org/10.1016/j.newton.2025.100300} {10.1016/j.newton.2025.100300}
		(\bibinfo {year} {2025})\BibitemShut {NoStop}%
		\bibitem [{\citenamefont {Sandvik}(2010)}]{Sandvik2010_ED}%
		\BibitemOpen
		\bibfield  {author} {\bibinfo {author} {\bibfnamefont {A.~W.}\ \bibnamefont
				{Sandvik}},\ }\href {https://doi.org/10.1063/1.3518900} {\bibfield  {journal}
			{\bibinfo  {journal} {AIP Conference Proceedings}\ }\textbf {\bibinfo
				{volume} {1297}},\ \bibinfo {pages} {135} (\bibinfo {year} {2010})},\ \Eprint
		{https://arxiv.org/abs/https://pubs.aip.org/aip/acp/article-pdf/1297/1/135/11407753/135_1_online.pdf}{}
		\BibitemShut {NoStop}%
		\bibitem [{\citenamefont {White}(1992)}]{White1992_dmrg}%
		\BibitemOpen
		\bibfield  {author} {\bibinfo {author} {\bibfnamefont {S.~R.}\ \bibnamefont
				{White}},\ }\href {https://doi.org/10.1103/PhysRevLett.69.2863} {\bibfield
			{journal} {\bibinfo  {journal} {Phys. Rev. Lett.}\ }\textbf {\bibinfo
				{volume} {69}},\ \bibinfo {pages} {2863} (\bibinfo {year}
			{1992})}\BibitemShut {NoStop}%
		\bibitem [{\citenamefont {Haegeman}\ \emph {et~al.}(2011)\citenamefont
			{Haegeman}, \citenamefont {Cirac}, \citenamefont {Osborne}, \citenamefont
			{Pi\ifmmode~\check{z}\else \v{z}\fi{}orn}, \citenamefont {Verschelde},\ and\
			\citenamefont {Verstraete}}]{Haegeman2011_tdvp}%
		\BibitemOpen
		\bibfield  {author} {\bibinfo {author} {\bibfnamefont {J.}~\bibnamefont
				{Haegeman}}, \bibinfo {author} {\bibfnamefont {J.~I.}\ \bibnamefont {Cirac}},
			\bibinfo {author} {\bibfnamefont {T.~J.}\ \bibnamefont {Osborne}}, \bibinfo
			{author} {\bibfnamefont {I.}~\bibnamefont {Pi\ifmmode~\check{z}\else
					\v{z}\fi{}orn}}, \bibinfo {author} {\bibfnamefont {H.}~\bibnamefont
				{Verschelde}},\ and\ \bibinfo {author} {\bibfnamefont {F.}~\bibnamefont
				{Verstraete}},\ }\href {https://doi.org/10.1103/PhysRevLett.107.070601}
		{\bibfield  {journal} {\bibinfo  {journal} {Phys. Rev. Lett.}\ }\textbf
			{\bibinfo {volume} {107}},\ \bibinfo {pages} {070601} (\bibinfo {year}
			{2011})}\BibitemShut {NoStop}%
		\bibitem [{\citenamefont {Weinberg}\ and\ \citenamefont
			{Bukov}(2019)}]{QuSpin2019}%
		\BibitemOpen
		\bibfield  {author} {\bibinfo {author} {\bibfnamefont {P.}~\bibnamefont
				{Weinberg}}\ and\ \bibinfo {author} {\bibfnamefont {M.}~\bibnamefont
				{Bukov}},\ }\href {https://doi.org/10.21468/SciPostPhys.7.2.020} {\bibfield
			{journal} {\bibinfo  {journal} {SciPost Phys.}\ }\textbf {\bibinfo {volume}
				{7}},\ \bibinfo {pages} {020} (\bibinfo {year} {2019})}\BibitemShut {NoStop}%
		\bibitem [{\citenamefont {Hauschild}\ \emph {et~al.}(2024)\citenamefont
			{Hauschild}, \citenamefont {Unfried}, \citenamefont {Anand}, \citenamefont
			{Andrews}, \citenamefont {Bintz}, \citenamefont {Borla}, \citenamefont
			{Divic}, \citenamefont {Drescher}, \citenamefont {Geiger}, \citenamefont
			{Hefel}, \citenamefont {Hémery}, \citenamefont {Kadow}, \citenamefont
			{Kemp}, \citenamefont {Kirchner}, \citenamefont {Liu}, \citenamefont
			{Möller}, \citenamefont {Parker}, \citenamefont {Rader}, \citenamefont
			{Romen}, \citenamefont {Scalet}, \citenamefont {Schoonderwoerd},
			\citenamefont {Schulz}, \citenamefont {Soejima}, \citenamefont {Thoma},
			\citenamefont {Wu}, \citenamefont {Zechmann}, \citenamefont {Zweng},
			\citenamefont {Mong}, \citenamefont {Zaletel},\ and\ \citenamefont
			{Pollmann}}]{Johannes_2024_tenpy}%
		\BibitemOpen
		\bibfield  {author} {\bibinfo {author} {\bibfnamefont {J.}~\bibnamefont
				{Hauschild}}, \bibinfo {author} {\bibfnamefont {J.}~\bibnamefont {Unfried}},
			\bibinfo {author} {\bibfnamefont {S.}~\bibnamefont {Anand}}, \bibinfo
			{author} {\bibfnamefont {B.}~\bibnamefont {Andrews}}, \bibinfo {author}
			{\bibfnamefont {M.}~\bibnamefont {Bintz}}, \bibinfo {author} {\bibfnamefont
				{U.}~\bibnamefont {Borla}}, \bibinfo {author} {\bibfnamefont
				{S.}~\bibnamefont {Divic}}, \bibinfo {author} {\bibfnamefont
				{M.}~\bibnamefont {Drescher}}, \bibinfo {author} {\bibfnamefont
				{J.}~\bibnamefont {Geiger}}, \bibinfo {author} {\bibfnamefont
				{M.}~\bibnamefont {Hefel}}, \bibinfo {author} {\bibfnamefont
				{K.}~\bibnamefont {Hémery}}, \bibinfo {author} {\bibfnamefont
				{W.}~\bibnamefont {Kadow}}, \bibinfo {author} {\bibfnamefont
				{J.}~\bibnamefont {Kemp}}, \bibinfo {author} {\bibfnamefont {N.}~\bibnamefont
				{Kirchner}}, \bibinfo {author} {\bibfnamefont {V.~S.}\ \bibnamefont {Liu}},
			\bibinfo {author} {\bibfnamefont {G.}~\bibnamefont {Möller}}, \bibinfo
			{author} {\bibfnamefont {D.}~\bibnamefont {Parker}}, \bibinfo {author}
			{\bibfnamefont {M.}~\bibnamefont {Rader}}, \bibinfo {author} {\bibfnamefont
				{A.}~\bibnamefont {Romen}}, \bibinfo {author} {\bibfnamefont
				{S.}~\bibnamefont {Scalet}}, \bibinfo {author} {\bibfnamefont
				{L.}~\bibnamefont {Schoonderwoerd}}, \bibinfo {author} {\bibfnamefont
				{M.}~\bibnamefont {Schulz}}, \bibinfo {author} {\bibfnamefont
				{T.}~\bibnamefont {Soejima}}, \bibinfo {author} {\bibfnamefont
				{P.}~\bibnamefont {Thoma}}, \bibinfo {author} {\bibfnamefont
				{Y.}~\bibnamefont {Wu}}, \bibinfo {author} {\bibfnamefont {P.}~\bibnamefont
				{Zechmann}}, \bibinfo {author} {\bibfnamefont {L.}~\bibnamefont {Zweng}},
			\bibinfo {author} {\bibfnamefont {R.~S.~K.}\ \bibnamefont {Mong}}, \bibinfo
			{author} {\bibfnamefont {M.~P.}\ \bibnamefont {Zaletel}},\ and\ \bibinfo
			{author} {\bibfnamefont {F.}~\bibnamefont {Pollmann}},\ }\href
		{https://doi.org/10.21468/SciPostPhysCodeb.41} {\bibfield  {journal}
			{\bibinfo  {journal} {SciPost Phys. Codebases}\ ,\ \bibinfo {pages} {41}}
			(\bibinfo {year} {2024})}\BibitemShut {NoStop}%
		\bibitem [{\citenamefont {Li}\ and\ \citenamefont
			{Haldane}(2008)}]{LiHaldane2008}%
		\BibitemOpen
		\bibfield  {author} {\bibinfo {author} {\bibfnamefont {H.}~\bibnamefont
				{Li}}\ and\ \bibinfo {author} {\bibfnamefont {F.~D.~M.}\ \bibnamefont
				{Haldane}},\ }\href {https://doi.org/10.1103/PhysRevLett.101.010504}
		{\bibfield  {journal} {\bibinfo  {journal} {Phys. Rev. Lett.}\ }\textbf
			{\bibinfo {volume} {101}},\ \bibinfo {pages} {010504} (\bibinfo {year}
			{2008})}\BibitemShut {NoStop}%
		\bibitem [{\citenamefont {Qi}\ \emph {et~al.}(2012)\citenamefont {Qi},
			\citenamefont {Katsura},\ and\ \citenamefont
			{Ludwig}}]{Qi2012_ES_edge_spectrum}%
		\BibitemOpen
		\bibfield  {author} {\bibinfo {author} {\bibfnamefont {X.-L.}\ \bibnamefont
				{Qi}}, \bibinfo {author} {\bibfnamefont {H.}~\bibnamefont {Katsura}},\ and\
			\bibinfo {author} {\bibfnamefont {A.~W.~W.}\ \bibnamefont {Ludwig}},\ }\href
		{https://doi.org/10.1103/PhysRevLett.108.196402} {\bibfield  {journal}
			{\bibinfo  {journal} {Phys. Rev. Lett.}\ }\textbf {\bibinfo {volume} {108}},\
			\bibinfo {pages} {196402} (\bibinfo {year} {2012})}\BibitemShut {NoStop}%
		\bibitem [{\citenamefont {Drescher}\ \emph {et~al.}(2026)\citenamefont
			{Drescher}, \citenamefont {Vanderstraeten}, \citenamefont {Moessner},
			\citenamefont {Pollmann},\ and\ \citenamefont
			{Knolle}}]{drescher2026dynamicaldimerstructurefactor}%
		\BibitemOpen
		\bibfield  {author} {\bibinfo {author} {\bibfnamefont {M.}~\bibnamefont
				{Drescher}}, \bibinfo {author} {\bibfnamefont {L.}~\bibnamefont
				{Vanderstraeten}}, \bibinfo {author} {\bibfnamefont {R.}~\bibnamefont
				{Moessner}}, \bibinfo {author} {\bibfnamefont {F.}~\bibnamefont {Pollmann}},\
			and\ \bibinfo {author} {\bibfnamefont {J.}~\bibnamefont {Knolle}},\ }\href
		{https://arxiv.org/abs/2604.24868} {\bibinfo {title} {Dynamical dimer
				structure factor of the triangular $s=1/2$ heisenberg antiferromagnet}}
		(\bibinfo {year} {2026}),\ \Eprint {https://arxiv.org/abs/2604.24868}
		{arXiv:2604.24868 [cond-mat.str-el]} \BibitemShut {NoStop}%
	\end{thebibliography}
	%

	
\end{document}